\documentclass[a4paper,11pt]{article}
\usepackage{jheppub}
\usepackage{physics}
\usepackage{subfigure}

\usepackage{color}

\title{\boldmath
Wasserstein Space of Quantum Chaos
}

\author[a]{Koji Hashimoto,}
\author[a]{Norihiro Tanahashi}
\author[b]{ and Kentaroh Yoshida}
\affiliation[a]{Department of Physics, Kyoto University, Kyoto 606-8502, Japan}
\affiliation[b]{Graduate School of Science and Engineering, Saitama University, Saitama 338-8570, Japan}
\emailAdd{koji@scphys.kyoto-u.ac.jp}
\emailAdd{tanahashi@gauge.scphys.kyoto-u.ac.jp}
\emailAdd{kenyoshida@mail.saitama-u.ac.jp}

\abstract{
We find that the effective dimension of the Wasserstein space of energy eigenstates decreases as a quantum system becomes more chaotic. To demonstrate this, we study a quantum coupled harmonic oscillator system using Husimi Q-representations, to which Sinkhorn-regularized optimal transport is applied to construct an embedding geometry via the Gram-spectrum method. We also demonstrate that exponential OTOC growth, referred to here as quantum scrambling even in the absence of chaos, induces a folding structure in the emergent Wasserstein space, which may underlie the chaotic reduction of the Wasserstein dimension. At the separatrix (the scrambling point) of the inverted harmonic oscillator, the Wasserstein distance correctly captures the Lyapunov exponent. Furthermore, we discover that a branching structure in the Wasserstein space signals quantum scar states within the chaotic sea of phase space. Our optimal transport approach thus provides a new diagnostic for quantum chaos, quantum scrambling, quantum scars, and quantum Lyapunov exponents. The observed chaotic dimensional reduction also supports the recent conjecture \cite{hashimoto2026holography} that the Wasserstein space serves as an emergent holographic space through the manifold hypothesis, since chaoticity is a characteristic signature of black holes in holography.
}

\begin{document}
\maketitle


\clearpage
\section{Introduction}
\label{sec:intro}

The holographic principle \cite{Maldacena:1997re, hooft1993dimensional, Susskind:1994vu}, in which gravitational spacetimes emerge from lower-dimensional non-gravitating quantum systems, is a central driving force in quantum gravity research, while the mechanism by which the holographic principle works has yet to be revealed. Two of the present authors (K.H. and N.T.) recently presented a novel view \cite{hashimoto2026holography} for the problem, using the notions of optimal transport \cite{villani2008optimal} and manifold hypothesis. The proposed viewpoint of \cite{hashimoto2026holography} can be summarized as follows: The holographic ``emergent" spacetime is a dimensionally reduced Wasserstein space obtained by the optimal transport among quantum states of the boundary quantum field theory. The Wasserstein space is normally infinite dimensional as the Hilbert space of quantum field theories is infinite dimensional. The manifold hypothesis in machine learning \cite{bengio2013representation,fefferman2016testing} states that the intrinsic dimensions of data in the total space are quite small, suggesting that this could happen to the holographic principle, to find a dimensionally reduced Wasserstein space as the emergent holographic spacetime. 

In \cite{hashimoto2026holography}, interesting structure of the Wasserstein space of a single quantum harmonic oscillator was reported, to support the conjecture by supplying the basic dictionary to interpret the Wasserstein space constructed by the Wasserstein distance between quantum states as an emergent spacetime: For example, the 1-Wasserstein space attains minimal dimensionality for the distance between Husimi Q-representations, and the Lindblad harmonic oscillator produces a Wasserstein spacetime which mimics a black hole spacetime. However, what is still missing is the mechanism of the dimensional reduction of the Wasserstein space. In what kind of boundary quantum systems does the dimensional reduction of the Wasserstein space occur? Any answer to this question would, in fact, single out the quantum field theories that allow the holographic description, a la the manifold hypothesis.

To this end, we recall that quantum chaos has been an important diagnostic for holography. The Maldacena-Shenker-Stanford bound for quantum chaos \cite{maldacena2016bound} suggests that for the emergence of the holographic black hole the boundary quantum system needs maximal quantum chaos. Therefore, in this paper, we examine the conjecture from the viewpoint of quantum chaos --- we will find that the simplest quantum chaotic model, which is the coupled harmonic oscillator, exhibits dimensional reduction in its Wasserstein space made of the energy eigenstates when the chaoticity parameter increases.

What is the origin of this dimensional reduction of the Wasserstein space?
Quantum scrambling, which could serve as a necessary condition for quantum chaos, was originally thought of as a nature of black holes \cite{sekino2008fast}, thus should be present in any candidate of the boundary quantum field theories admitting a holographic description. In this paper, we look at a simple scrambling model, which is the quantum double-well system in one dimension, to discover that the Wasserstein space is folded at the scrambling energy. Since repeated scrambling may seed quantum chaos, the folding structure of the Wasserstein space could be the origin of its dimensional reduction.

In this manner, our examination of the simplest quantum chaotic model shows the dimensional reduction of the Wasserstein space obtained from the optimal transport among quantum states. The coupled harmonic oscillator is known to capture the chaotic property of Yang-Mills theory \cite{matinyan1981classical,matinyan1981stochasticity,savvidy1984classical,bir1994chaos} and the Yang-Mills theory is holographic in its strong coupling and the large-$N$ limits, thus even though our model is still a simple model, the connection to the holographic principle is there.

In the course of this journey, we encounter a strange structure of the Wasserstein space --- a branching structure. The quantum states on the branch formed in the emergent Wasserstein space of the chaotic model are well separated from the other states, suggesting that they survive if the system is heated. We look at the phase space structure of these states and find that they are quantum scar states. Quantum many-body scars \cite{turner2018weak,shiraishi2017systematic} provide a mechanism for weak violation of the eigenstate thermalization hypothesis \cite{deutsch1991quantum,srednicki1994chaos}, in which atypical eigenstates coexist with otherwise thermal eigenstates. The terminology originates from single-particle quantum scars in classically chaotic systems~\cite{heller1984bound}, and in our two-dimensional chaotic model we can spot the quantum scar as a branch structure of the Wasserstein space of the quantum states.

Why can the Wasserstein distance capture the quantum chaos so well? In fact, it was proposed by Zyczkowski in 1993 \cite{zyczkowski1993generalize} that the quantum Lyapunov exponent can be measured by Wasserstein distances between Husimi Q-representation of quantum states, and later in~\cite{wang2021quantum} it was demonstrated concretely in some chaotic models. Our thought is along the line, and to demonstrate the connection to ours, we evaluate quantum Lyapunov exponents in the time evolution of the inverted harmonic oscillator, as it mimics the scrambling point of the double-well model. We find that three measures of the quantum Lyapunov exponent, (i) microcanonical out-of-time-ordered correlator \cite{Hashimoto:2017oit}, (ii) half-probability contour length~\cite{toda1986quantal,toda1987quantal} and (iii) Wasserstein distance, all agree to give the same value for the Lyapunov exponent.

The organization of this paper is as follows. First in Sec.~\ref{sec:oned-models}, we describe that the quantum scrambling causes the folding structure of the Wasserstein space of the double-well system. Then in Sec.~\ref{sec:coupled-ho} we study the coupled harmonic oscillator system to find the dimensional reduction of the Wasserstein space as the chaos parameter increases. In Sec.~\ref{sec:scar} a branch structure of the Wasserstein space is found to correspond to quantum scar states. In Sec.~\ref{sec:lyap} we examine the quantum Lyapunov exponents measured by the Wasserstein distance. The final section is for a summary and discussion. Appendix \ref{sec:2-w} summarizes additional numerical results using 2-Wasserstein distance. Appendix \ref{app:numerical-precision} establishes the validity of the numerical analysis of the Wasserstein distance conducted in Secs.~\ref{sec:oned-models} and \ref{sec:coupled-ho}.


\section{Scrambling and optimal transport: Folded Wasserstein space}
\label{sec:oned-models}

In this section we reveal the relation between the scrambling phenomenon and the structure of the Wasserstein space
emergent from the energy eigenstates. We analyze one-dimensional quantum systems with three kinds of potentials:
the harmonic oscillator, the double-well potential, and the partly flattened oscillator.
For all three cases, we obtain the Euclidean embedding of the 1-Wasserstein distance matrix\footnote{Using the 2-Wasserstein distance instead of the 1-Wasserstein distance does not modify our conclusions presented in this section and the next section. We stick to the 1-Wasserstein in this paper as we follow the findings of \cite{hashimoto2026holography}, which demonstrated that the 1-Wasserstein distance better suits the manifold hypothesis and the analogy to the Krylov complexity for the harmonic oscillator. In App.~\ref{sec:2-w}, we show the numerical results using the 2-Wasserstein distance for the content of Secs.~\ref{sec:oned-models} and \ref{sec:coupled-ho}.} for the Husimi Q-representations of the energy eigenstates, and will find that for the energy at which the scrambling is observed, the Wasserstein space exhibits a folding structure. We show that the folding structure comes from a separatrix.

\subsection{Quantum models with scrambling}

The scrambling is the phenomenon characterized by the exponential growth in the out-of-time order correlators. Even when the system does not possess chaos, the scrambling can occur, and the cause of it is thought to be saddle point structure 
in phase space. In this paper we study three models, for a comparison between scrambling and non-scrambling models, and for a check of the universality of our claim on the folding of the Wasserstein space. The potential $V(x)$ for these three models, with the Hamiltonian $H=-\partial_x^2+V(x)$, is shown in Fig.~\ref{fig:oned-potentials}. 

\begin{figure}[t]
\centering
\subfigure[Double-well]{\includegraphics[width=0.49\textwidth]{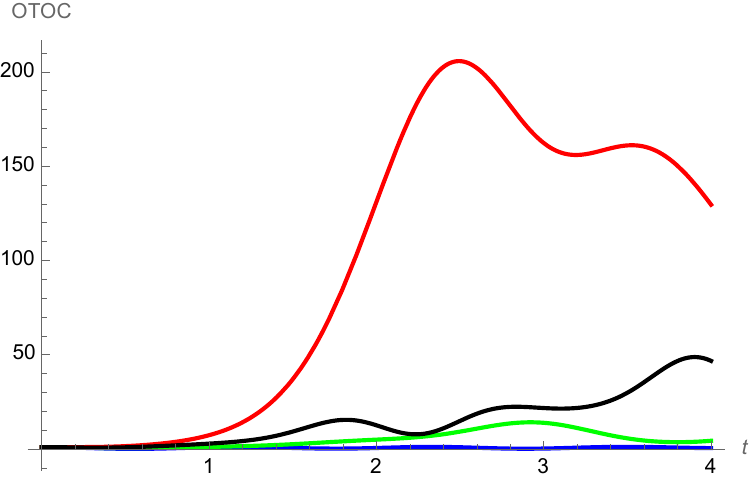}}
\subfigure[Partly flattened oscillator]{\includegraphics[width=0.49\textwidth]{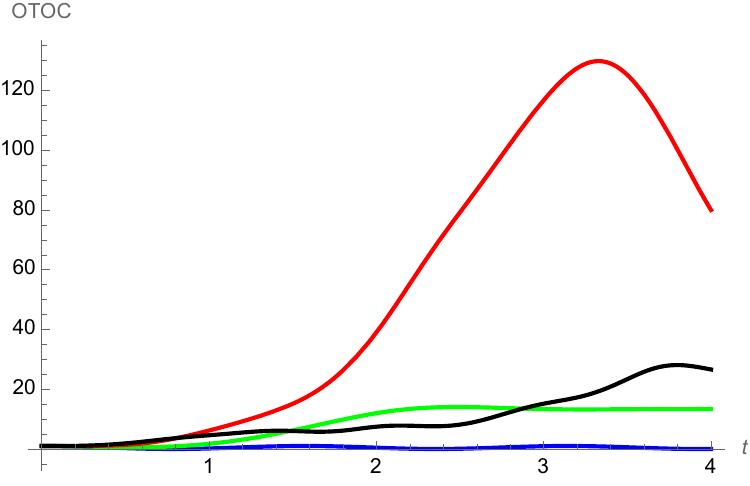}}
\caption{The time evolution of the microcanonical out-of-time-order correlators (OTOCs) for the double-well case (a) and the partly flattened oscillator case (b). In both cases (a) and (b), the red lines grow exponentially in time, which is the scrambling phenomenon. For (a), the plotted one is for $n=0$ (blue), $n=2$ (green), $n=5$ (red) and $n=9$ (black), where $n=0,1,2,\cdots$ labels the energy level of even parity states counted from the ground state. The $n=5$ state has $E=0.088$, which is close to the top of the potential hill. For (b), the plotted one is for $n=0$ (blue), $n=2$ (green), $n=4$ (red) and $n=9$ (black). The $n=4$ state has $E=13.07$, which is close to the plateau part of the potential. }
\label{fig:otoc}
\end{figure}

\begin{figure}[t]
\centering
\includegraphics[width=0.98\textwidth]{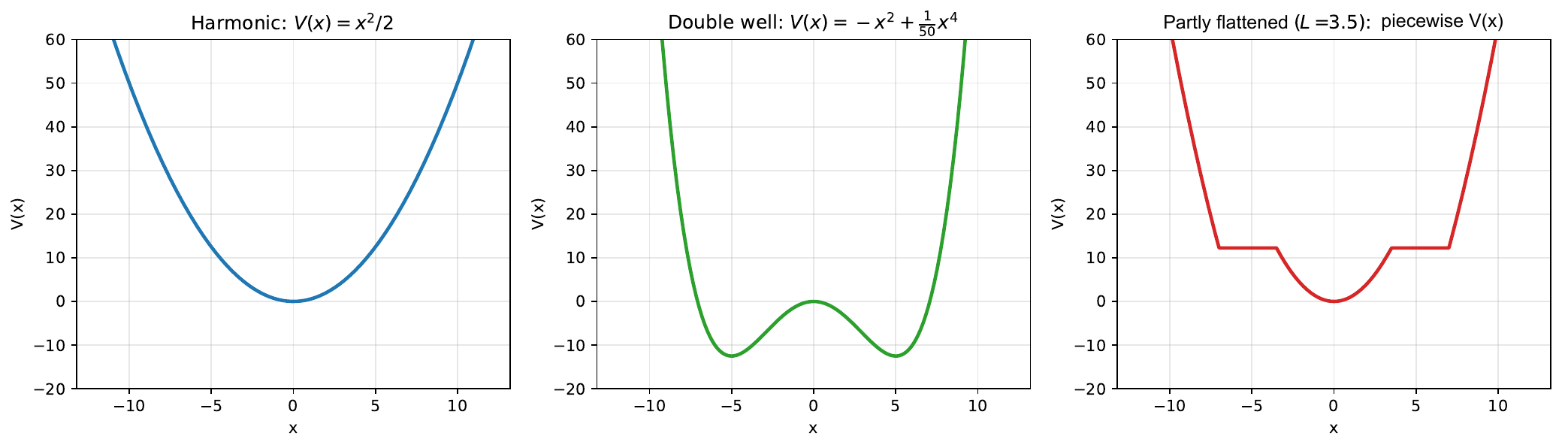}
\caption{One-dimensional potentials used in this section: harmonic, double-well, and partly flattened oscillator.}
\label{fig:oned-potentials}
\end{figure}

The first model is just the ordinary quantum harmonic oscillator in one dimension, which is what was studied in \cite{hashimoto2026holography}. This does not exhibit scrambling. The second model is a one-dimensional double-well model. Since it has a potential hill whose top is at the center, thus in fact it exhibit the scrambling, as shown in \cite{Hashimoto:2020xfr}. Note that in one dimension there is no quantum chaos. The third model is what we call a partly flattened oscillator, see Fig.~\ref{fig:oned-potentials} for the shape of the potential. We cut the harmonic potential and introduce flat plateau parts. These serve as the ``saddle" points. In fact, it exhibits the scrambling, see Fig.~\ref{fig:otoc}. In this figure, we present numerical evaluation of the following microcanonical OTOC \cite{Hashimoto:2017oit},
\begin{align}
    c_n(t) \equiv \langle n | [x(t),p(0)]^2|n\rangle
\end{align}
for the $n$-th energy eigenstates ($n=0,1,2,\cdots$, see the definition of the states below). The red lines for the state energy close to the saddle grow exponentially in time, as opposed to other lines for the states whose energy is not close to the saddle. This is scrambling. So these models are essential for our purpose.

With these models, we proceed to the numerical evaluation of the optimal transport (1-Wasserstein distance) between Husimi Q-representation of the low-lying energy eigenstates. To recapitulate, here we show the definition of the Husimi Q-representation \cite{husimi1940some}:
For a wavefunction $\psi(x)$ in one dimension, we define the coherent-state overlap as
\begin{equation}
\langle q,p|\psi\rangle
=\pi^{-1/4}\int dx\,
\exp\!\left[-\frac{(x-q)^2}{2}-i\,p\!\left(x-\frac{q}{2}\right)\right]\psi(x),
\end{equation}
and the Husimi Q-representation by
\begin{equation}
Q_\psi(q,p)=\frac{1}{2\pi}\left|\langle q,p|\psi\rangle\right|^2.
\end{equation}

Let us introduce the three models, with some numerical details necessary to perform the optimal transport analyses. 
For the optimal transport we need to discretize the phase space on a finite grid, then the Husimi Q-representation $Q_\psi$ is flattened to a normalized vector. 

\paragraph{Harmonic oscillator.}
We use the harmonic potential
\begin{equation}
V(x)=x^2.
\end{equation}
The states are the first 16 energy eigenstates counted from the ground state, 
$n=0,\dots,15$ (16 states).
The Husimi Q-representation is evaluated analytically in the number basis,
\begin{equation}
Q_n(q,p)=\frac{1}{2\pi}\,e^{-\alpha}\,\frac{\alpha^n}{n!},
\qquad
\alpha=\frac{q^2+p^2}{2}.
\end{equation}
The phase-space window is chosen from the analytic support estimate:
$q,p\in[-q_{\max},q_{\max}]$ with
$q_{\max}=p_{\max}=\max(6.5,\sqrt{2(n_{\max})}+2)\simeq 7.48$ for $n_{\max}=15$,
on a $37\times 37$ grid.

\paragraph{Double-well.}
We use
\begin{equation}
V(x)=-x^2+\frac{1}{50}x^4 .
\label{eq:dwp}
\end{equation}
Wavefunctions are obtained numerically from the finite-difference Schr\"odinger operator
$-\partial_x^2+V(x)$ on $x\in[-11,11]$ with $1001$ points.
From the lowest 40 candidates, we keep the lowest 16 even-parity states. The state whose energy is close to the saddle of the potential is for $n=5$.
Husimi and transport grids are $q,p\in[-12,12]$ with $37\times37$ points.

\paragraph{Partly flattened oscillator.}
With $L=3.5$, we use
\begin{equation}
V(x)=
\begin{cases}
x^2, & |x|<L,\\
L^2, & L<|x|<2L,\\
x^2-3L^2, & |x|>2L.
\end{cases}
\label{eq:kinkyp}
\end{equation}
This has the plateau parts in the potential, to produce the scrambling, while it does not have any tunneling potential hill as opposed to the double-well model.
Again we solve $(-\partial_x^2+V)\psi=E\psi$ numerically by finite difference on
$x\in[-16,16]$ with $1401$ points, keep 16 even-parity states from 40 candidates.
The state whose energy is close to the saddle of the potential is for $n=4$.
We use the Husimi grid $q,p\in[-12,12]$ with $37\times37$ points.

\subsection{Optimal transport and Wasserstein distance matrix}
\label{subsec:otwdm}

For the numerical evaluation of the 1-Wasserstein distance, we employ the Sinkhorn regularization, which we briefly review below.
First we discretize the phase space in a proper manner onto a Cartesian grid.
Let $Q_1,Q_2$ be two discrete Husimi distributions on the same grid, and let $C_{ij}$ be the Euclidean ground cost between phase-space points specified by $i$ and $j$. The $1$-Wasserstein problem seeks
\begin{equation}
W_1(Q_1, Q_2)=\min_{\pi\in\Pi(Q_1,Q_2)} \sum_{ij}\pi_{ij} C_{ij},
\label{eq:W1-exact}
\end{equation}
where $\pi(Q_1,Q_2) \in \Pi(Q_1, Q_2)$ denotes a coupling (or a transport plan) with marginals $Q_1$ and $Q_2$, meaning the projection equation $\sum_j \pi(Q_1,Q_2)_{ij}=(Q_1)_i$ and $\sum_i \pi(Q_1,Q_2)_{ij}=(Q_2)_j$;
$\pi_{ij}$ represents the fraction of ``mass'' moved from location $i$ to
location $j$.
Direct linear programming is expensive for repeated pairwise computations, so we use entropic regularization~\cite{cuturi2013sinkhorn}:
\begin{equation}
\min_{\pi\in\Pi(\mu,\nu)}\ \sum_{ij}\pi_{ij} C_{ij}
\;+\;\varepsilon\sum_{ij}\pi_{ij}\left(\log \pi_{ij}-1\right).
\label{entropic-regularization}
\end{equation}
The second term is the regularization term with $\varepsilon>0$ being a hyperparameter.
The entropy term smooths the
feasible set and makes the optimization strictly convex, and the unique solution takes
the scaling form\footnote{The scaling form follows from the Lagrange multiplier equations.  Adding
multipliers $a_i,b_j$ for the marginal constraints and varying with respect to
$\pi_{ij}$ gives $C_{ij}+\epsilon\log\pi_{ij}-a_i-b_j=0$.  Thus
$\pi_{ij}=e^{a_i/\epsilon}e^{-C_{ij}/\epsilon}e^{b_j/\epsilon} \equiv u_iK_{ij}v_j$, with $K_{ij}=e^{-C_{ij}/\epsilon}$.  The Sinkhorn updates are the alternating normalizations of $u$ and $v$ required by the two marginals.}
\begin{equation}
\pi=\mathrm{diag}(u)\,K\,\mathrm{diag}(v), \qquad
K_{ij}=e^{-C_{ij}/\varepsilon}.
\label{eq:sinkhorn-form}
\end{equation}
The vectors $u,v$ are found by alternating normalization (Sinkhorn iterations)
under the marginal constraints
$\mathrm{diag}(u)\,K\,\mathrm{diag}(v)\,\mathbf{1} = Q_1$ and
$\mathrm{diag}(v)\,K^\top\,\mathrm{diag}(u)\,\mathbf{1} = Q_2$, where $\mathbf{1}$ is a vector all of whose elements are unity.
More explicitly, the vectors $u,v$ are obtained by iterating element-wise division as
\begin{equation}
  u^{(n+1)} = \frac{Q_1}{K v^{(n)}}, \qquad
  v^{(n+1)} = \frac{Q_2}{K^\top u^{(n+1)}}
  \label{eq:sinkhorn-iter}
\end{equation}
at the $n$-th iteration, starting from $u^{(0)} = \mathbf{1}$, $v^{(0)} = \mathbf{1}$.
In this manner, we can numerically evaluate 
(the entropically-regularized approximation of)
the distance $W_1$.

For any pair of states $(n,m)$, we compute the entropically regularized transport distance
\begin{equation}
D_{nm}=W_1^{(\varepsilon)}(Q_n,Q_m)
\end{equation}
with Sinkhorn iterations:
\begin{equation}
\varepsilon=0.05,\qquad N_{\mathrm{iter}}=300.
\end{equation}
The cost matrix is the Euclidean distance between phase-space grid points $(q_a,p_a)$ and $(q_b, p_b)$,
$C_{ab}=\sqrt{(q_a-q_b)^2 + (p_a-p_b)^2}$.

\begin{figure}[th]
\centering
\includegraphics[width=0.95\textwidth]{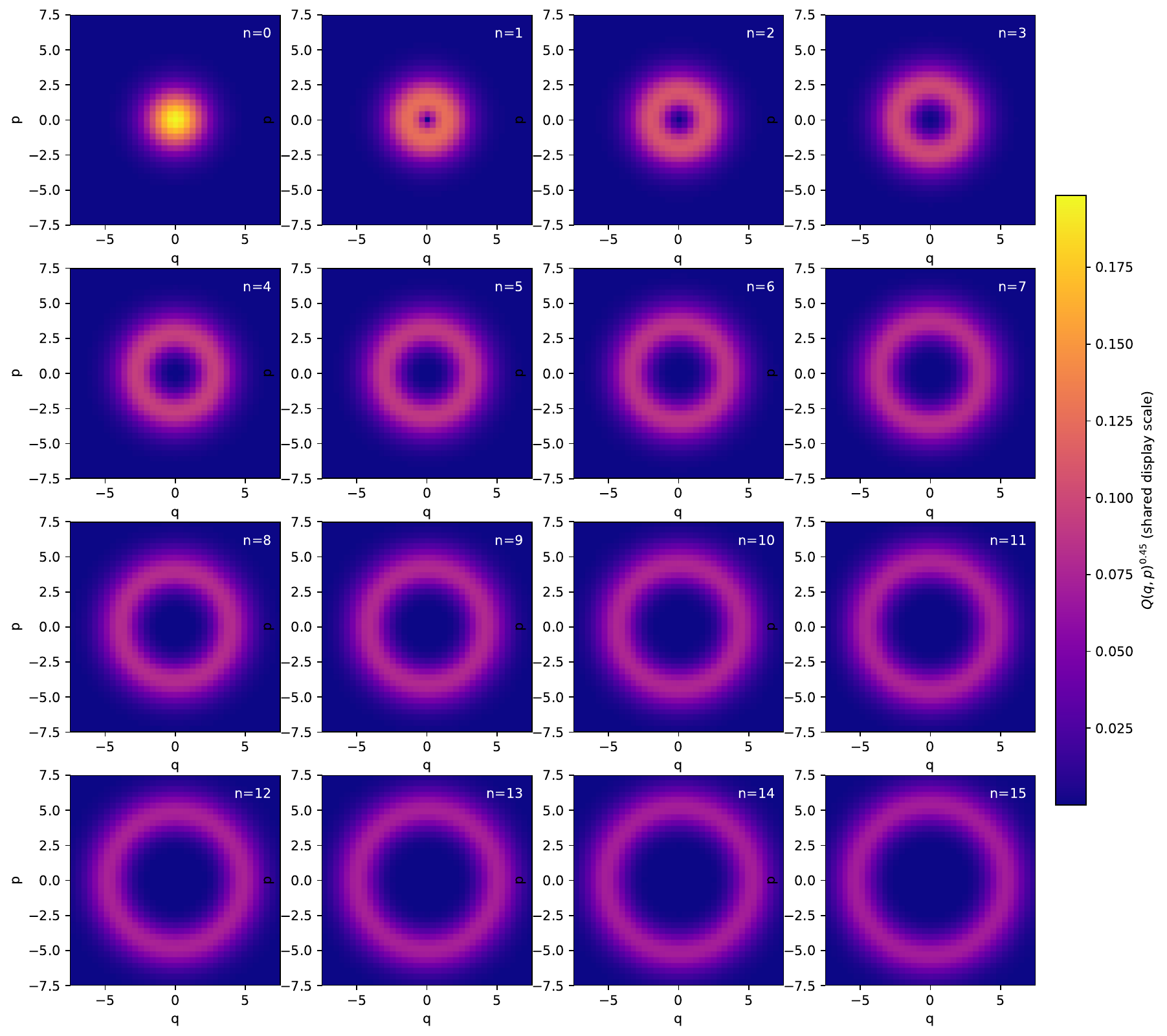}
\caption{Husimi Q-representation for 16 energy eigenstates: harmonic oscillator. From top-left to bottom right, the energy grows, and we labels them as $n=0,1,\cdots,15.$}
\label{fig:oned-husimi-ho}
\end{figure}

\begin{figure}[th]
\centering
\includegraphics[width=0.95\textwidth,page=1]{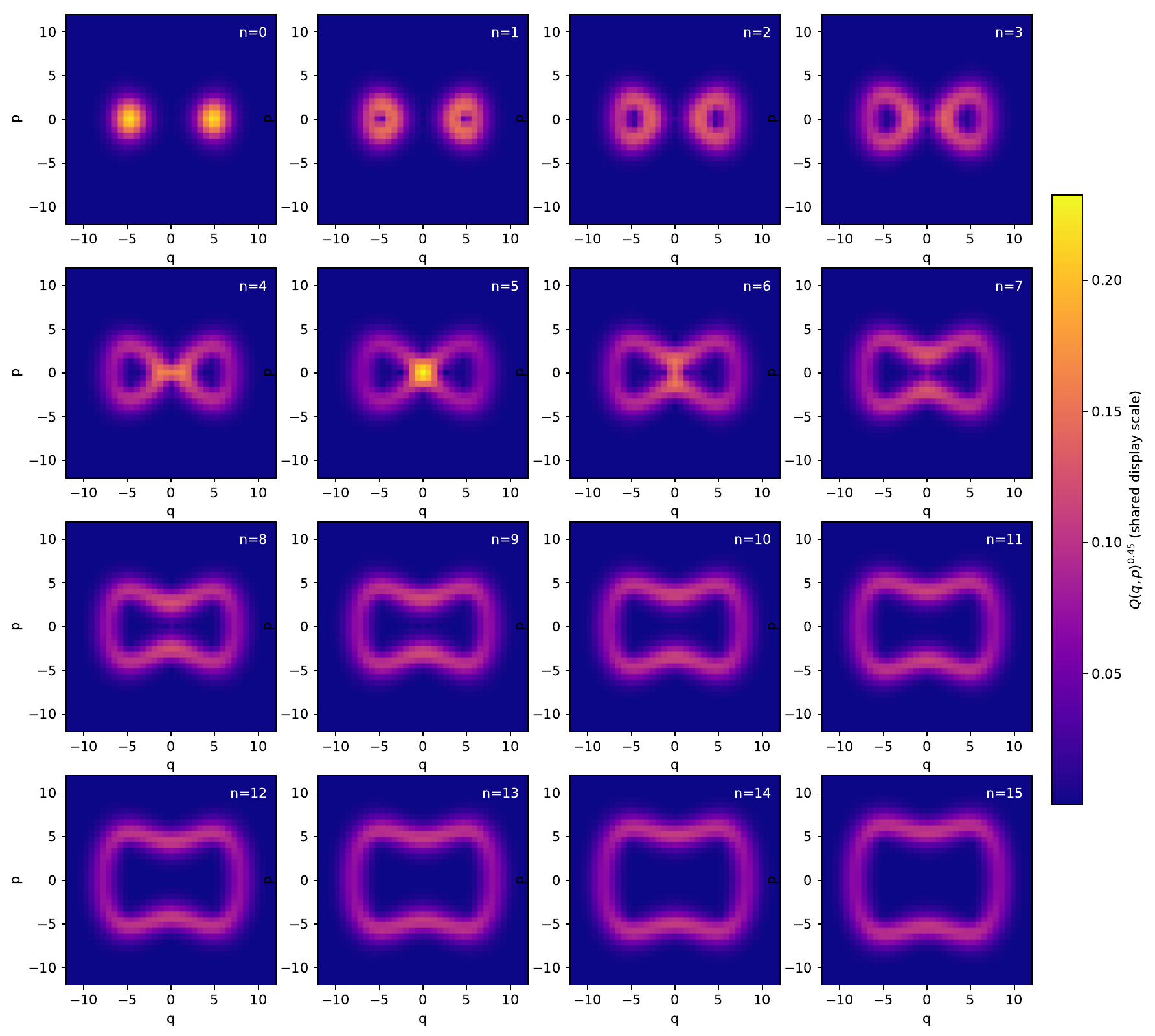}
\caption{Husimi Q-representation for 16 energy eigenstates: double-well.}
\label{fig:oned-husimi-dw}
\end{figure}

\begin{figure}[th]
\centering
\includegraphics[width=0.95\textwidth]{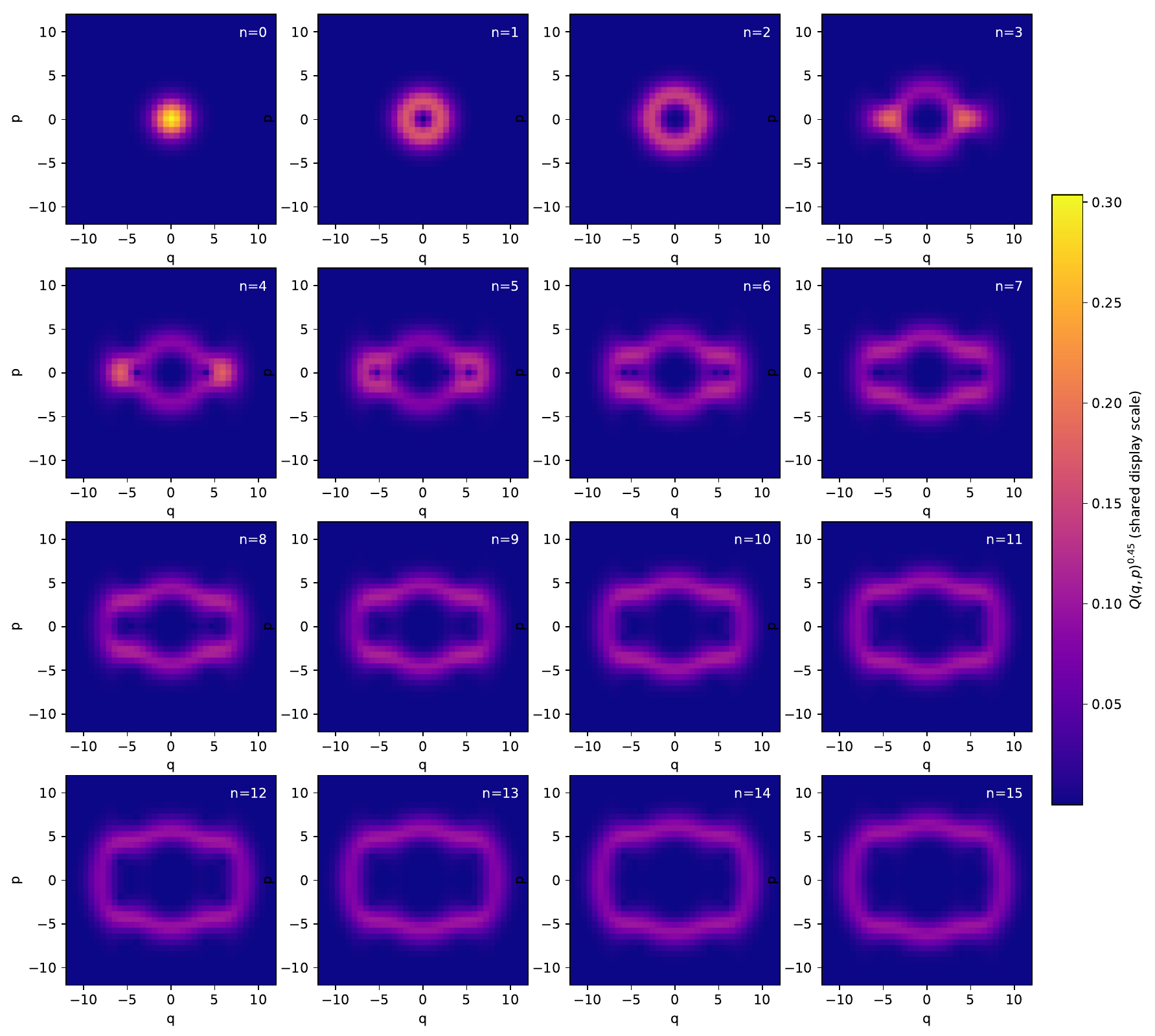}
\caption{Husimi Q-representation for 16 energy eigenstates: partly flattened oscillator.}
\label{fig:oned-husimi-kinky}
\end{figure}

\begin{figure}[th]
\centering
\subfigure[Harmonic oscillator: distance matrix]{
\includegraphics[width=0.47\textwidth]{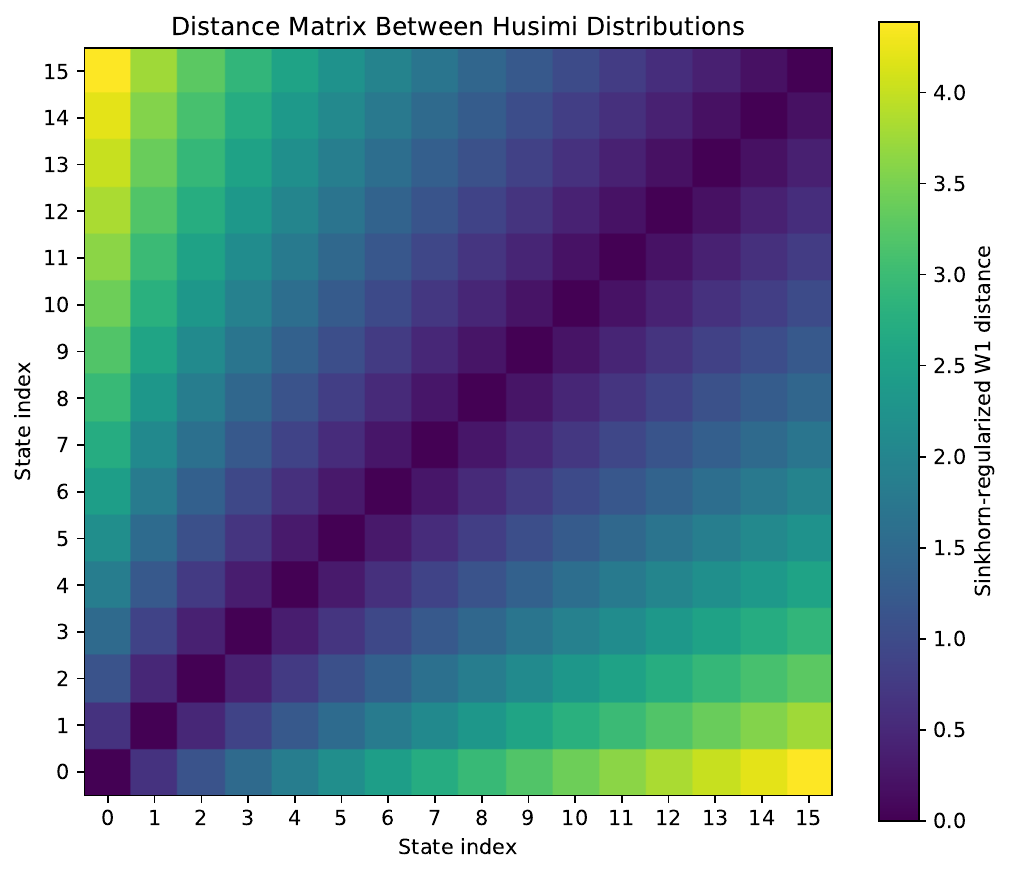}
}
\subfigure[Double-well: distance matrix]{
\includegraphics[width=0.47\textwidth,page=2]{double_well_husimi_wasserstein_figures.pdf}
}
\subfigure[Partly flattened oscillator: distance matrix]{
\includegraphics[width=0.47\textwidth]{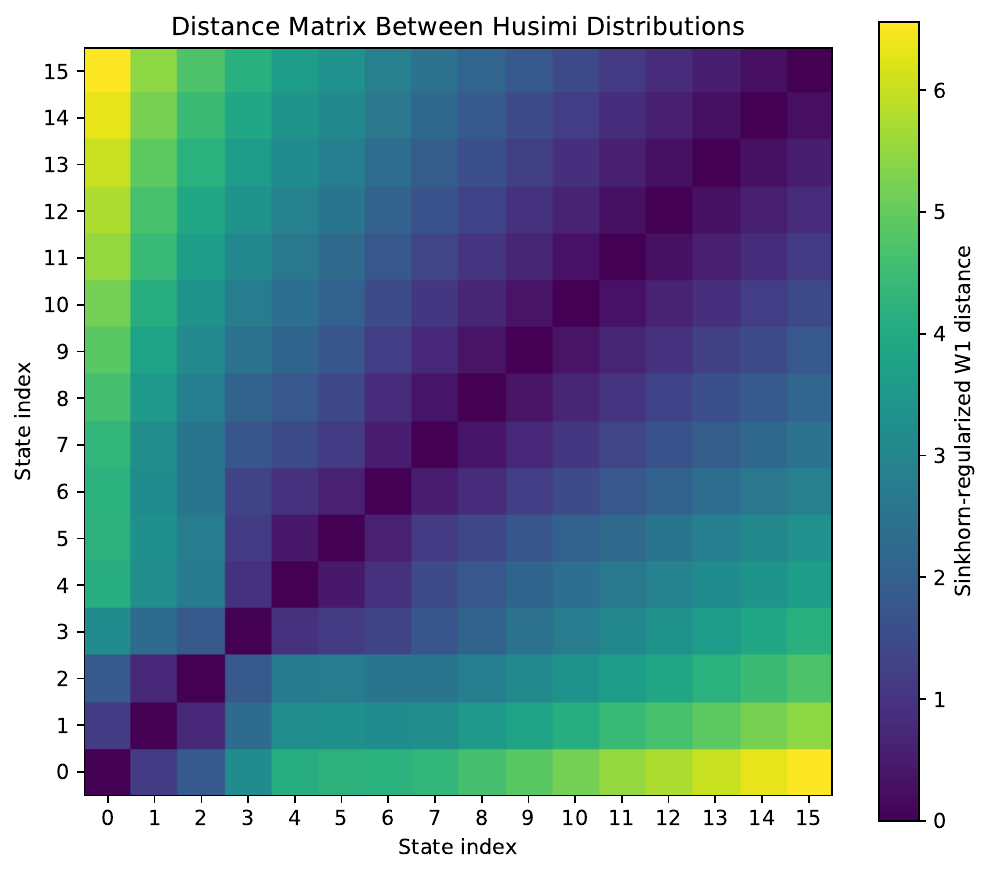}
}
\caption{Pairwise Wasserstein distance matrices constructed from the Husimi Q-representation. Each panel shows the distance matrix $D$ for (a) the harmonic oscillator, (b) the double-well, and (c) the partly flattened oscillator.}
\label{fig:oned-distance}
\end{figure}

Below we show our numerical results. The Husimi Q-representations for the first 16 energy eigenstates 
are shown in
Figs.~\ref{fig:oned-husimi-ho}, \ref{fig:oned-husimi-dw}, and \ref{fig:oned-husimi-kinky},
each of which is for the harmonic oscillator, the double-well, and the partly flattened oscillator. Then the computed 
1-Wasserstein distance matrices are shown in Fig.~\ref{fig:oned-distance}.

From the heatmaps of the distance matrices, we can immediately observe that in the case of (b) the double-well the bright lines at $n=5$ appear, meaning that the state $|n=5\rangle$ is far away from the other states. As for the partly flattened oscillator, the structure is not that obvious compared to the double-well case, but there is a line structure in the heatmap between $n=3$ and $n=4$. In the next subsection, we embed this distance matrix into a Euclidean space and plot points representing the states, to clarify the geometric structure of the Wasserstein space for three models.

\subsection{Folded structure of Wasserstein space}
\label{sec:fold}

Let us embed the distance matrix into a point cloud in a Euclidean space as its realization.
The method follows that of \cite{hashimoto2026holography}, and here we briefly review it.
From the pairwise Wasserstein distances between the lowest $16$ states, we form a distance matrix $D$ which is $16 \times 16$ real-valued symmetric matrix. Using the distance matrix for each potential model, we calculate the centered Gram matrix
\begin{equation}
G\equiv -\frac12\,J D^{\circ 2}J,\qquad
J\equiv I-\frac{1}{n}\mathbf{1}\mathbf{1}^{\mathsf T},\quad n=16.
\end{equation}
Here $D^{\circ 2}$ means elementwise squared of the matrix $D$, and $\mathbf{1}\mathbf{1}^{\mathsf T}$ is the matrix with all its entries equal to unity.
The effective dimensions of the point cloud embedded in the Euclidean space can be seen from the eigenvalues of the Gram matrix $G$. More precisely, the dimensions can be evaluated by the decay profile of the sorted Gram eigenvalues $\lambda_i$. 

\begin{table}[htbp]
\centering
\begin{tabular}{lrrrrr}
\hline
Potential & $\lambda_1$ & $\lambda_2$ & $\lambda_3$ & $\lambda_4$ & $\lambda_5$ \\
\hline
Harmonic oscillator & 26.34 & 0.10 & 0.05 & 0.02 & 0.01 \\
Double-well & 37.74 & 7.46 & 0.09 & -0.07 & 0.05 \\
Partly flattened oscillator & 54.48 & 7.43 & 0.24 & 0.13 & -0.07 \\
\hline
\end{tabular}
\caption{Top five Gram-matrix eigenvalues for the one-dimensional W1 calculations, ordered by decreasing absolute value.}
\label{tab:one-dimensional-w1-gram-eigenvalues-top5}
\end{table}

Table \ref{tab:one-dimensional-w1-gram-eigenvalues-top5} shows the eigenvalues of the Gram matrix for each case of the potential. As is clearly seen, the harmonic oscillator case has only a single nonzero eigenvalue $\lambda_1$, while the others are vanishing (with small numerical errors or order $0.1$). In fact, this numerical result confirms what was found in \cite{hashimoto2026holography}, that is, the Husimi Q-representation can be embedded in exact 1-dimensional space for the 1-Wasserstein distance. Then looking at the other cases, the double-well and the partly flattened oscillator, the first two eigenvalues $\lambda_1$ and $\lambda_2$ are nonvanishing while the others are small. This means that these two cases can be embedded effectively into two-dimensional Euclidean space.

\begin{figure}[t]
\centering
\subfigure[Harmonic oscillator]{
\includegraphics[width=0.45\textwidth]{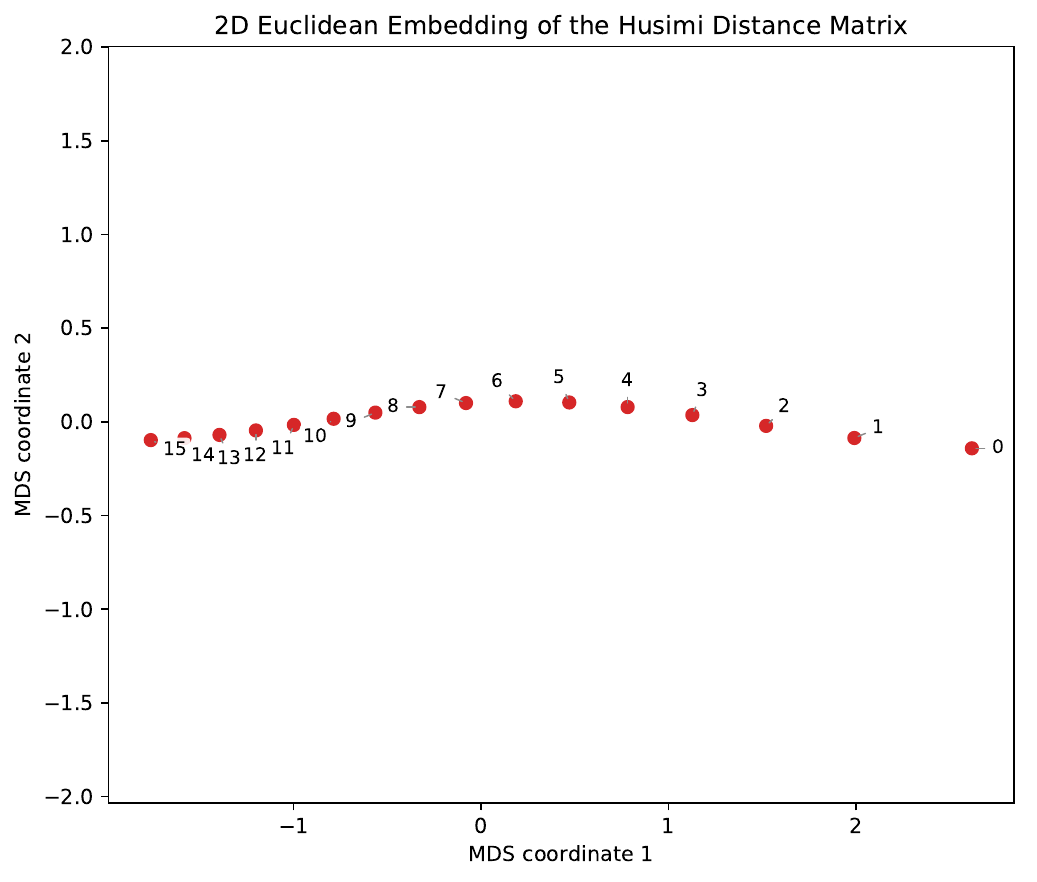}
}
\subfigure[Double-well]{
\includegraphics[width=0.45\textwidth,page=3]{double_well_husimi_wasserstein_figures.pdf}
}
\subfigure[Partly flattened oscillator]{
\includegraphics[width=0.45\textwidth]{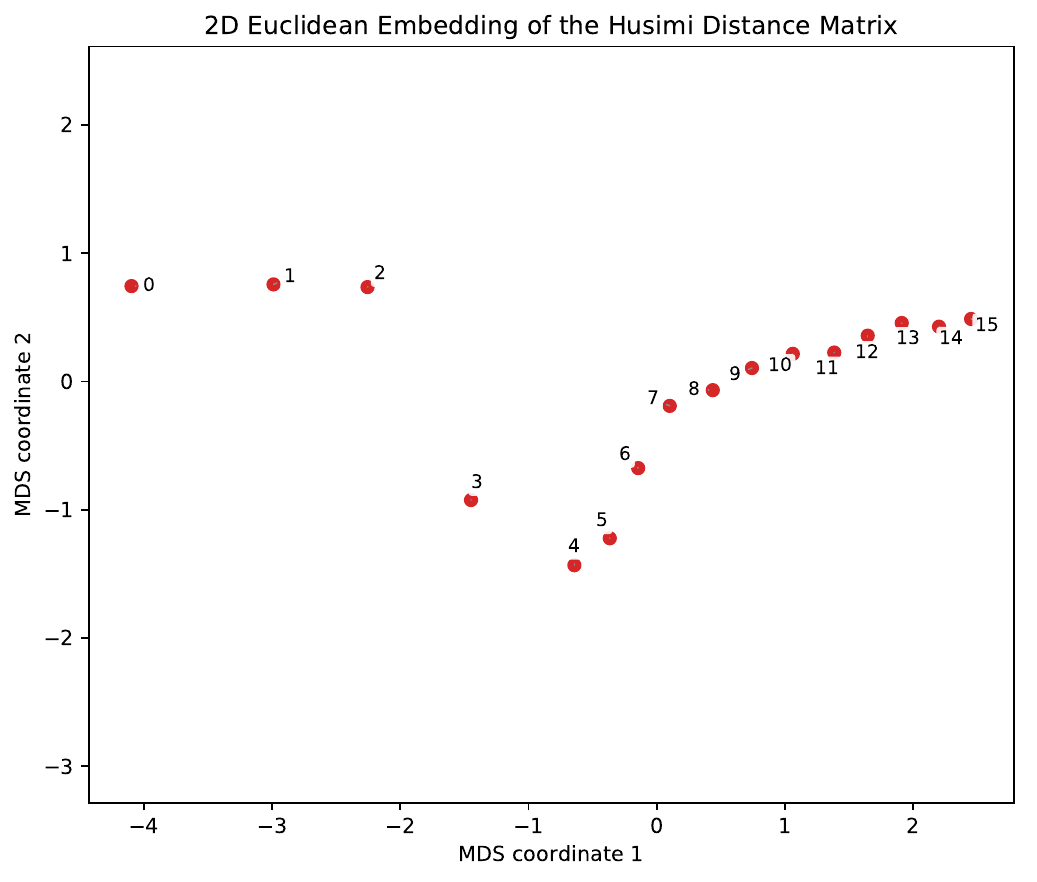}
}
\caption{2-dimensional Euclidean embeddings of the Wasserstein distance matrices.}
\label{fig:oned-embedding-final}
\end{figure}

We shall graphically view the embedded point cloud in the following manner. 
Let $G v_k = \lambda_k v_k$ $(k=1,\dots,n)$ with eigenvalues ordered as $\lambda_1 \ge \lambda_2 \ge \cdots \ge \lambda_n$. Define
\begin{align}
\Lambda_2 :=
\begin{pmatrix}
\lambda_1 & 0 \\
0 & \lambda_2
\end{pmatrix},
\qquad
V_2 := \begin{pmatrix} v_1 & v_2 \end{pmatrix},
\end{align}
where $v_1,v_2\in\mathbb{R}^n$ are orthonormal eigenvectors corresponding to $\lambda_1,\lambda_2$.
Then the two-dimensional classical multidimensional scaling (MDS) embedding is
\begin{align}
X = V_2 \Lambda_2^{1/2}.
\end{align}
This $X$ shows the coordinates of the points which approximately respect the 1-Wasserstein distance matrix as Euclidean distances\footnote{Strictly speaking, $X$ is the two-dimensional classical MDS embedding constructed from the leading positive eigenmodes of the centered Gram matrix $G$. When $G$ is positive semidefinite this gives an exact Euclidean realization, but in our numerics small negative eigenvalues can appear. Thus the plotted point cloud should be interpreted as an approximate two-dimensional representation of the Wasserstein geometry.} on this plane $\mathbb{R}^2$.
Figure~\ref{fig:oned-embedding-final} shows the embeddings for all three models,
constructed from the distance matrices in Fig.~\ref{fig:oned-distance}.

The point cloud of the harmonic oscillator energy eigenstates is aligned on a line, as expected. The slight curving of the line is due to the numerical error of the Sinkhorn optimization on the phase space grid. The notable feature of the 1-Wasserstein space is for the double-well and the partly flattened oscillator. The points are aligned on a line but the line has a folding structure. The folded points ($n=5$ for the double-well model and $n=4$ for the partly flattened oscillator) are exactly the energy eigenstates at which the scrambling occurs, that is, the microcanonical OTOC grows exponentially in time. Thus here we conclude that {\it the scrambling (Exponential growth of OTOC) is correlated with the folding of the emergent 1-Wasserstein space of the Husimi Q-representation of the energy eigenstates}.

\subsection{Folding from separatrix}
\label{subsec:separatrix}

In this subsection we describe the origin of the folding of the Wasserstein space. Let us focus on the double-well potential case. The folded structure occurs at the energy eigenstate $n=5$, at which the OTOC grows exponentially in time. The energy value of the state is close to $E=0$ which is the top of the potential barrier separating the two wells. Therefore, around there, the potential is approximated by an inverted harmonic oscillator, $H = p_x^2 -x^2$.\footnote{See \cite{takahashi1986wigner} for a classical limit of the Husimi Q-representation of states near the potential hill of the double-well model.}

In the terminology of dynamical systems, the top of the hill is a separatrix that divides the phase space into two regions: the motion on either side of the potential, or the motion that goes over the potential. The classical orbits of the former are $x^2-p_x^2 = c$ with $c>0$, while those of the latter are with $c<0$. The separatrix is $c=0$, which amounts to two intersecting lines $x=\pm p_x$. 
In the parametric representation of the orbits in terms of time $t$ as the parameter,
each orbit is written as
\begin{align}
    (x,p_x) = (\sqrt{|c|} \cosh t, \sqrt{|c|} \sinh t) 
\end{align}
for the orbit which does not go over the potential hill (with $c>0$), while
\begin{align}
    (x,p_x) = (\sqrt{|c|} \sinh t, \sqrt{|c|} \cosh t) 
\end{align}
for the orbit which goes over the potential hill (with $c<0$). The separatrix orbits are
\begin{align}
    (x,p_x) = \pm (e^t, e^t) \qquad \mbox{and} \qquad \pm (e^{-t}, -e^{-t}) 
\label{c0}
\end{align}
which are straight half-lines. 

Note that these separatrix orbits \eqref{c0} are half lines which end at the origin of the phase space.
The classical orbits never intersect, and the origin of the phase space can be reached only after an infinite amount of time, so it is the time-like boundary of the orbits.
This fact leads to an important observation: the staying duration on the orbit per unit time diverges at the origin, since it needs infinite time to reach the origin. \eqref{c0} suggests that the density on the separatrix line is given by $d\mu_{\rm separatrix} \sim dt = dx/x$, which diverges at the origin $x=0$.
In fact, if we look at the Husimi Q-representation of the $n=5$ state in Fig.~\ref{fig:oned-husimi-dw}, there exists a bright spot at the origin. This signifies the character of the separatrix.

And indeed this explains the folding structure of the Wasserstein space. It is obviously easier to transport the Husimi Q-representation of the $n=3$ to that of $n=7$ than to that of $n=5$, because the states $n=3$ and $n=7$ are almost uniform on the classical orbits while the state $n=5$ is completely localized at the origin. Transporting everything on the $n=3$ orbit to the origin is costly, and it is easier to transport it to the $n=7$ orbit. Therefore the state $n=3$ is close to the state $n=7$, while the state $n=5$ is far distant.
This causes the folded structure of the Wasserstein space.

This reminds us of the Sieber-Richter encounter \cite{sieber2001correlations} in the chaos theory. The quantum chaos is dictated by this encounter at which two classical orbits close to each other recombine. 
Correlations between periodic orbits related by the Sieber-Richter pairs generate the universal random-matrix corrections to spectral statistics~\cite{muller2004semiclassical, muller2005periodic, heusler2007periodic}.
The separatrix we have is this kind of recombination of classical orbits, and whenever two orbits recombine the intersecting orbit should appear, and any intersecting orbit takes infinite amount of time to reach the intersection point. This recombination phenomenon is one of the classical mechanisms relevant to chaotic behavior, and it is in this sense that the separatrix structure is related to the exponential growth of the microcanonical OTOC near the barrier-top energy.
Once the intersection point exists, the orbit ``forgets" when the motion started, and the chaos shows up. The exponential growth of the microcanonical OTOC for the energy equal to the separatrix is due to this in the classical picture. Therefore, we conclude that the folded structure of the Wasserstein space and the exponential growth of the microcanonical OTOC share the same origin, which is the separatrix.

In the next section, we study a truly chaotic system and will find that the Wasserstein space is dimensionally reduced, and the origin of this dimensional contraction may be understood as the cumulative effect of multiple folding structures of the type found in this section.


\section{Chaos and optimal transport: Dimensionally reduced Wasserstein space}
\label{sec:coupled-ho}

In this section, we treat a quantum system which is known to exhibit quantum chaos, and will find that the effective dimensions of the emergent Wasserstein space given by the Husimi Q-representation of the energy eigenstates {\it decrease} as the Lyapunov exponent
(or, equivalently, the coupling parameter $g$ controlling the non-integrability)
grows.\footnote{One can check that the classical Lyapunov exponent actually grows monotonically as the coupling constant $g$ increases, within our model.} This behavior is quite consistent with 
the philosophy of the paper \cite{hashimoto2026holography} stating the manifold hypothesis that the holography can be viewed as a dimensional reduction of the emergent Wasserstein space, as the quantum chaos, in particular the maximal quantum chaos, is expected to be a necessary condition for the holography through the dynamics on black hole horizons.
Specifically, we introduce a chaoticity parameter, which is a coupling constant $g$ in a coupled harmonic oscillator in two dimensions. Our goal is to quantify how the Wasserstein geometry of low-lying quantum states in phase space changes with $g$. Indeed, we will find that the effective embedding dimensions of the Wasserstein space decrease as we take a large value of $g$.

\subsection{Coupled harmonic oscillator and Husimi Q-representation}
\label{sec:coupled-HO_Husimi-Q}

The quantum system we treat is a two-dimensional quantum coupled harmonic oscillator (called Pullen Edmonds Hamiltonian) whose Hamiltonian is given by \cite{pullen1981comparison}
\begin{equation}
H = p_x^2 + p_y^2 + V(x,y), \qquad
V(x,y)=\frac{1}{4}\left(x^2+y^2\right)+g\,x^2y^2 .
\label{eq:hamcho}
\end{equation}
Here $g$ is the coupling constant. When $g$ vanishes, the system reduces to just uncoupled two harmonic oscillators.
This system is well-known in two perspectives; first, the coupling term stems from an essential part of Yang-Mills quantum mechanics and thus has been known to exhibit classical chaos for high energy, and thus, second, the quantum chaos of this system has been studied with the out-of-time-ordered correlators (OTOCs) to exhibit the exponential growth in time \cite{akutagawa2020out}.
Therefore this model is suitable for finding a possible relation between the quantum chaos and the optimal transport.

\begin{figure}[th]
\centering
\includegraphics[width=0.95\textwidth]{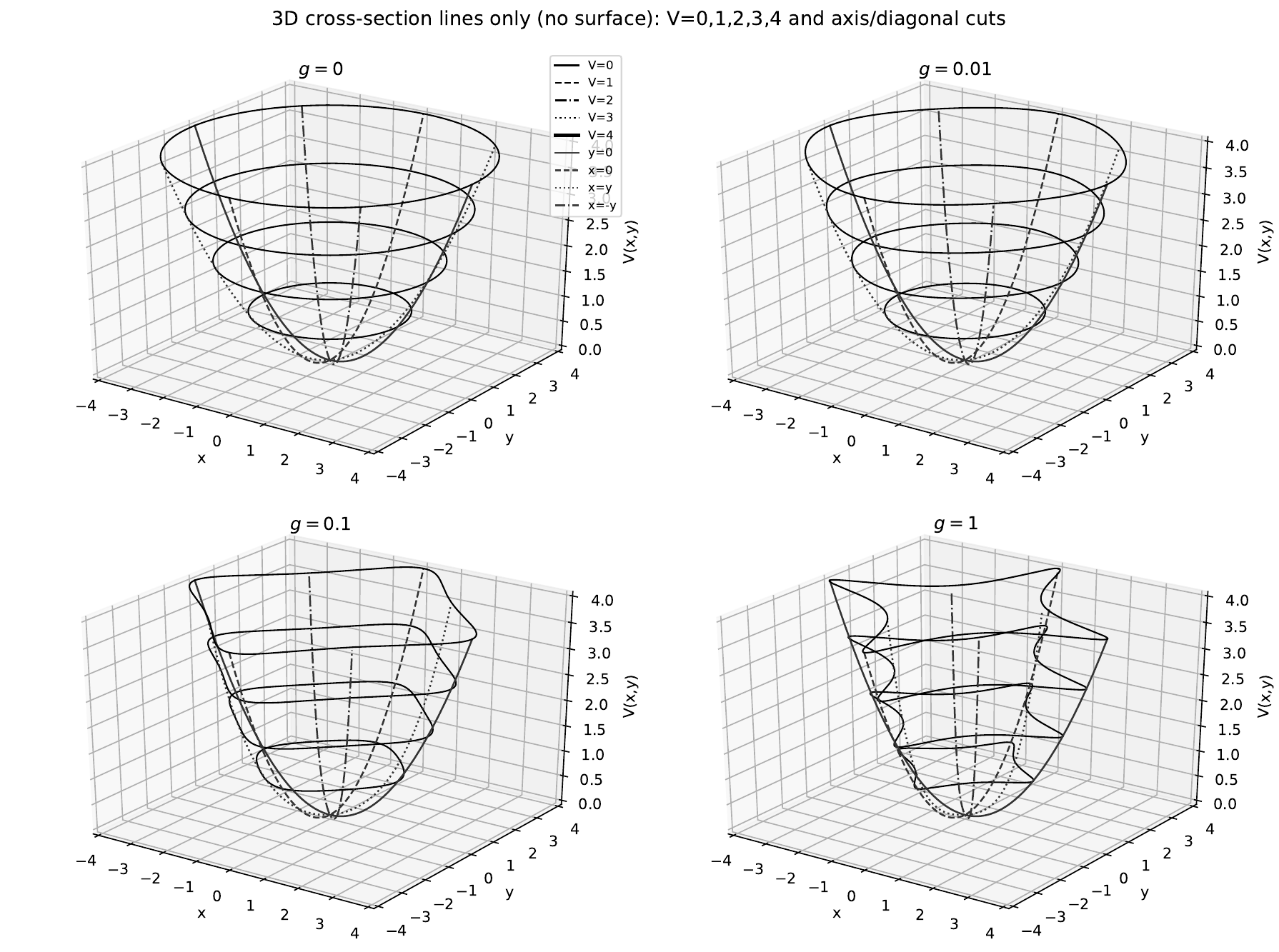}
\caption{Three-dimensional line-only cross-section visualization (no surface rendering) of $V(x,y)=\frac14(x^2+y^2)+g x^2y^2$ for $g=0,\ 1/100,\ 1/10,\ 1$, on $-4<x<4$, $-4<y<4$ with display range $0<V<4$. The lines show level sets $V=0,1,2,3,4$ together with guide cross sections $y=0$, $x=0$, $x=y$, $x=-y$.}
\label{fig:coupled-2d-potentials}
\end{figure}

For our numerical calculation of the optimal transport, we choose the values of the coupling constant $g$ as
\begin{equation}
g\in\left\{0,\ \frac{1}{100},\ \frac{1}{10},\ 1\right\}.
\end{equation}
The quartic mixing term controls non-integrability; increasing $g$ is taken as increasing chaoticity.
The quantum system with $g=0$ is integrable with no chaoticity.
The corresponding potential landscapes are shown in Fig.~\ref{fig:coupled-2d-potentials}. 

\begin{figure}[th]
\centering
\includegraphics[width=0.6\textwidth]{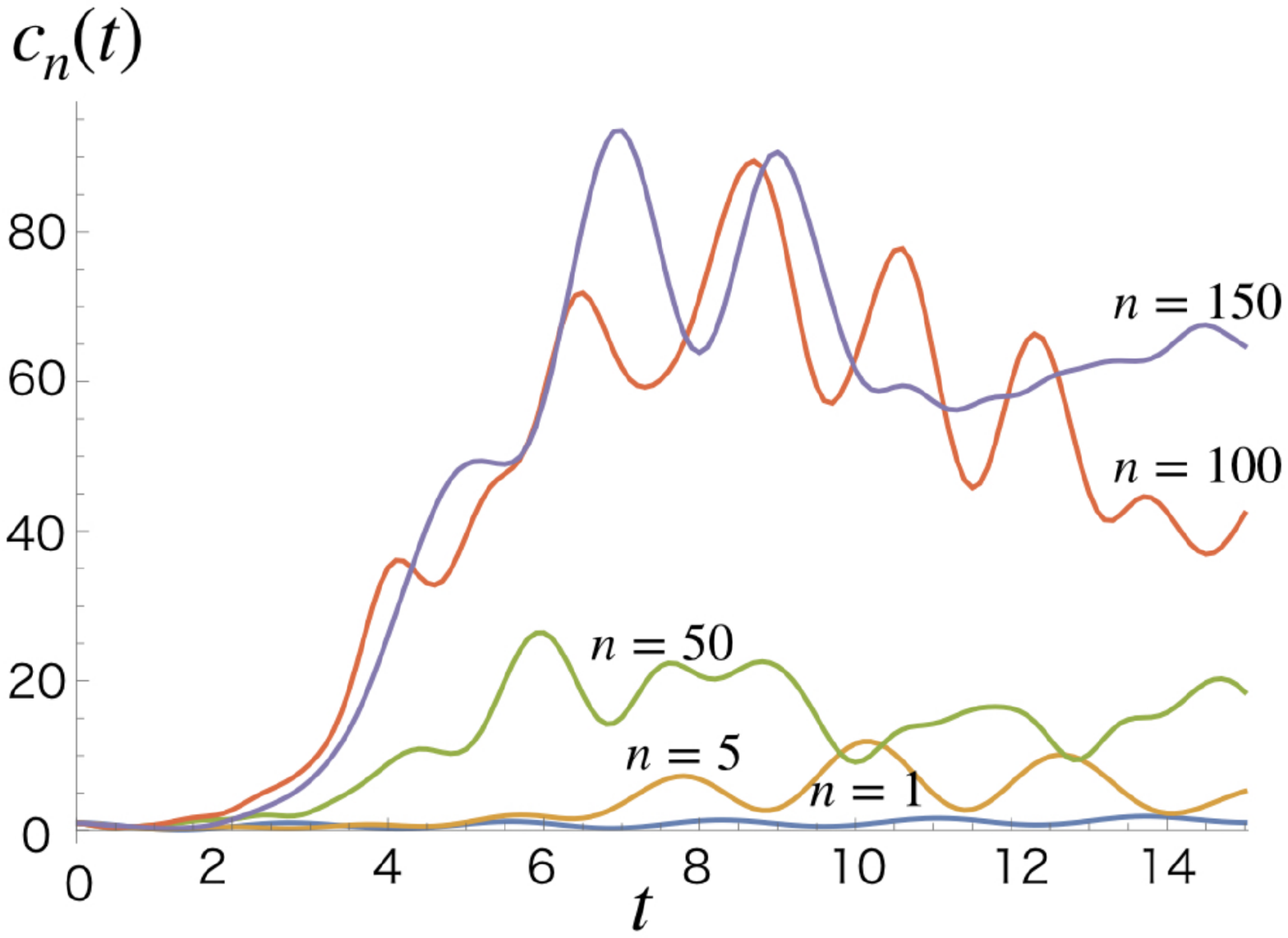}
\caption{An excerpt, Fig.~4 of \cite{akutagawa2020out}. The exponential growth of a microcanonical OTOC $c_n(t) \equiv \langle n| [x(t),p(0)]^2 |n\rangle$ is numerically observed for the $n$-th energy eigenstate $|n\rangle$ with $n=100$ and $n=150$. For lower levels the exponential growth is not observed.}
\label{fig:CHO2020}
\end{figure}

In \cite{akutagawa2020out} it was shown that a microcanonical OTOC grows exponentially in time, for the choice of $g=1/10$
for the energy eigenstates of the level of order ${\cal O}(100)$, see Fig.~\ref{fig:CHO2020} which is an excerpt from Fig.~4 of \cite{akutagawa2020out}. For lower energy levels, the harmonic oscillator terms $x^2+y^2$ in the potential dominate relative to the chaotic coupling $gx^2y^2$, thus the system does not show the exponential growth of the OTOC.

Let us prepare the distributions for the optimal transport. As we follow the methodology proposed in \cite{hashimoto2026holography}, we first prepare the Husimi Q-representation. 
For each eigenstate $|\psi\rangle$ we define the Husimi Q-representation 
in the four-dimensional phase space as
\begin{equation}
Q_\psi(q_x,q_y,p_x,p_y)
=\frac{1}{(2\pi)^2}\left|
\langle \alpha_x,\alpha_y|\psi\rangle
\right|^2 ,
\qquad
\alpha_x=\frac{q_x}{2}+i p_x,\ \ 
\alpha_y=\frac{q_y}{2}+i p_y .
\end{equation}
Here $|\alpha_x,\alpha_y\rangle=|\alpha_x\rangle\otimes|\alpha_y\rangle$ is a product coherent state.
The coherent state is just to prepare a complete basis upon which the true energy eigenstates are expanded. 
Any normalizable pure state can be expanded in this two-dimensional harmonic-oscillator number basis,
\begin{equation}
|\psi\rangle=\sum_{n_x,n_y} c_{n_x n_y}\,\ket{n_x}\otimes \ket{n_y} ,
\end{equation}
with the coherent-state overlap evaluated analytically as
\begin{equation}
\langle \alpha_x,\alpha_y|\psi\rangle
=e^{-(|\alpha_x|^2+|\alpha_y|^2)/2}
\sum_{n_x,n_y}
\frac{c_{n_x n_y}}{\sqrt{n_x!\,n_y!}}
(\alpha_x^*)^{n_x}(\alpha_y^*)^{n_y},
\end{equation}
using
\begin{equation}
\langle \alpha|n_x\rangle=e^{-|\alpha|^2/2}\frac{(\alpha^*)^{n_x}}{\sqrt{n_x!}},
\end{equation}
for the energy eigenstate of a one-dimensional single harmonic oscillator $\ket{n_x}$.

Therefore the Husimi Q-representation is computed with the formula
\begin{equation}
Q_\psi(q_x,q_y,p_x,p_y)
=\frac{1}{(2\pi)^2}
e^{-(|\alpha_x|^2+|\alpha_y|^2)}
\left|
\sum_{n_x,n_y}
\frac{c_{n_x n_y}}{\sqrt{n_x!\,n_y!}}
(\alpha_x^*)^{n_x}(\alpha_y^*)^{n_y}
\right|^2 .
\end{equation}
In our numerics, this analytic harmonic-oscillator-basis expression is used directly to evaluate
the Husimi distribution on the grids of the phase space.

\begin{figure}[th]
\centering
\subfigure[$g=0$]{
\includegraphics[width=0.75\textwidth]{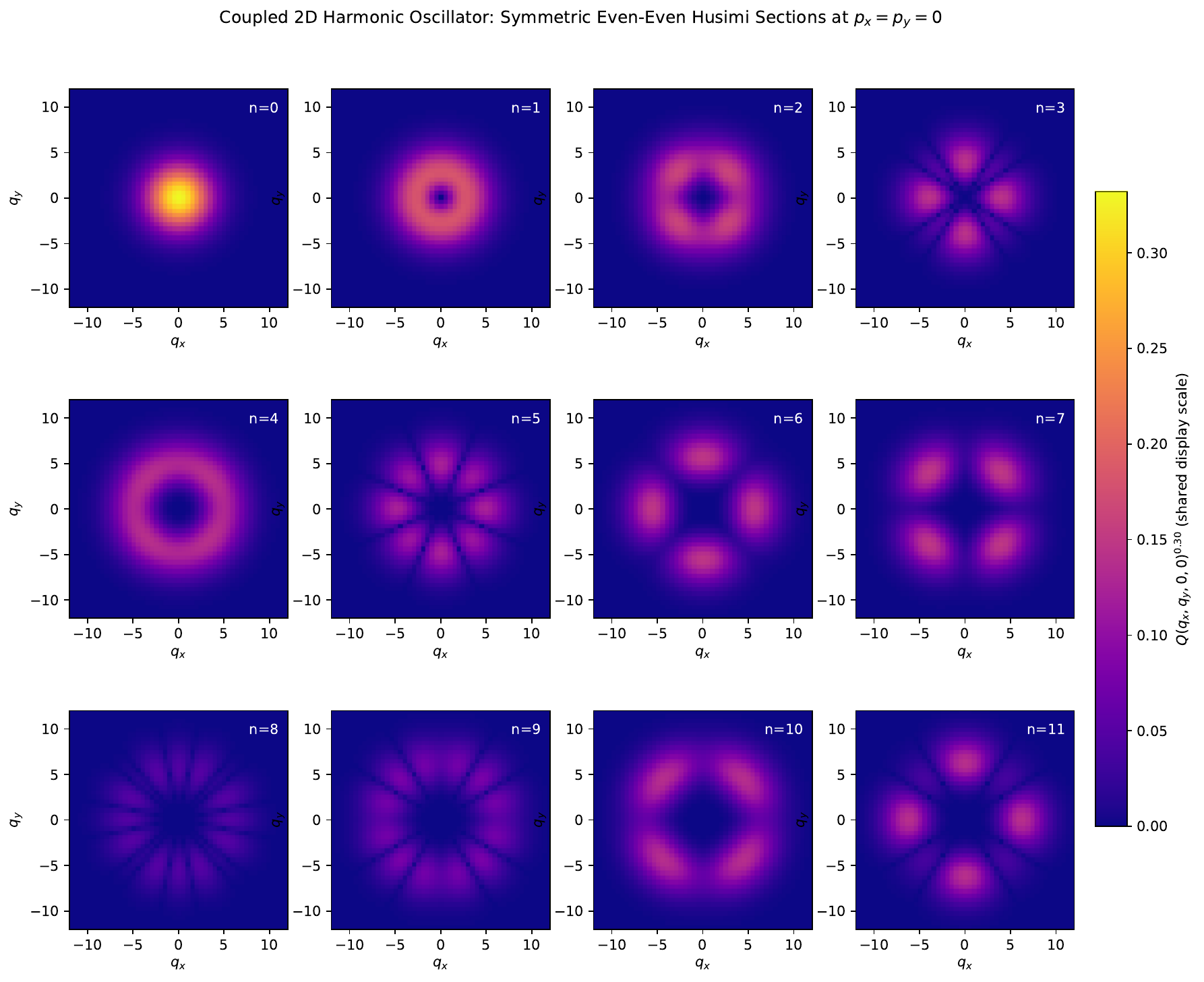}
}
\vspace{0.2cm}
\subfigure[$g=1/100$]{
\includegraphics[width=0.75\textwidth]{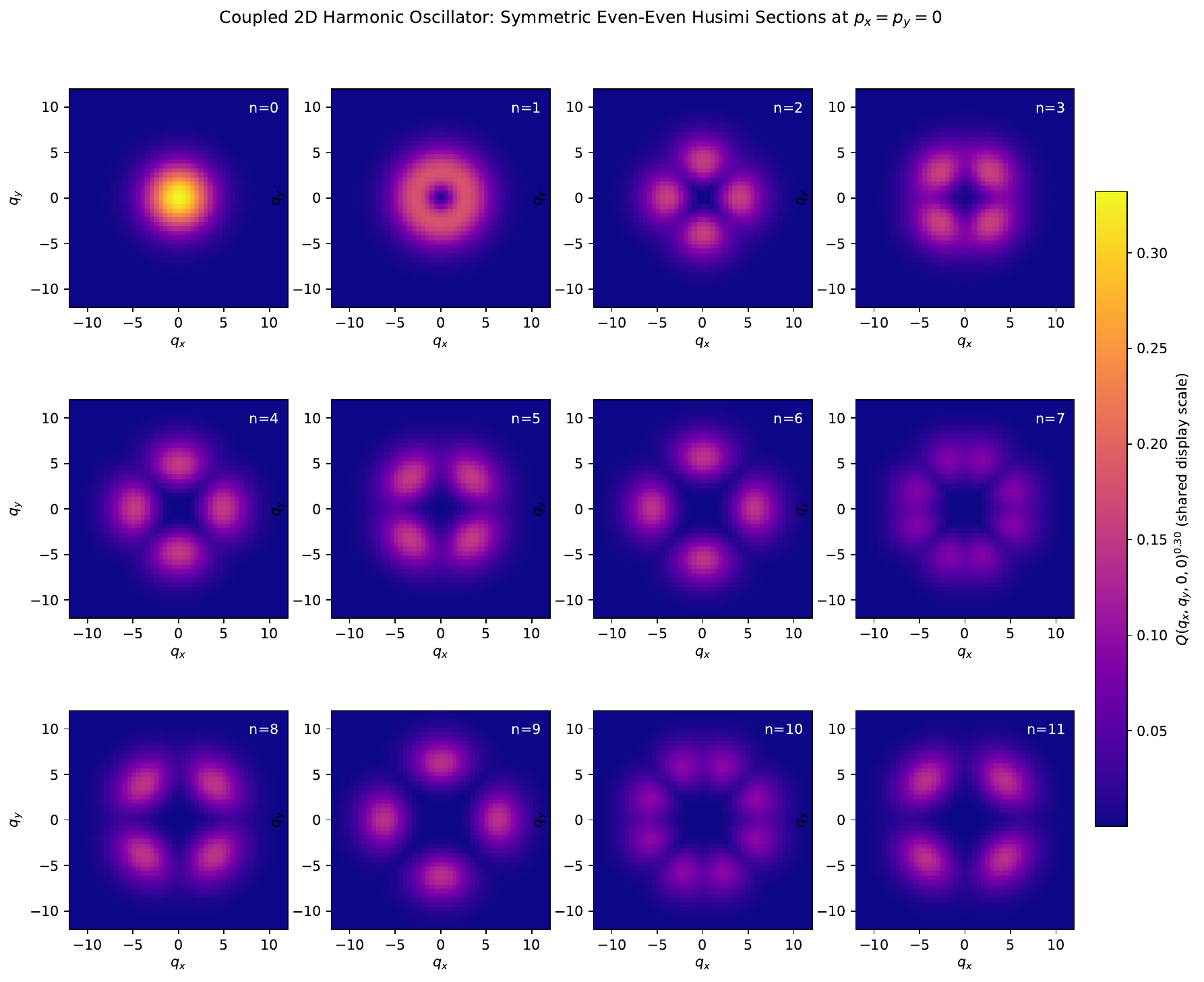}
}
\caption{Husimi section plots for the energy eigenstates for $g=0$ and $g=1/100$.}
\label{fig:husimi-four-g-a}
\end{figure}

\begin{figure}[th]
\centering
\subfigure[$g=1/10$]{
\includegraphics[width=0.75\textwidth]{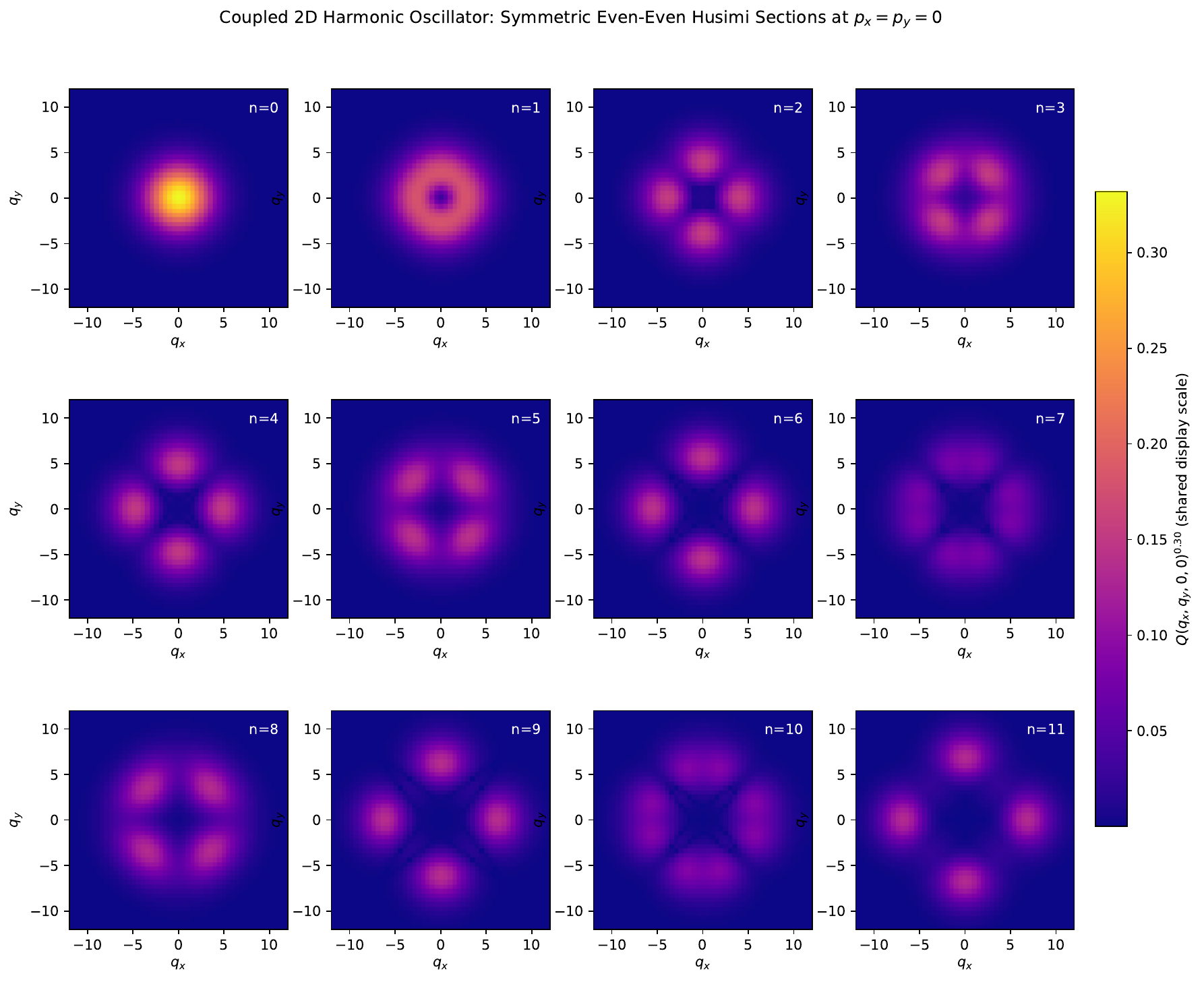}
}
\vspace{0.6cm}
\subfigure[$g=1$]{
\includegraphics[width=0.75\textwidth]{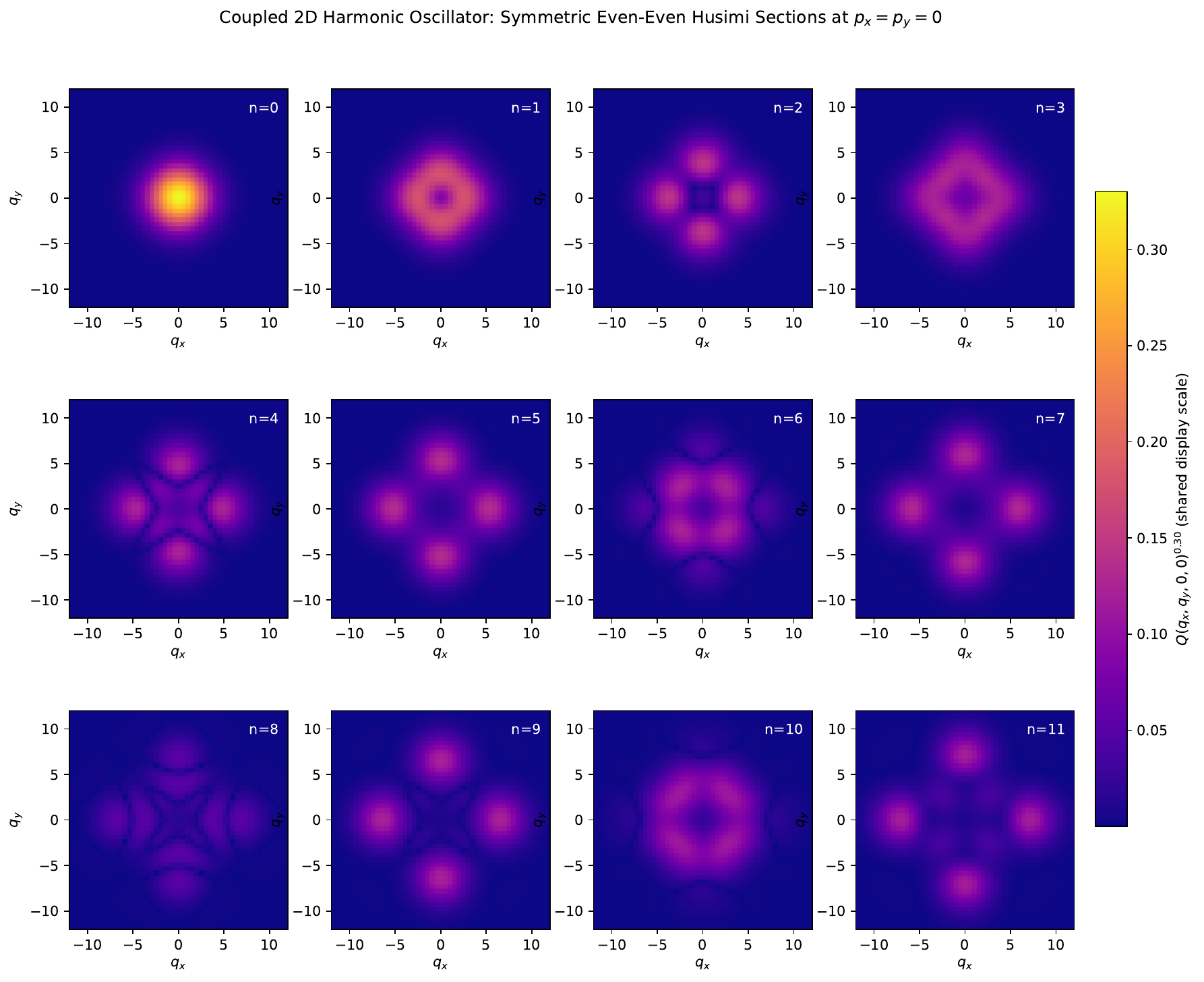}
}
\caption{Husimi section plots for the energy eigenstates for $g=1/10$ and $g=1$.}
\label{fig:husimi-four-g-b}
\end{figure}

To show how the energy eigenstates look like, we visualize the 2-dimensional section
\begin{equation}
Q_\psi(q_x,q_y,0,0)
\end{equation}
for the energy eigenstates of the coupled harmonic oscillator.
Figures~\ref{fig:husimi-four-g-a} and \ref{fig:husimi-four-g-b} collect Husimi section plots used in the comparison, for
$g=0,\;1/100,\;1/10,\;1$. We plot only the lowest 12 energy eigenstates which are parity-even in $x$, $y$ and $x=y$ axes, to ``unfold" the spectra (which is commonly used in the study of quantum chaos).\footnote{For $g=1/10$, the lowest 12 states in the $(P_x,P_y,S_{xy})=(+,+,+)$ sector are not low-lying in the unreduced spectrum: they correspond approximately to the full-spectrum levels
up to ${\cal O}(70)$. Hence the top of our analysis window in energy levels is close to the $n \sim 100,150$ window where the OTOC analysis in \cite{akutagawa2020out} directly observed exponential growth.}
The left-top plot is for the ground state, and going to the right then to the next lower row corresponds to the energy increase for the energy eigenstates.
Obviously, when the coupling constant is changed, the Husimi Q-representation changes.
Note that for the optimal transport
all the distances are computed from the full $4$-dimensional discretized distributions, not the cross section.\footnote{
Note also that the system with $g=0$ is special, is the sense that it admits degeneracy in energy so 
the plots in the panel may not be ordered in the same way as the plots in the other panels.}
For the Husimi section plots, we plot the region $q_x,q_y\in[-12,12]$ with 49 $\times$ 49 points at $p_x=p_y=0$.

More specifically in our numerical study, 
we work in the sector even under $x\to -x$, even under $y\to -y$, and symmetric under $x\leftrightarrow y$,
and we keep the lowest $12$ states in the selected symmetry sector.
First, the one-dimensional oscillator basis is truncated to even levels 
($n$=$0,2,\ldots,18$: 10 basis states per axis).\footnote{
The one-dimensional basis size is chosen with reference to the
uncoupled case \(g=0\).  For \(g=0\), writing
\(n_x=2a\) and \(n_y=2b\) with \(0\leq a\leq b\), the energy in the
chosen \((P_x,P_y,S_{xy})=(+,+,+)\) sector is ordered by
\(s=a+b\).  The first 12 states in this sector are exhausted by
\(s\leq 5\), and hence the largest one-dimensional quantum number
required is \(n_{\max}=2s_{\max}=10\).  Thus, 6 one-dimensional bases,
namely \(n=0,2,\ldots,10\), would already be sufficient for \(g=0\).
For finite coupling \(g>0\), however, the interaction \(g x^2 y^2\)
mixes different harmonic-oscillator number states, and a larger
basis is required.  We therefore use 10 bases per axis uniformly for all values of \(g\) in the numerical calculation. The convergence with respect to this basis
truncation is examined in Appendix~\ref{app:numerical-precision}.}
Then
the two-dimensional Hamiltonian is assembled in the product basis and projected to the
$(P_x,P_y,S_{xy})=(+,+,+)$ sector, then diagonalized numerically by a sparse Hermitian eigensolver.\footnote{In practice, instead of computing all eigenvalues of the projected Hamiltonian
matrix at once, we extract only the lowest 12 energy eigenstates by a standard numerical method that improves the approximation step by step. Since our analysis uses only this low-energy part of the spectrum, this procedure is sufficient, and for the present matrix size the resulting eigenvalues and eigenvectors are stable and reproducible at double precision.
}

\subsection{Optimal transport and Wasserstein distance matrix}

With the Husimi Q-representation of the energy eigenstates at our hands, we proceed to evaluate the optimal transport distance between those states with fixed coupling constant $g$. 
For the numerical evaluation of the 1-Wasserstein distance, we follow the method explained in Sec.~\ref{subsec:otwdm}.\footnote{See App.~\ref{sec:2-w} for the similar results using 2-Wasserstein distance.} 

The numerical setup for our evaluation is as follows. 
We take the phase-space discretization
\begin{align}
&\text{OT grid:}\quad q\in[-12,12],\ p\in[-6,6]\ \text{with}\ 9\times 9\times 9\times 9\ \text{points}.
\label{resolution}
\end{align}
We check that this range of the phase space is large enough to accommodate all the Husimi Q-representation of the energy eigenstates of our concern (the lowest 12 states in our chosen unfolded sector). That is, this grid is chosen after explicit boundary diagnostics to ensure negligible support loss at the phase-space boundary.
As for the Sinkhorn hyperparameter, in the present analysis we use
\begin{equation}
\varepsilon=0.08,\qquad N_{\rm iter}=500 .
\label{hyperparameters}
\end{equation}
This balances numerical stability and geometric fidelity for the present grid size.\footnote{Of course one can make the hyperparameters more minute with a finer grid, which may cost a lot more for the numerical evaluation of the Wasserstein distance matrix.
See Appendix~\ref{app:numerical-precision} for the convergence of the results with respect to $\varepsilon$.}

\begin{figure}[t]
\centering
\subfigure[$g=0$]{
\includegraphics[width=0.47\textwidth]{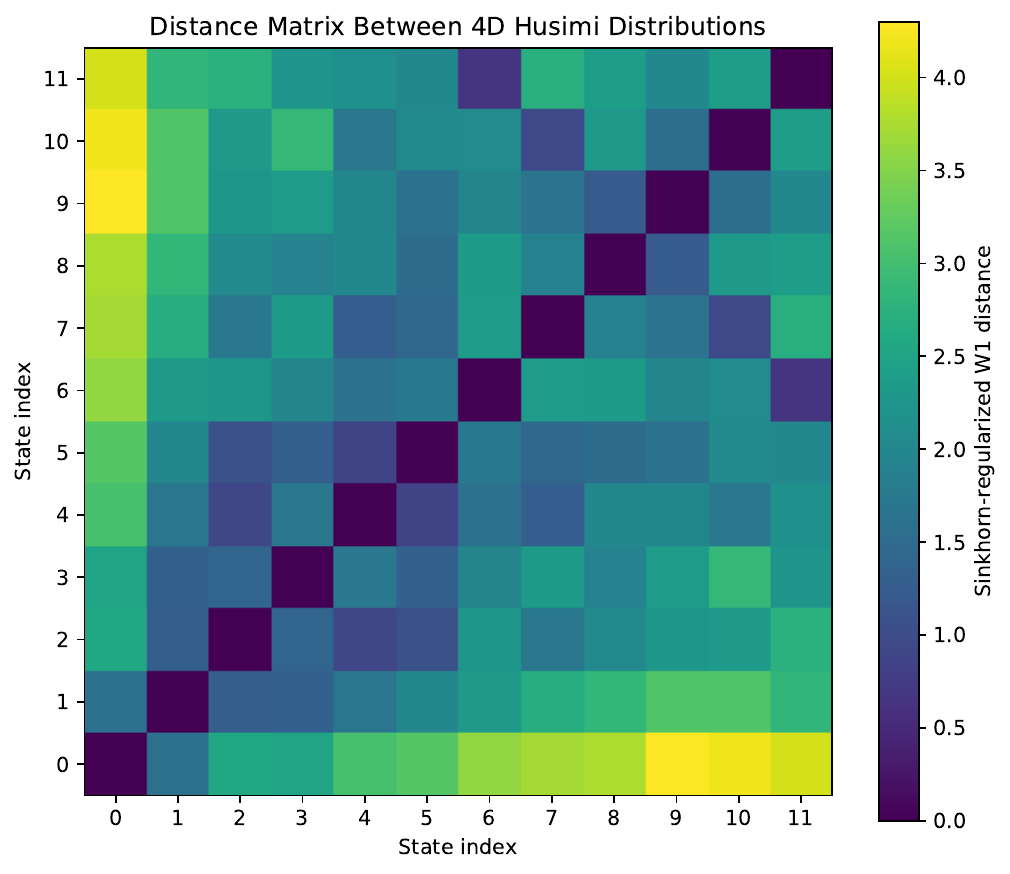}
}
\subfigure[$g=1/100$]{
\includegraphics[width=0.47\textwidth]{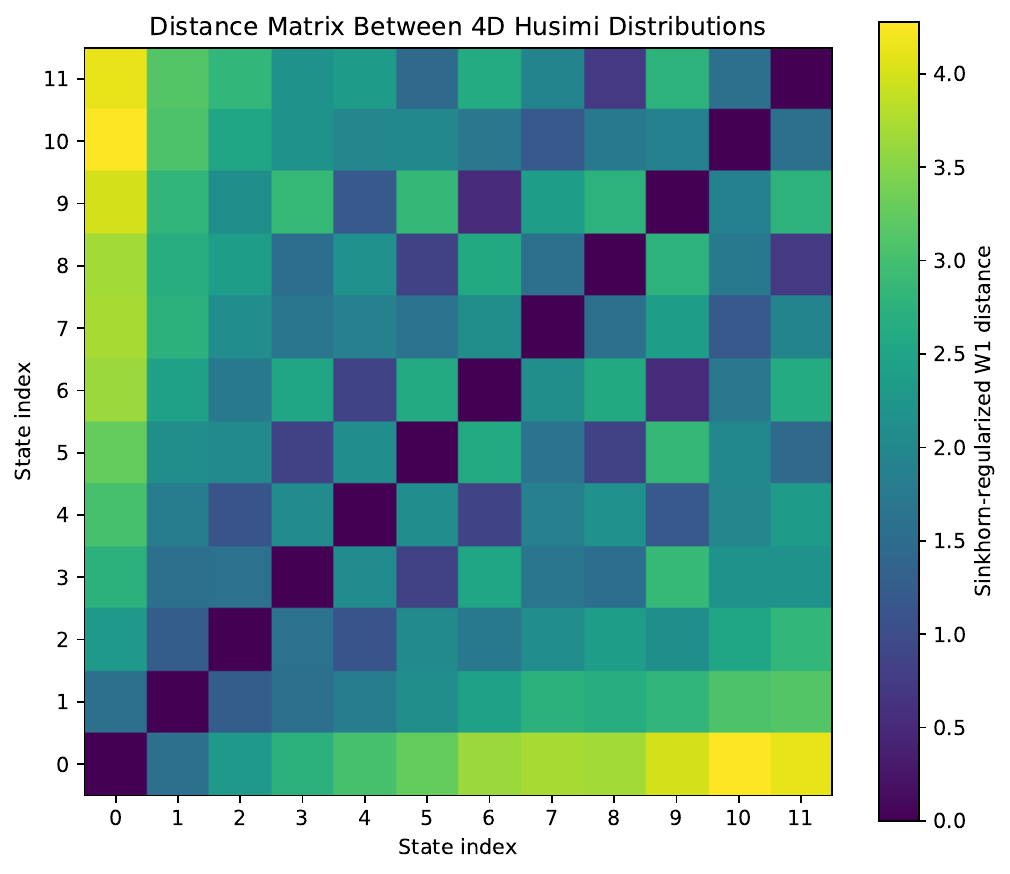}
}
\subfigure[$g=1/10$]{
\includegraphics[width=0.47\textwidth]{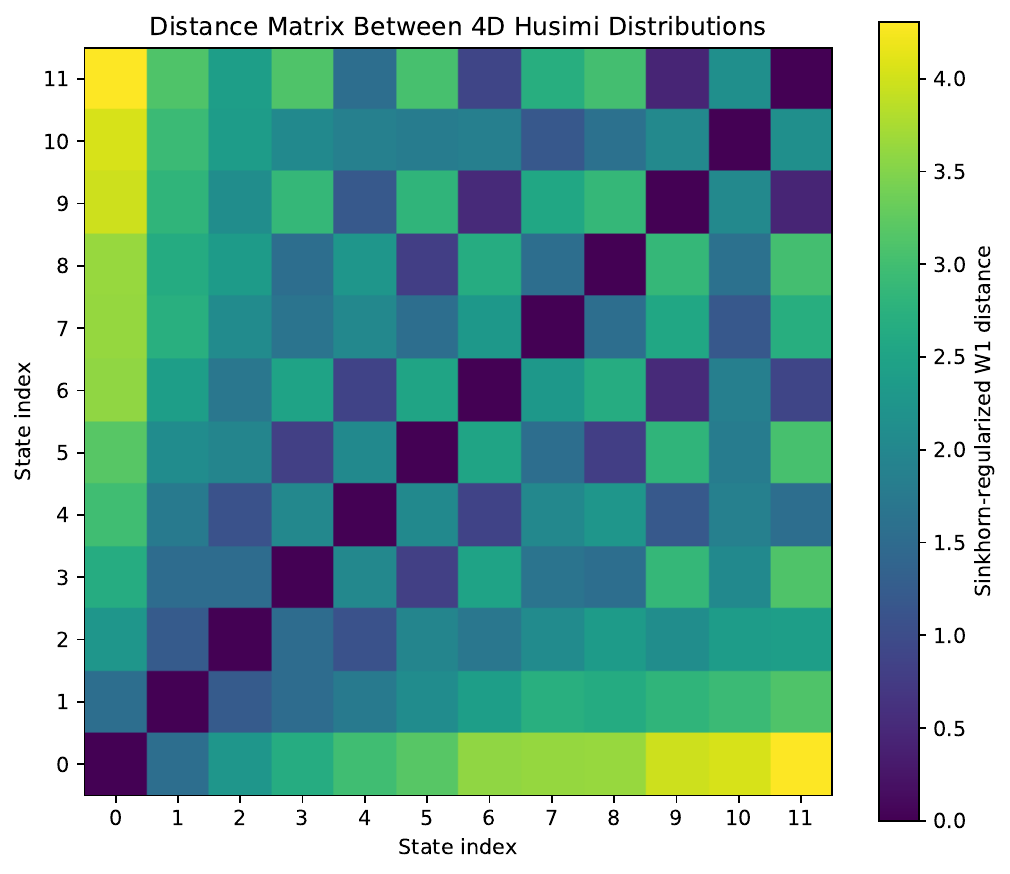}
}
\subfigure[$g=1$]{
\includegraphics[width=0.47\textwidth]{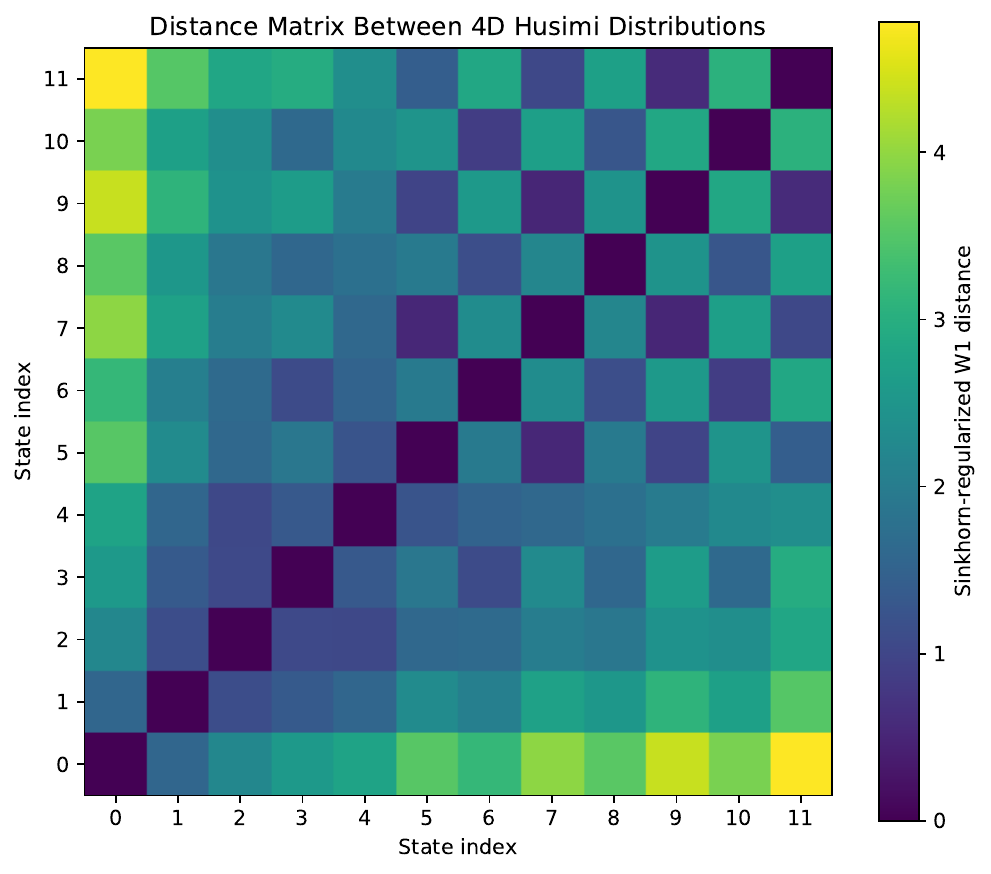}
}
\caption{Distance-matrix heatmaps for the four values of the coupling $g$.}
\label{fig:distmat-four-g}
\end{figure}

Let us present our numerical results.
Figure~\ref{fig:distmat-four-g} shows the full pairwise distance matrices for the four values of the coupling $g$.
The complete numerical matrices $D_{ij}$ (all entries, all couplings) are listed in
Tables~\ref{tab:dist-full-g-0}--\ref{tab:dist-full-g-1}, rounded to two decimals for brevity, so that the visual patterns in Fig.~\ref{fig:distmat-four-g} are therefore directly traceable to the explicit entries in these tables.

\begin{table}[p]
\centering
\tiny
\setlength{\tabcolsep}{5pt}
\begin{tabular}{c|cccccccccccc}
\hline
i/j & 0 & 1 & 2 & 3 & 4 & 5 & 6 & 7 & 8 & 9 & 10 & 11 \\
\hline
0 & 0.00 & 1.51 & 2.44 & 2.38 & 2.88 & 3.00 & 3.41 & 3.52 & 3.57 & 4.00 & 3.98 & 3.81 \\
1 & 1.51 & 0.00 & 1.23 & 1.26 & 1.60 & 1.90 & 2.19 & 2.53 & 2.69 & 2.97 & 2.97 & 2.68 \\
2 & 2.44 & 1.23 & 0.00 & 1.32 & 0.86 & 1.01 & 2.15 & 1.64 & 1.94 & 2.15 & 2.19 & 2.60 \\
3 & 2.38 & 1.26 & 1.32 & 0.00 & 1.64 & 1.23 & 1.87 & 2.22 & 1.82 & 2.23 & 2.71 & 2.12 \\
4 & 2.88 & 1.60 & 0.86 & 1.64 & 0.00 & 0.82 & 1.53 & 1.20 & 1.90 & 1.87 & 1.64 & 2.04 \\
5 & 3.00 & 1.90 & 1.01 & 1.23 & 0.82 & 0.00 & 1.64 & 1.34 & 1.44 & 1.53 & 1.94 & 1.90 \\
6 & 3.41 & 2.19 & 2.15 & 1.87 & 1.53 & 1.64 & 0.00 & 2.23 & 2.22 & 1.87 & 1.94 & 0.63 \\
7 & 3.52 & 2.53 & 1.64 & 2.22 & 1.20 & 1.34 & 2.23 & 0.00 & 1.77 & 1.57 & 0.90 & 2.55 \\
8 & 3.57 & 2.69 & 1.94 & 1.82 & 1.90 & 1.44 & 2.22 & 1.77 & 0.00 & 1.17 & 2.19 & 2.25 \\
9 & 4.00 & 2.97 & 2.15 & 2.23 & 1.87 & 1.53 & 1.87 & 1.57 & 1.17 & 0.00 & 1.45 & 1.90 \\
10 & 3.98 & 2.97 & 2.19 & 2.71 & 1.64 & 1.94 & 1.94 & 0.90 & 2.19 & 1.45 & 0.00 & 2.28 \\
11 & 3.81 & 2.68 & 2.60 & 2.12 & 2.04 & 1.90 & 0.63 & 2.55 & 2.25 & 1.90 & 2.28 & 0.00 \\
\hline
\end{tabular}
\caption{Full $12\times12$ distance matrix $D_{ij}$ for $g=0$ (rounded to two decimals).}
\label{tab:dist-full-g-0}
\end{table}

\begin{table}[p]
\centering
\tiny
\setlength{\tabcolsep}{5pt}
\begin{tabular}{c|cccccccccccc}
\hline
i/j & 0 & 1 & 2 & 3 & 4 & 5 & 6 & 7 & 8 & 9 & 10 & 11 \\
\hline
0 & 0.00 & 1.51 & 2.19 & 2.60 & 2.88 & 3.10 & 3.46 & 3.53 & 3.50 & 3.81 & 4.00 & 3.93 \\
1 & 1.51 & 0.00 & 1.20 & 1.51 & 1.69 & 2.00 & 2.33 & 2.60 & 2.53 & 2.68 & 2.93 & 2.98 \\
2 & 2.19 & 1.20 & 0.00 & 1.51 & 1.07 & 1.94 & 1.64 & 2.00 & 2.28 & 2.01 & 2.41 & 2.69 \\
3 & 2.60 & 1.51 & 1.51 & 0.00 & 1.94 & 0.79 & 2.39 & 1.60 & 1.45 & 2.72 & 2.10 & 2.10 \\
4 & 2.88 & 1.69 & 1.07 & 1.94 & 0.00 & 2.00 & 0.82 & 1.77 & 2.07 & 1.13 & 1.90 & 2.23 \\
5 & 3.10 & 2.00 & 1.94 & 0.79 & 2.00 & 0.00 & 2.47 & 1.57 & 0.79 & 2.71 & 1.90 & 1.41 \\
6 & 3.46 & 2.33 & 1.64 & 2.39 & 0.82 & 2.47 & 0.00 & 2.00 & 2.46 & 0.48 & 1.64 & 2.49 \\
7 & 3.53 & 2.60 & 2.00 & 1.60 & 1.77 & 1.57 & 2.00 & 0.00 & 1.51 & 2.25 & 1.13 & 1.84 \\
8 & 3.50 & 2.53 & 2.28 & 1.45 & 2.07 & 0.79 & 2.46 & 1.51 & 0.00 & 2.63 & 1.64 & 0.67 \\
9 & 3.81 & 2.68 & 2.01 & 2.72 & 1.13 & 2.71 & 0.48 & 2.25 & 2.63 & 0.00 & 1.77 & 2.63 \\
10 & 4.00 & 2.93 & 2.41 & 2.10 & 1.90 & 1.90 & 1.64 & 1.13 & 1.64 & 1.77 & 0.00 & 1.51 \\
11 & 3.93 & 2.98 & 2.69 & 2.10 & 2.23 & 1.41 & 2.49 & 1.84 & 0.67 & 2.63 & 1.51 & 0.00 \\
\hline
\end{tabular}
\caption{Full $12\times12$ distance matrix $D_{ij}$ for $g=1/100$ (rounded to two decimals).}
\label{tab:dist-full-g-1-100}
\end{table}

\begin{table}[p]
\centering
\tiny
\setlength{\tabcolsep}{5pt}
\begin{tabular}{c|cccccccccccc}
\hline
i/j & 0 & 1 & 2 & 3 & 4 & 5 & 6 & 7 & 8 & 9 & 10 & 11 \\
\hline
0 & 0.00 & 1.45 & 2.13 & 2.52 & 2.82 & 3.01 & 3.40 & 3.43 & 3.44 & 3.76 & 3.82 & 4.00 \\
1 & 1.45 & 0.00 & 1.17 & 1.44 & 1.66 & 1.97 & 2.28 & 2.57 & 2.49 & 2.66 & 2.78 & 2.94 \\
2 & 2.13 & 1.17 & 0.00 & 1.44 & 1.04 & 1.87 & 1.61 & 1.94 & 2.23 & 2.00 & 2.23 & 2.28 \\
3 & 2.52 & 1.13 & 1.44 & 0.00 & 1.90 & 0.77 & 2.36 & 1.57 & 1.45 & 2.71 & 1.93 & 2.94 \\
4 & 2.82 & 1.66 & 1.04 & 1.90 & 0.00 & 1.93 & 0.82 & 1.90 & 2.13 & 1.13 & 1.77 & 1.45 \\
5 & 3.01 & 1.97 & 1.87 & 0.77 & 1.93 & 0.00 & 2.38 & 1.45 & 0.75 & 2.66 & 1.69 & 2.88 \\
6 & 3.40 & 2.28 & 1.61 & 2.36 & 0.82 & 2.38 & 0.00 & 2.19 & 2.52 & 0.48 & 1.74 & 0.83 \\
7 & 3.43 & 2.57 & 1.94 & 1.57 & 1.90 & 1.45 & 2.19 & 0.00 & 1.45 & 2.44 & 1.10 & 2.55 \\
8 & 3.44 & 2.49 & 2.23 & 1.45 & 2.13 & 0.75 & 2.52 & 1.45 & 0.00 & 2.71 & 1.51 & 2.85 \\
9 & 3.76 & 2.66 & 2.00 & 2.71 & 1.13 & 2.66 & 0.48 & 2.44 & 2.71 & 0.00 & 1.93 & 0.42 \\
10 & 3.82 & 2.78 & 2.23 & 1.93 & 1.77 & 1.69 & 1.74 & 1.10 & 1.51 & 1.93 & 0.00 & 2.01 \\
11 & 4.00 & 2.94 & 2.28 & 2.94 & 1.45 & 2.88 & 0.83 & 2.55 & 2.85 & 0.42 & 2.01 & 0.00 \\
\hline
\end{tabular}
\caption{Full $12\times12$ distance matrix $D_{ij}$ for $g=1/10$ (rounded to two decimals).}
\label{tab:dist-full-g-1-10}
\end{table}

\begin{table}[p]
\centering
\tiny
\setlength{\tabcolsep}{5pt}
\begin{tabular}{c|cccccccccccc}
\hline
i/j & 0 & 1 & 2 & 3 & 4 & 5 & 6 & 7 & 8 & 9 & 10 & 11 \\
\hline
0 & 0.00 & 1.32 & 1.90 & 2.19 & 2.36 & 3.01 & 2.71 & 3.38 & 3.03 & 3.73 & 3.25 & 4.00 \\
1 & 1.32 & 0.00 & 0.94 & 1.13 & 1.32 & 1.94 & 1.74 & 2.33 & 2.15 & 2.65 & 2.33 & 3.00 \\
2 & 1.90 & 0.94 & 0.00 & 0.90 & 0.90 & 1.38 & 1.41 & 1.72 & 1.61 & 2.07 & 2.00 & 2.41 \\
3 & 2.19 & 1.13 & 0.90 & 0.00 & 1.13 & 1.61 & 0.91 & 1.94 & 1.34 & 2.23 & 1.38 & 2.52 \\
4 & 2.36 & 1.32 & 0.90 & 1.13 & 0.00 & 1.07 & 1.28 & 1.38 & 1.51 & 1.69 & 1.93 & 2.00 \\
5 & 3.01 & 1.94 & 1.38 & 1.61 & 1.07 & 0.00 & 1.66 & 0.44 & 1.66 & 0.82 & 2.12 & 1.20 \\
6 & 2.71 & 1.74 & 1.41 & 0.91 & 1.28 & 1.66 & 0.00 & 1.97 & 0.98 & 2.19 & 0.72 & 2.44 \\
7 & 3.38 & 2.33 & 1.72 & 1.94 & 1.38 & 0.44 & 1.97 & 0.00 & 1.87 & 0.44 & 2.28 & 0.90 \\
8 & 3.03 & 2.15 & 1.61 & 1.34 & 1.51 & 1.66 & 0.98 & 1.87 & 0.00 & 2.10 & 1.09 & 2.33 \\
9 & 3.73 & 2.65 & 2.07 & 2.23 & 1.69 & 0.82 & 2.19 & 0.44 & 2.10 & 0.00 & 2.44 & 0.50 \\
10 & 3.25 & 2.33 & 2.00 & 1.38 & 1.93 & 2.12 & 0.72 & 2.28 & 1.09 & 2.44 & 0.00 & 2.60 \\
11 & 4.00 & 3.00 & 2.41 & 2.52 & 2.00 & 1.20 & 2.44 & 0.90 & 2.33 & 0.50 & 2.60 & 0.00 \\
\hline
\end{tabular}
\caption{Full $12\times12$ distance matrix $D_{ij}$ for $g=1$ (rounded to two decimals).}
\label{tab:dist-full-g-1}
\end{table}

A physical observation on the distance matrix and the chaoticity follows here. Looking at Fig.~\ref{fig:distmat-four-g}, we notice that the heatmap for $g=1$ case has a more uniform top-right corner compared to the one for $g=0$.
In these heatmaps, toward the top-right corners, the energy grows. All the models are almost integrable at the ground state, but when the energy increases, the ordered phase changes to the chaotic phase. So we expect some signature of chaos at the top-right corner. The homogeneous heatmap for the $g=1$ case is interpreted as indicating that all the high-energy eigenstates of the $g=1$ system look quite similar to each other: the cost of transport between adjacent energy levels is not so different from that between energy levels far apart from each other. In other and simpler words, chaotic states
look mutually more similar.


\subsection{Dimensional reduction, manifold hypothesis and quantum chaos}

Let us proceed to the evaluation of the effective dimensions of the Wasserstein space. We will find that the effective dimensions decrease when the chaoticity grows. This means that a chaotic quantum system follows the manifold hypothesis --- the effective quantum state space is shrunken to a low-dimensional manifold.

We embed the energy eigenstates in a Euclidean space, following the strategy of \cite{hashimoto2026holography}. Then we look at the distribution of those points representing the states, to see the effective dimensions of the structure of the point cloud in the Euclidean space. The dimensions can be evaluated by the decay profile of the sorted Gram eigenvalues $\lambda_i$. To obtain the Gram matrix, see Sec.~\ref{sec:fold} for its definition and numerical evaluation.

\begin{figure}[t]
\centering
\includegraphics[width=0.85\textwidth]{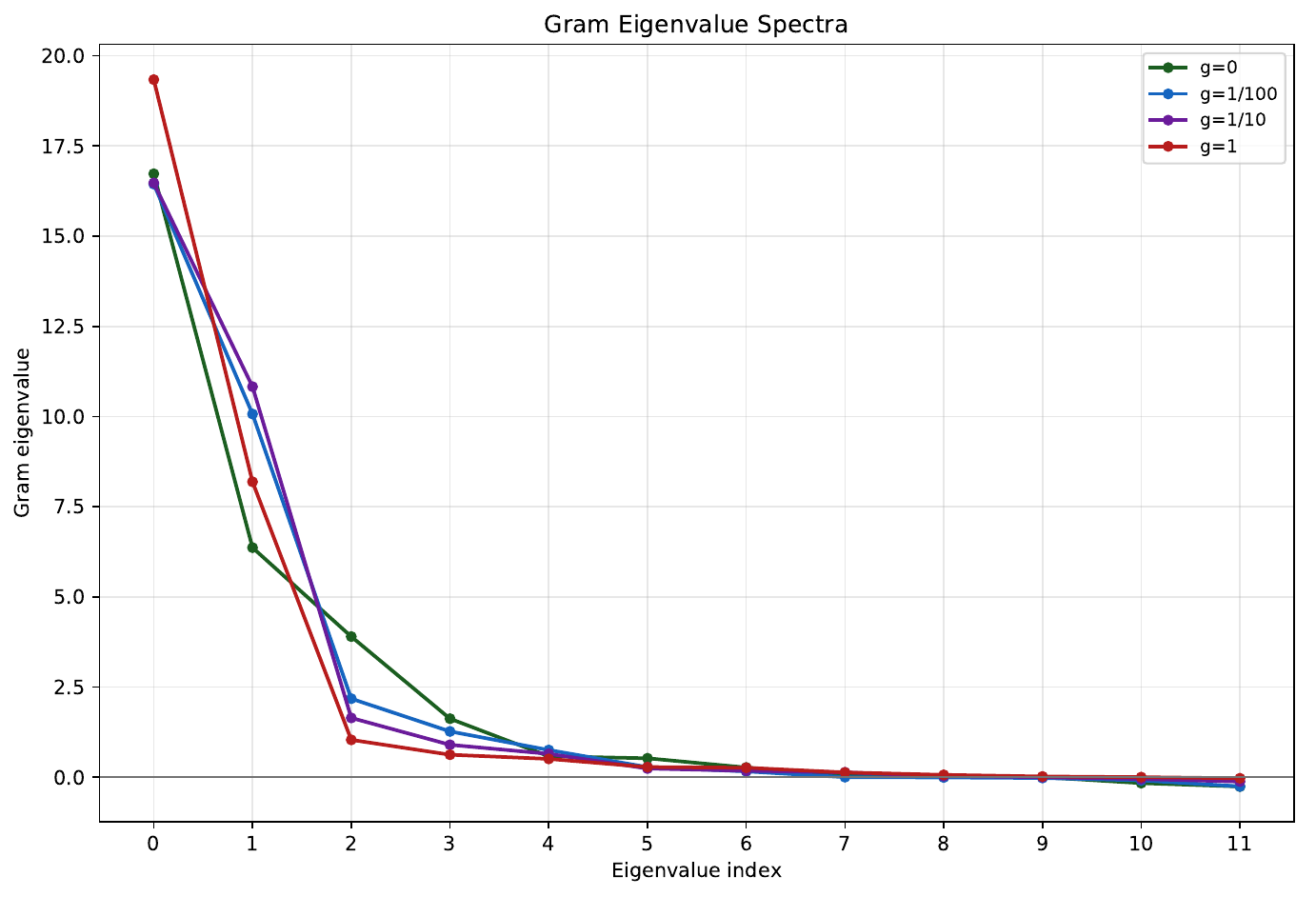}
\caption{Linear-scale Gram eigenvalue spectra for $g=0,1/100,1/10,1$.}
\label{fig:gram-linear}
\end{figure}

Fig.~\ref{fig:gram-linear} shows our results of the eigenvalues of the Gram matrix, for each case of $g=0,1/100,1/10,1$. A clear pattern is observed: when the coupling constant $g$ increases, the decay of the eigenvalues grows, meaning that the effective dimensions of the Wasserstein space decrease. Thus, we conclude that in the coupled harmonic oscillator system the Wasserstein space of Husimi Q-representation of the energy eigenstates achieves a lower-dimensional realization when the chaoticity parameter $g$ grows.

\begin{figure}[t]
\centering
\includegraphics[width=0.85\textwidth]{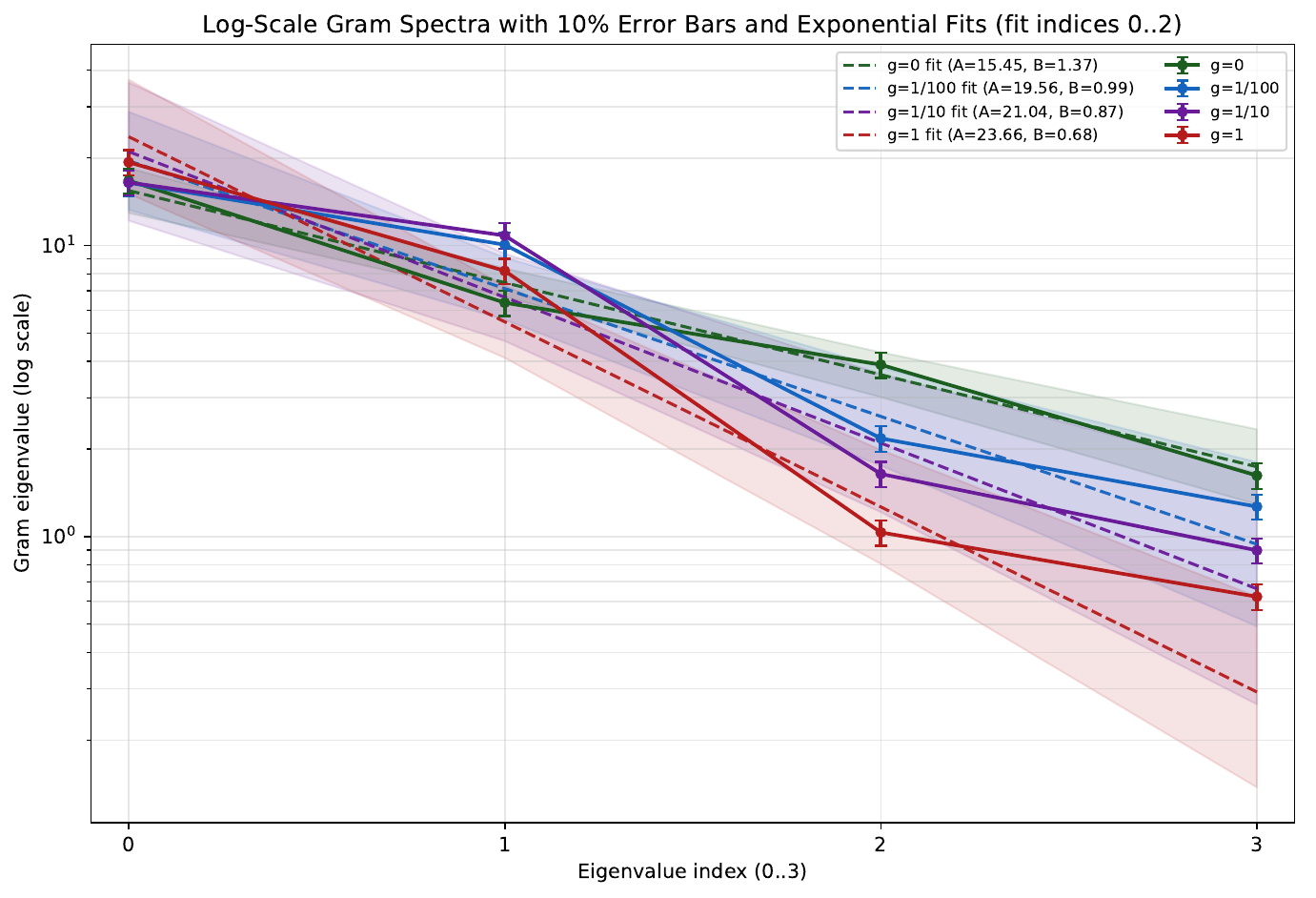}
\caption{Log-scale spectra (indices $0$--$3$ shown) with exponential-fit curves and $10\%$ error bars. Fits are performed using $k=0,1,2$.}
\label{fig:gram-logfit}
\end{figure}

To quantify the leading decay numerically, we fit
\begin{equation}
\lambda_k \simeq A\,e^{-k/B}
\end{equation}
using only first three major eigenvalues $k=0,1,2$ (with a $10\%$ relative uncertainty model on these points).
The selection of the first three is due to the fact that the sub-leading eigenvalues fluctuate due to our numerical discretization, as seen in Fig.~\ref{fig:gram-linear}. The value $10\%$ for the error bar is our hand-waving set, estimated by the fact that in fact one axis of the phase space is discretized into 9 points. 
See Fig.~\ref{fig:gram-logfit} for the logarithmic fitting of the Gram eigenvalues. The numerical values of the fitted slope are summarized in Table \ref{tab:gram-fit}.
As a goodness-of-fit indicator, we also report in Table~\ref{tab:gram-fit} the coefficient of determination
\begin{equation}
R^2 \equiv 1 - \frac{\sum_{k=0}^{2}\bigl(\log\lambda_k - \log\hat\lambda_k\bigr)^{2}}{\sum_{k=0}^{2}\bigl(\log\lambda_k - \overline{\log\lambda}\bigr)^{2}},
\label{eq:gram-R2}
\end{equation}
where $\hat\lambda_k = A\,e^{-k/B}$ are the fitted values and
$\overline{\log\lambda} = \tfrac{1}{3}\sum_{k=0}^{2}\log\lambda_k$ is the mean of the log eigenvalues used in the fit.

The central trend is the monotonic decrease of $B$ as $g$ increases:
\begin{equation}
B(g=0) > B(g=1/100) > B(g=1/10) > B(g=1).
\end{equation}
Hence, as chaoticity increases, Gram spectra become steeper and the state cloud is compressed into fewer dominant geometric directions.

\begin{table}[th]
\centering
\begin{tabular}{c|ccc}
\hline
$g$ & $A$ & $B$ & $R^2$ \\
\hline
$0$ & $15.4533$ & $1.3729$ & $0.9685$ \\
$1/100$ & $19.5609$ & $0.9889$ & $0.8179$ \\
$1/10$ & $21.0352$ & $0.8674$ & $0.6564$ \\
$1$ & $23.6611$ & $0.6829$ & $0.8466$ \\
\hline
\end{tabular}
\caption{Exponential fit parameters for Gram-spectrum decay, $\lambda_k\sim A e^{-k/B}$, fitted at $k=0,1,2$. A smaller $B$ means faster decay and stronger dimensional reduction.}
\label{tab:gram-fit}
\end{table}

Within the coupled-oscillator family studied here, larger $g$ (stronger chaotic mixing) correlates with stronger effective dimensional reduction in the Gram-spectrum sense.
Operationally, the leading Husimi-distance geometry is captured by fewer dominant modes as chaoticity increases.
This provides a concrete, transport-based quantitative signature of chaos-driven geometric compression in low-energy quantum states.

Let us emphasize the connection to the holographic principle. The endeavor started with the popularly known fact that maximum quantum chaos of the boundary quantum system generally corresponds to black holes in the bulk in holography, and combining it with the claim of \cite{hashimoto2026holography} that the dimensional reduction in the Wasserstein space along the manifold hypothesis can be a working principle for the holography, we were led to study the relation between the quantum chaos and the dimensional reduction of the Wasserstein space. The coupled harmonic oscillator, which is quantum chaotic, captures the essential part of Yang-Mills theory  \cite{bir1994chaos} or matrix models, and we know that the strong coupling limit needs to be taken for the holography to work in those models. Here in this section, the demonstrated dimensional reduction of the state space confirms that the optimal transport and the emergent Wasserstein space are consistent with the holographic principle in this sense. 


\section{Quantum scars as Wasserstein branch}
\label{sec:scar}

The dimensional reduction which we observed in the previous section shows that the effective dimensions of the Wasserstein space with the strong coupling constant $g=1$ reduce to two. Then how is the two-dimensional space structured? In this section we find a peculiar structure in the Wasserstein space for the model of the coupled harmonic oscillator in Sec.~\ref{sec:coupled-ho}.  We call it ``Wasserstein branch." And we argue that the physical meaning of the Wasserstein branch is quantum scar.

\begin{figure}[t]
\centering
\subfigure[$g=1/100$ embedding]{
\includegraphics[width=0.47\textwidth]{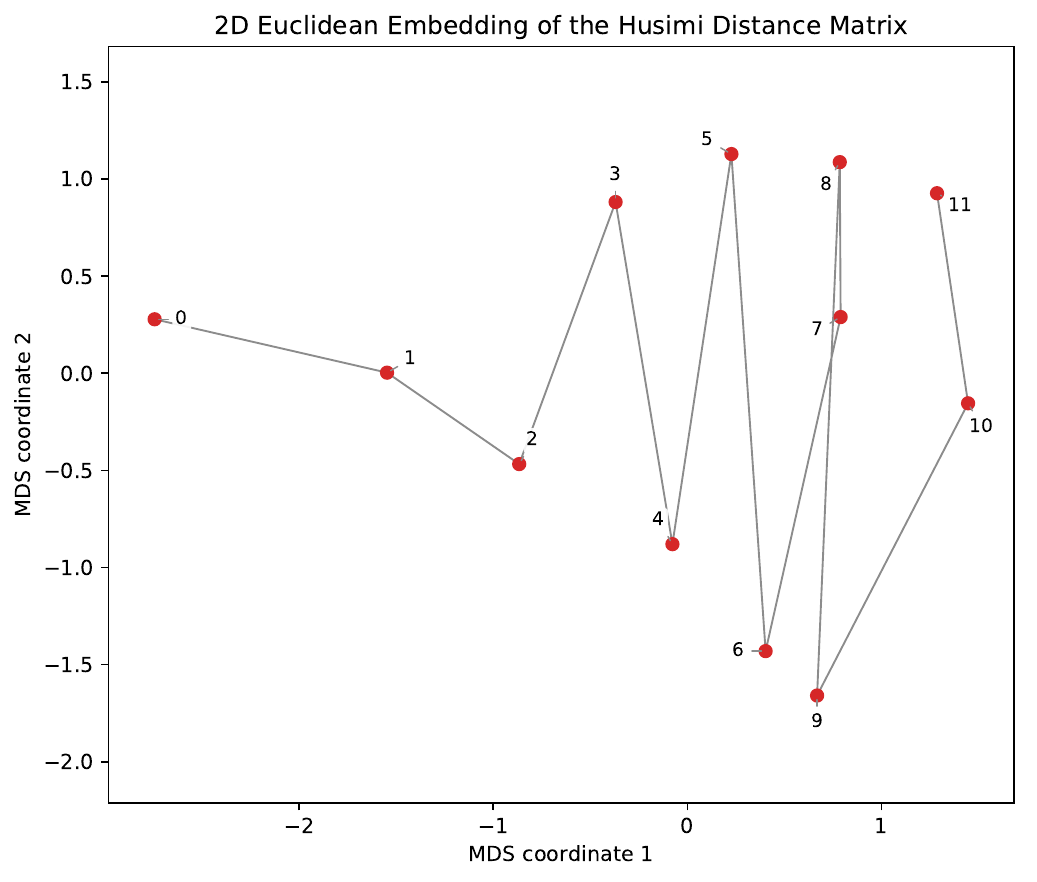}
}
\subfigure[$g=1$ embedding]{
\includegraphics[width=0.47\textwidth]{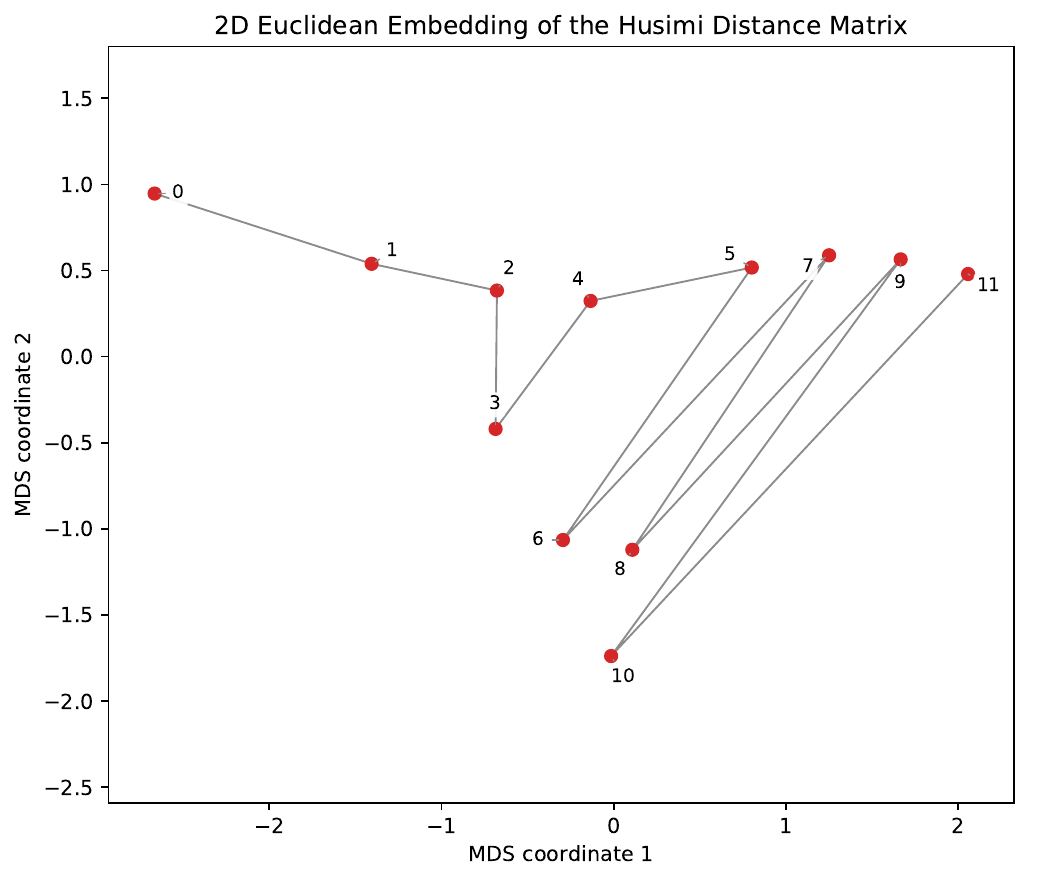}
}
\caption{Two-dimensional point-cloud embeddings from the distance matrices, shown for $g=1/100$ and $g=1$.}
\label{fig:embed-g1100-g1}
\end{figure}

First, see the 2-dimensional projection of the point cloud in the embedding Euclidean space, Fig.~\ref{fig:embed-g1100-g1}. It compares the state-point embeddings for a weakly coupled case ($g=1/100$) and a strongly coupled case ($g=1$), of the Wasserstein space of the coupled harmonic oscillator in Sec.~\ref{sec:coupled-ho}. To our eyes, as a whole, they look rather similar to each other. This means that the dimensional reduction occurs rather at higher dimensions, not in the projected one or two dimensions. This tendency is already observed in Fig.~\ref{fig:gram-linear}; the dimensional reduction is apparent for the eigenvalue index $k=2$ and $k=3$, which amounts to the effective dimensions three and four. 

If we look at the figure closer, we find that the $g=1$ embedding in Fig.~\ref{fig:embed-g1100-g1} shows an interesting structure in two dimensions: there appears a branching of the Wasserstein space, and the branch formed by the energy eigenstates $n=5, 7, 9, 11$, which we call a Wasserstein branch, seems to form a specific line structure which mimics the Wasserstein space of a 1-dimensional harmonic oscillator (remember \cite{hashimoto2026holography} and panel (a) of Fig.~\ref{fig:oned-embedding-final}). This observation suggests that there exists a quantum subsector similar to the 1-dimensional harmonic oscillator, hidden in this chaotic model. This subsector is separated from the rest, meaning that the states that belong to this Wasserstein branch are almost independent from the rest.

In fact, in the context of research on quantum scars, the model has been studied \cite{santhanam1998chaos} to result in the existence of highly localized quantum states which are along classically periodic orbits. These states are considered to be a certain version of the quantum scars. Motivated by this, let us explore the branch in the Wasserstein space from the viewpoint of the quantum scars.

We look at the Husimi Q-representations of the relevant energy eigenstates $n=5,7,9,11$ in the panel (b) ($g=1$ case) of Fig.~\ref{fig:husimi-four-g-b}. We see a clear resemblance among those states: the $p_x=p_y=0$ section of the Husimi Q-representation of those states all have four dots, the strong localization on the $x$ and $y$ axes a little away from the origin. Due to this similarity to each other, the Wasserstein distance among those gets shorter, forming a branch in the Wasserstein space. 

\begin{figure}[t]
\centering
\includegraphics[width=0.95\textwidth]{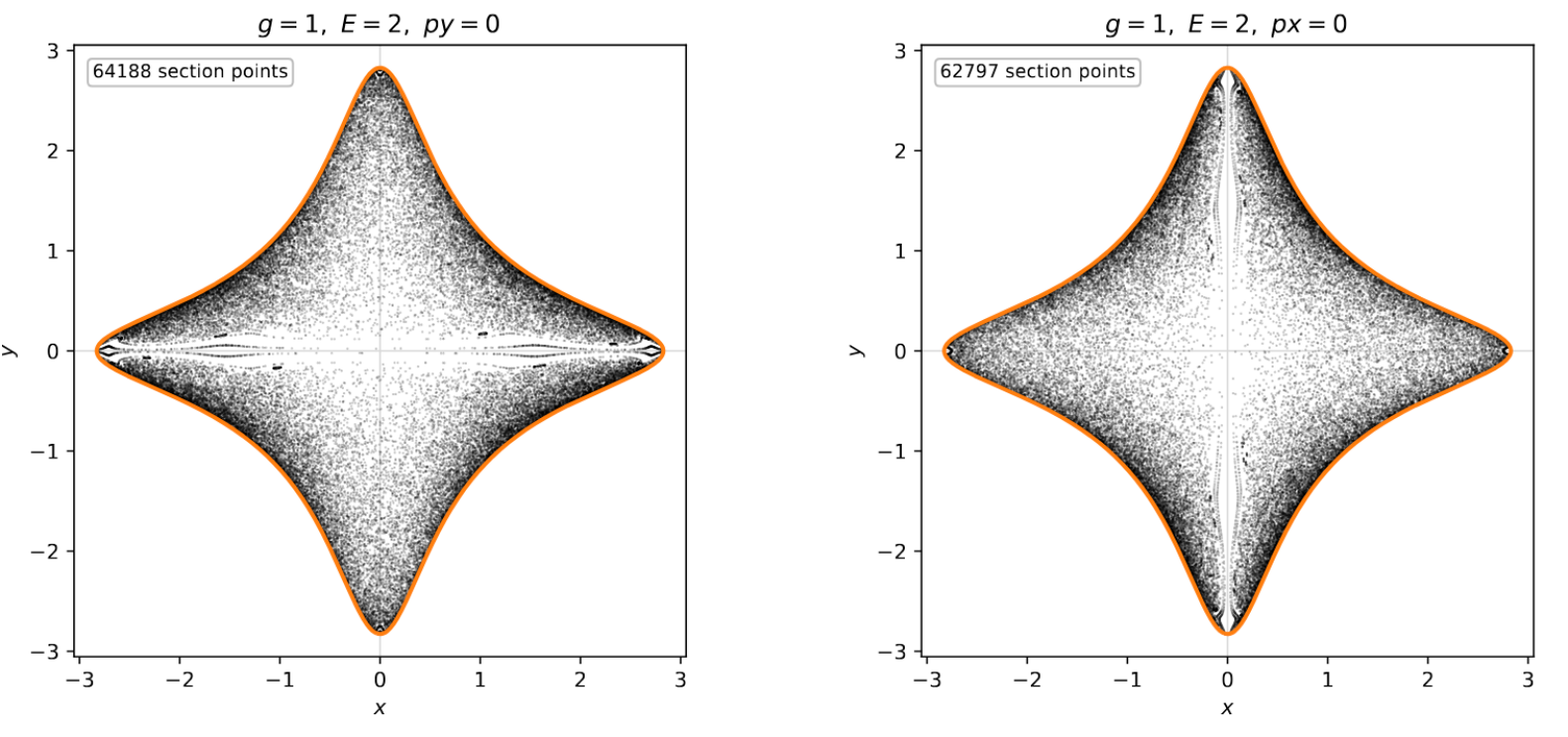}
\caption{Classical Poincare sections for $g=1$ at energy $E=2$. Left: $p_y=0$ section projected onto $x$-$y$ plane. Right: $p_x=0$ section projected onto $x$-$y$ plane.}
\label{fig:poincare}
\end{figure}

The structure in the Husimi Q-representation suggests the existence of periodic stable orbits and the regular islands in the chaotic sea of the classical model of the coupled harmonic oscillator. Let us study the Poincar\'e section of the classical coupled harmonic oscillator. The result is shown in Fig.~\ref{fig:poincare}. To draw these figures,
we fix the total energy to $H=E=2$ and compute two section plots in the configuration plane $(x,y)$, namely the crossings with $p_y=0$ and with $p_x=0$.\footnote{  
Initial conditions are sampled randomly on the energy shell $H=2$.  
For each section, we use 208 random initial conditions and integrate each orbit for 24000 steps, discarding the first 900 steps as burn-in.  
Whenever the chosen momentum component changes sign between two consecutive steps, we linearly interpolate the crossing point to estimate the section condition $p_y=0$ or $p_x=0$. The corresponding configuration-space point $(x,y)$ is then recorded and plotted.}  
The displayed point clouds show regular island structures near the coordinate axes. The pattern of the location of the regular islands looks quite similar to those of our Husimi Q-representation, the $n=5,7,9,11$ states in (b) of Fig.~\ref{fig:husimi-four-g-b}. Thus this similarity suggests that the Wasserstein branch came from the regular islands of the Poincar\'e section. The quantum states are in the subsector of the total Hilbert space and the subsector keeps a certain integrability which mimics the structure of the one-dimensional harmonic oscillator: a regular fluctuation around the periodic orbits.

\begin{figure}[t]
\centering
\includegraphics[width=0.45\textwidth]{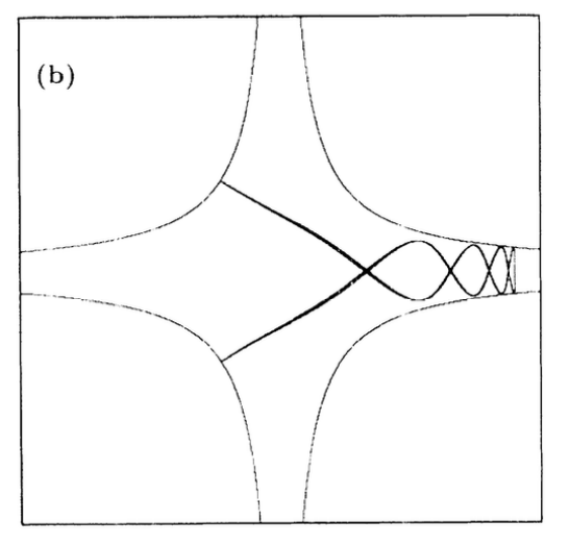}
\caption{A periodic orbit found in \cite{dahlqvist1990existence} for the model $H=(1/2)(p_x^2+p_y^2+x^2y^2)$. Reproduced from Fig.~1(b) in \cite{dahlqvist1990existence}, Copyright (1990) by the American Physical Society.}
\label{fig:regularo}
\end{figure}

The existence of such regular orbits of the model (without the harmonic term $x^2+y^2$) was shown in \cite{dahlqvist1990existence,marcinek1994yang} in the context of the non-ergodicity proof of the model. See Fig.~\ref{fig:regularo} for such an orbit. The orbit bounces many times at the valley region near the axes, and tends to stay near that part of the $x$-$y$ plane for a long time, suggesting the quantum wave function to be enhanced there.

\begin{figure}[t]
\centering
\subfigure[Poincar\'e section for $E=12$]{
\includegraphics[width=0.95\textwidth]{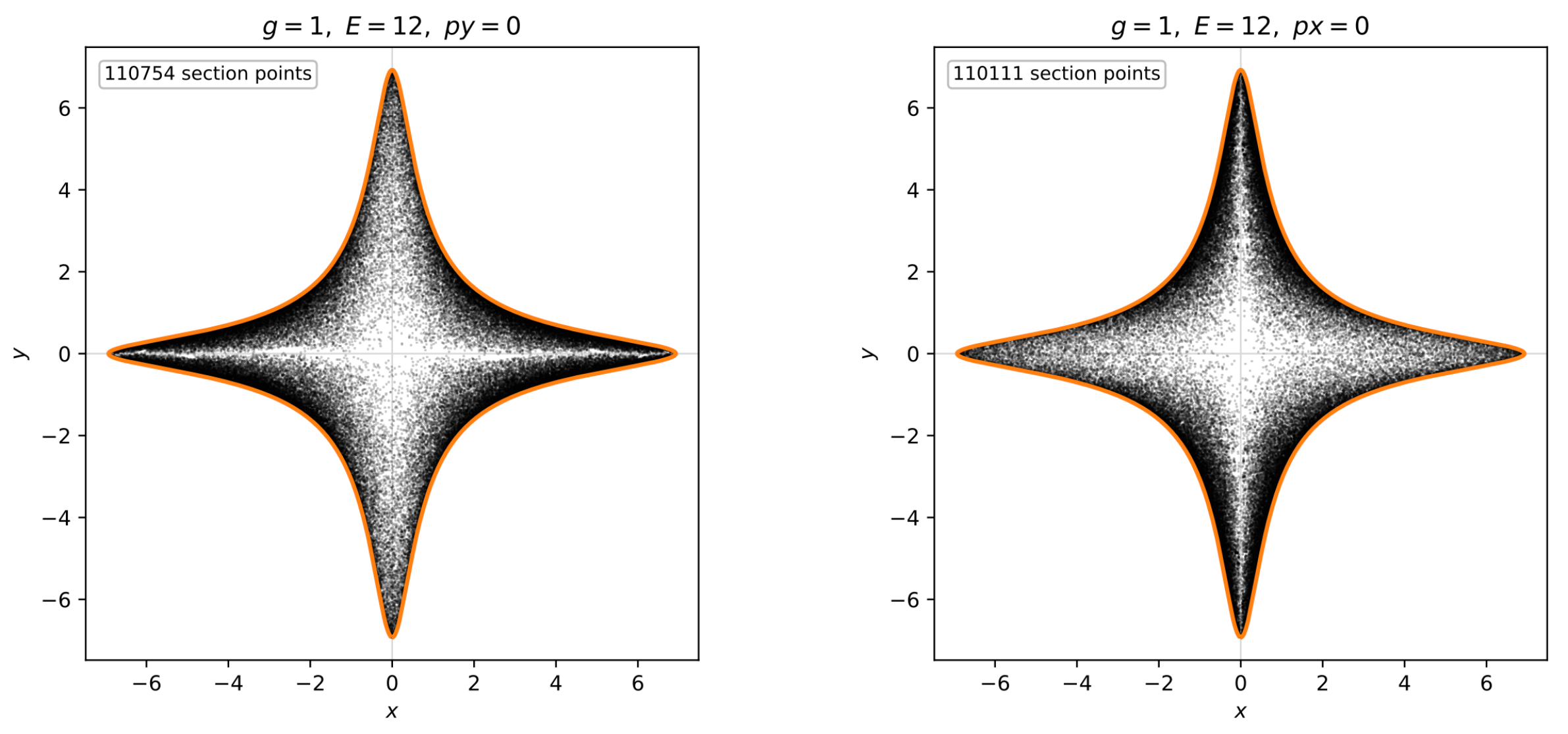}
}
\subfigure[Magnified plot of (a)]{
\includegraphics[width=0.95\textwidth]{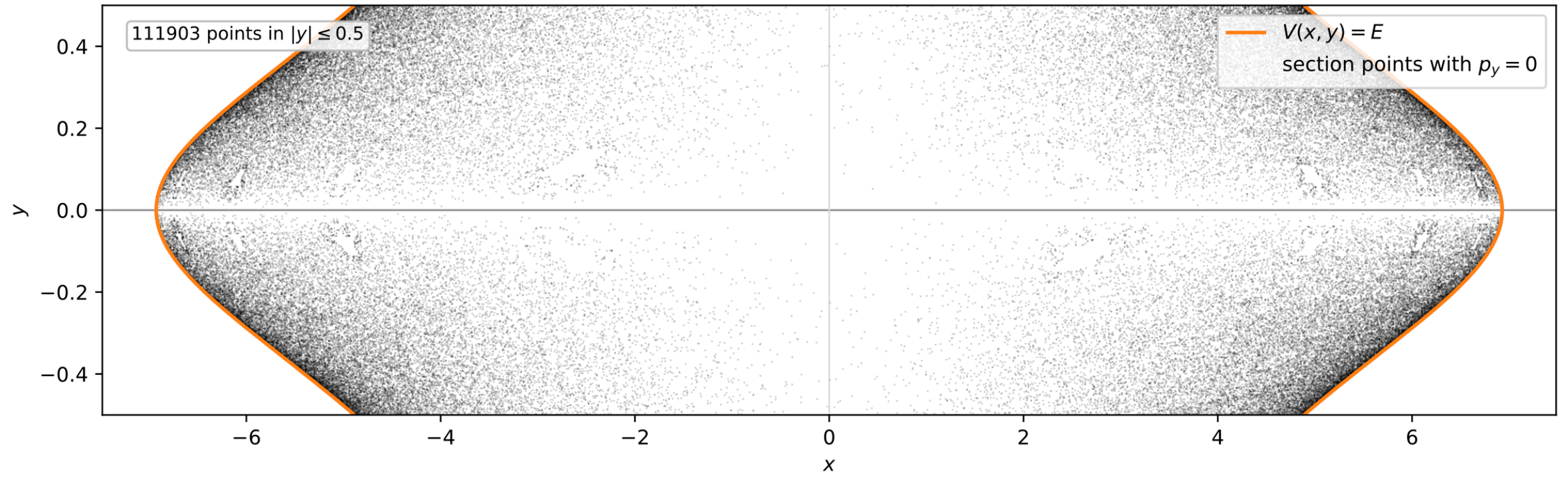}
}
\caption{Classical Poincare sections for $g=1$ at energy $E=12$. (a) Left: $p_y=0$ section projected onto $x$-$y$ plane. (a) Right: $p_x=0$ section projected onto $x$-$y$ plane. (b) Magnified plot of Figure (a) Left.}
\label{fig:poincare2}
\end{figure}

However, this interpretation of regular islands producing the Wasserstein branch there leaves a subtlety. The states in the Wasserstein branch are $n=5, 7, \cdots$ and the energy eigenvalues of those are $E=10.40, 12.61, \cdots$,
while the Poincar\'e section we drew in Fig.~\ref{fig:poincare} is much lower energy, $E=2$. At higher energy the chaotic sea grows, and in Fig.~\ref{fig:poincare2} we plot the Poincar\'e section for $E=12$. As we expected, the regular islands cannot be seen clearly, and most of the region is chaotic. Tiny regular islands can be spotted in the magnified plots of the Poincar\'e section, see panel (b) of Fig.~\ref{fig:poincare2}. These islands may be too small to host the whole Wasserstein branch.

So what is the interpretation of this Wasserstein branch? Our tentative conclusion is that it may be detecting quantum scars.\footnote{Here we use the term ``quantum scar'' in a phenomenological sense, based on the observed localization pattern and its correspondence with the classical axial orbits, rather than as a proof by an exact null-state construction.} Quantum scars are anomalous enhancements of eigenfunction density along short classical periodic orbits in classically chaotic systems. Regular orbits are stable classical periodic orbits, while the quantum scars can emerge even from unstable periodic orbits. \cite{santhanam1998chaos} showed that such channel-scarred states can be exponentially localized with localization correlated with the orbit’s local phase-space stability and pitchfork bifurcations. In our case, the $E=2$ regular islands are split into smaller $E=12$ islands, thus from $E=2$ to $E=12$ the bifurcation of the stable orbits is expected. Therefore our Wasserstein branch may be interpreted as a quantum scar.

\begin{figure}[t]
\centering\subfigure[The quantum scar state found in \cite{santhanam1998chaos}.]{
\includegraphics[width=0.45\textwidth]{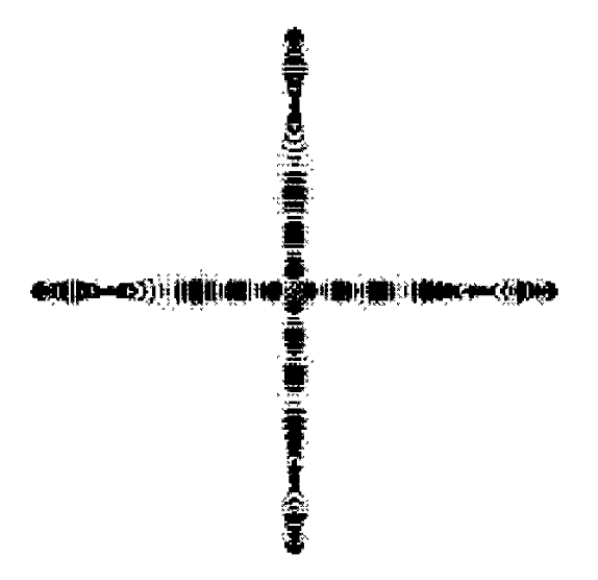}}
\centering\subfigure[The probability distribution, $g=1$.]{
\includegraphics[width=0.9\textwidth]{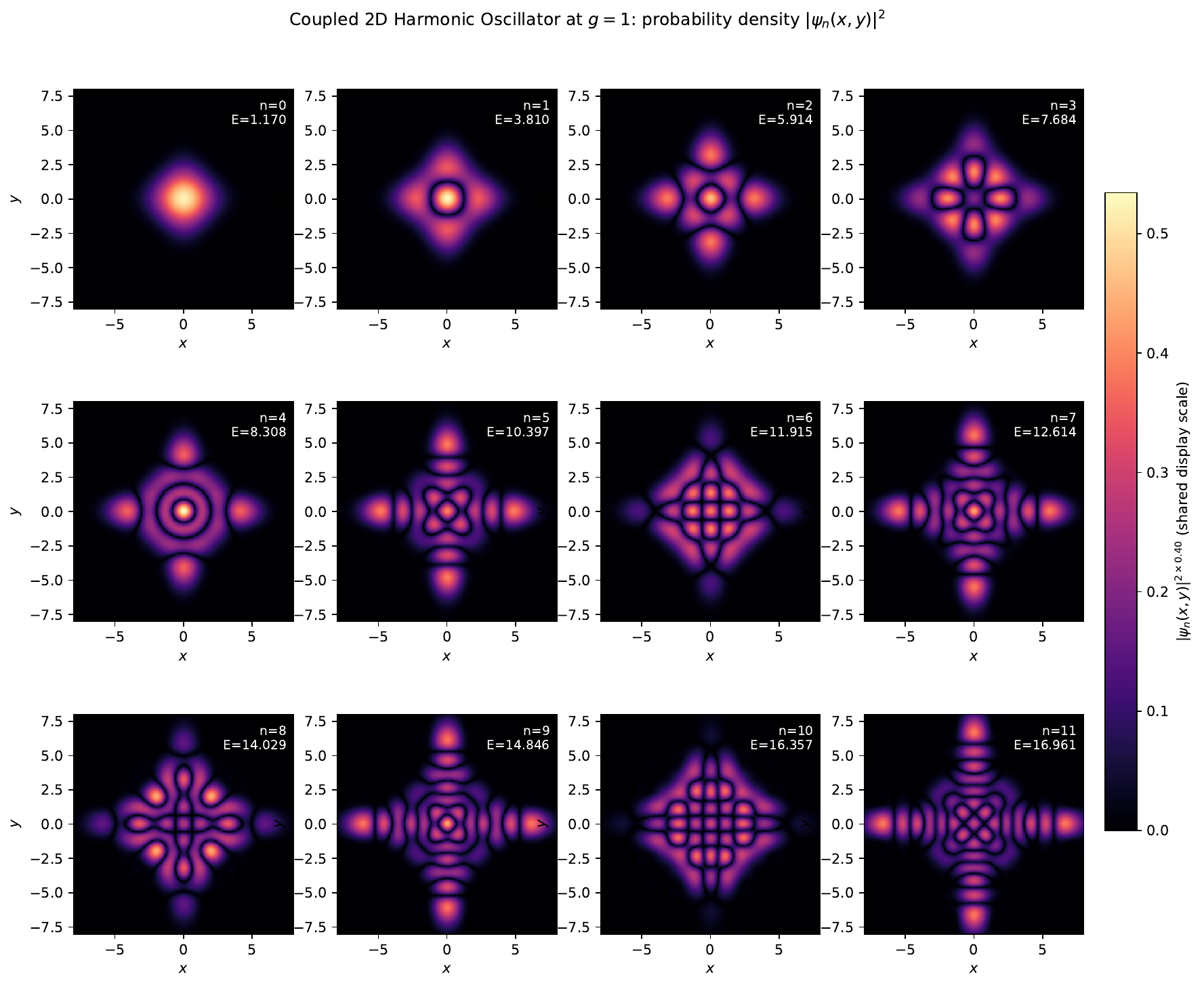}}
\caption{(a) The probability distribution of the quantum scar state found in \cite{santhanam1998chaos}. Reproduced from Fig.~4 Right in \cite{santhanam1998chaos}, Copyright (1998) by the American Physical Society. (b) The probability distribution of our model with $g=1$. The states in the Wasserstein branch, $n=5,7,9,11$, look quite similar to (a).}
\label{fig:scar}
\end{figure}

In \cite{santhanam1998chaos}, the study of a similar model, $H=(1/2)(p_x^2+p_y^2)+x^4+y^4+90x^2y^2$ revealed that there exists a well-localized quantum state, see Fig.~\ref{fig:scar} (a). The wave function is localized along the lines $x=0$ or $y=0$, and these lines are the flat direction of the potential, thus are the classical periodic orbits. And in Fig.~\ref{fig:scar} (b) we plot the probability distribution of our energy eigenstates for $g=1$. The states which belong to the Wasserstein branch, $n=5,7,9,11$, in fact have the same structure. So these states are made along the classical periodic orbits and are highly localized, thus are quantum scars.

Our interpretation is along the lines of methods for constructing quantum scars. In \cite{shibata2020onsager}, a quantum scar was constructed by perturbing an integrable system by a non-integrable deformation $H_{\rm pert}$, while finding a state which satisfies $H_{\rm pert}\ket{\psi}=0$. In our case, the integrable Hamiltonian is the harmonic oscillator, while the perturbation $H_{\rm pert}=gx^2y^2$ is non-integrable, and the states localized at $x=0$ or $y=0$ satisfy $H_{\rm pert}\ket{\psi}=0$.\footnote{More precisely, the branch states are not exact solutions of $H_{\rm pert}\ket{\psi}=0$. Since they have finite transverse width, the operator $H_{\rm pert}=g x^2 y^2$ does not annihilate them exactly. Rather, they are strongly localized near the coordinate axes, where $x^2y^2$ is small, so the perturbation is comparatively weak on these states. In this sense the relation to \cite{shibata2020onsager} is heuristic rather than exact.
}
Therefore, our results suggest that the Wasserstein branch is formed by scar-like states, and that this branch structure may persist to higher energy.
The number of states other than those on this Wasserstein branch will grow at higher energy, and the system will be saturated by those typical chaotic states. The main structure of the Wasserstein space is made of those chaotic states, while the Wasserstein branch will stay as a tiny but structured branch. The fact that the Wasserstein branch is away from the other states is consistent with the fact that quantum scar states are non-thermal, since even in a thermal bath the branch states are far away from the rest so it will take a lot of cost for heat perturbation to reach those states.

In summary, our study in this section shows that the Wasserstein space can be one of the diagnostics for spotting the quantum scar states. Further demonstration of the connection between the Wasserstein branches and quantum scars in many quantum models is necessary to test this hypothesis. If the dimensional reduction is generic in chaotic systems, the quantum scars can be spotted in this reduced Wasserstein space as branches.


\section{Quantum Lyapunov and Wasserstein distance}
\label{sec:lyap}

So far in this paper we have seen the relations between the structure of the Wasserstein space and the quantum scrambling, the quantum chaos or the quantum scar. To explore the origin of these relations, in this section we study a time evolving state rather than the energy eigenstates to see the Lyapunov exponent directly in the optimal transport of quantum states. To this end, we consider a Gaussian state in the inverted harmonic oscillator, which allows us to calculate everything analytically in its time evolution. This state exhibits a positive Lyapunov exponent --- indeed it is calculated in three different manners: (i) microcanonical OTOC \cite{Hashimoto:2017oit}, (ii) half-probability contour length \cite{toda1986quantal,toda1987quantal}, and (iii) Wasserstein distance. We show that they coincide, meaning that the Wasserstein distance captures the quantum Lyapunov exponent properly in this case.


As a preparation, for an inverted harmonic oscillator, we calculate the time evolution of a Gaussian wave packet, its Husimi representation and the associated Gaussian covariance data. These are to be used for (ii) the half-probability contour length and (iii) the Wasserstein distance between initial and evolved Husimi distributions later.

The Hamiltonian for the inverted harmonic oscillator is
\begin{equation}
H=\frac{p^2}{2m}-\frac{1}{2}m\Omega^2 x^2.
\end{equation}
To obtain an analytic description of the time evolution of a quantum state, we consider an ansatz
\begin{equation}
\psi(t,x)=f(t)\exp[-x^2q(t)],
\end{equation}
with complex functions $f(t), q(t)$.
Substituting into the Schr\"odinger equation yields
\begin{equation}
i\frac{\dot f}{f}=\frac{1}{m}q,
\qquad
-i\dot q+\frac{2}{m}q^2+\frac{m}{2}\Omega^2=0.
\end{equation}
Introducing $h(t)$ with
\begin{equation}
q=-\frac{im}{2}\frac{\dot h}{h},
\end{equation}
we find that this $h(t)$ satisfies a simple equation $\ddot h-\Omega^2h=0$, which can be solved as 
\begin{equation}
h=A\cosh(\Omega t)+B\sinh(\Omega t).
\end{equation}
Thus we obtain
\begin{equation}
q(t)
=-\frac{im\Omega}{2}
\frac{A\sinh(\Omega t)+B\cosh(\Omega t)}
{A\cosh(\Omega t)+B\sinh(\Omega t)},
\end{equation}
and from $i\dot f/f=q/m$, we find a solution
\begin{equation}
f=C\,h^{-1/2}
\end{equation}
where $C$ is the overall normalization of the wave function.

Let us impose the initial condition $q(0)=1/(2\sigma_0^2)$ where $\sigma_0$ is the initial width of the Gaussian wave function, then it fixes the integration constant as
\begin{equation}
\frac{A}{B}=-im\Omega\sigma_0^2.
\end{equation}
Hence we obtain the explicit form of the time evolution of the Gaussian wave function as
\begin{align}
q(t)
&=
-\frac{im\Omega}{2}
\frac{-im\Omega\sigma_0^2\sinh(\Omega t)+\cosh(\Omega t)}
{-im\Omega\sigma_0^2\cosh(\Omega t)+\sinh(\Omega t)},
\\
f(t)
&=
\tilde C\,
\left(-im\Omega\sigma_0^2\cosh(\Omega t)+\sinh(\Omega t)\right)^{-1/2}.
\end{align}

Let us calculate the Husimi Q-representation of this state. With coherent-state wavefunction
\begin{equation}
\langle x|\alpha\rangle=
\left(\frac{m\omega}{\pi}\right)^{1/4}
\exp\!\left[
-\frac{m\omega}{2}(x-x_0)^2
+ip_0x-\frac{i}{2}x_0p_0
\right],
\end{equation}
\begin{equation}
x_0=\sqrt{\frac{2}{m\omega}}\Re\alpha,
\qquad
p_0=\sqrt{2m\omega}\Im\alpha,
\end{equation}
the Gaussian integration gives
\begin{align}
\langle\psi|\alpha\rangle
&=
\sqrt{\pi}\,
\left(q^*+\frac{m\omega}{2}\right)^{-1/2}
f^*\left(\frac{m\omega}{\pi}\right)^{1/4}
\nonumber\\
&\hspace{0.6cm}\times
\exp\!\left[
\frac{1}{4}\frac{(m\omega x_0+ip_0)^2}{q^*+m\omega/2}
-\frac{m\omega}{2}x_0^2-\frac{i}{2}x_0p_0
\right].
\end{align}
Therefore the Husimi Q-representation is
\begin{align}
Q(\alpha)
&=
\left|q+\frac{m\omega}{2}\right|^{-1}|f|^2\left(\frac{m\omega}{\pi}\right)^{1/2}
\nonumber\\
&\hspace{0.2cm}\times
\exp\!\left[
\frac{1}{4}\frac{(m\omega x_0+ip_0)^2}{q^*+m\omega/2}
+\frac{1}{4}\frac{(m\omega x_0-ip_0)^2}{q+m\omega/2}
-m\omega x_0^2
\right].
\label{eq:husimitg}
\end{align}
To extract geometric information on the shape of this Gaussian, it is convenient to define
\begin{equation}
\frac{2}{m\omega}q=G_1+iG_2,
\end{equation}
which means, with $a\equiv m\Omega\sigma_0^2$,
\begin{equation}
G_1=
a\frac{\Omega}{\omega}\frac{1}{\sinh^2\Omega t+a^2\cosh^2\Omega t},
\qquad
G_2=
-(1+a^2)\frac{\Omega}{\omega}
\frac{\cosh\Omega t\,\sinh\Omega t}{\sinh^2\Omega t+a^2\cosh^2\Omega t}.
\label{eq:note-g1g2}
\end{equation}
Then the exponent of the Husimi Q-representation \eqref{eq:husimitg} becomes a quadratic form
\begin{equation}
\text{exponent}=-\frac12\,\vec\alpha^{\,T}\Sigma^{-1}\vec\alpha,
\qquad
\vec\alpha=({\rm Re}\, \alpha, {\rm Im} \, \alpha)^T,
\end{equation}
with
\begin{equation}
\Sigma^{-1}
=
\frac{4}{(1+G_1)^2+G_2^2}
\begin{pmatrix}
G_1(1+G_1)+G_2^2 & G_2\\
G_2 & 1+G_1
\end{pmatrix},
\end{equation}
or equivalently, the covariance matrix is given by
\begin{equation}
\Sigma=
\frac{1}{4G_1}
\begin{pmatrix}
1+G_1 & -G_2\\
-G_2 & G_1(1+G_1)+G_2^2
\end{pmatrix}.
\label{eq:note-sigma-explicit}
\end{equation}
We find that the Husimi Q-representation of the time-evolved quantum state keeps the Gaussian form whose covariance matrix is given by \eqref{eq:note-sigma-explicit}. At late time, this is dominantly written by
\begin{equation}
G_1\sim 4\frac{\Omega}{\omega}\frac{a}{1+a^2}e^{-2\Omega t},
\qquad
G_2\sim-\frac{\Omega}{\omega}+\mathcal O(e^{-2\Omega t}).
\end{equation}

\begin{figure}[t]
\centering
\includegraphics[width=0.62\textwidth]{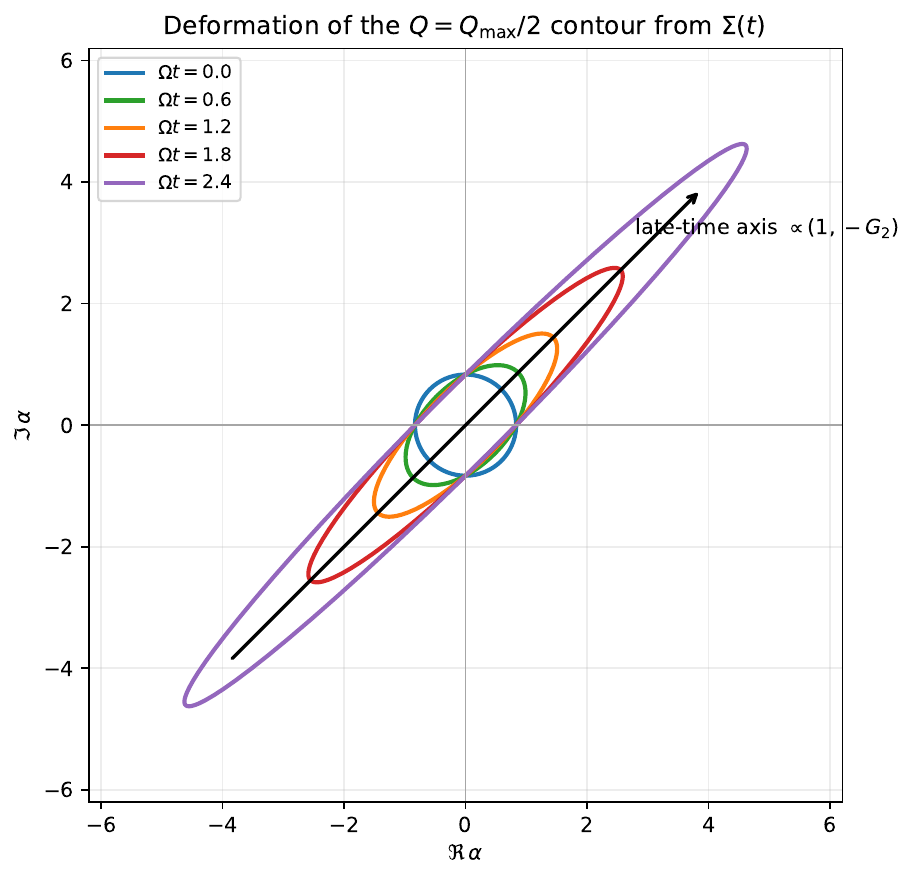}
\caption{The contour $Q=Q_{\max}/2$ in the complex $\alpha$-plane is plotted for increasing times using $\Sigma(t)$ from
Eqs.~\eqref{eq:note-g1g2}--\eqref{eq:note-sigma-explicit} (numerically shown with
$a=1,\ \Omega/\omega=1$). The distribution is progressively elongated along the late-time
principal direction.}
\label{fig:note-p8-deformation}
\end{figure}

Fig.~\ref{fig:note-p8-deformation} visualizes the corresponding deformation of the
half-maximum contour in the $\alpha$-plane using this covariance matrix. From the plot, it is obvious that the Gaussian is deformed to an elongated shape, while the width is kept finite.


Using these analytic expressions for the state, we analytically analyze three methods for the evaluation of the quantum Lyapunov exponent: (i) microcanonical OTOC, (ii) half-probability contour length, and (iii) Wasserstein distance. The last (iii) is our prime target. To the best of our knowledge, the idea of using the Wasserstein distance as a measure of the quantum Lyapunov exponent was first proposed in \cite{zyczkowski1993generalize}, and later it was developed in \cite{wang2021quantum}. 

\begin{description}

\item {\bf Microcanonical OTOC}.

The out-of-time-order correlator \cite{larkin1969quasiclassical,kitaev2015} can measure the Lyapunov exponent in quantum systems, and the microcanonical OTOC \cite{Hashimoto:2017oit} suits our purpose.
It was explicitly described in \cite{morita2022extracting} (see \cite{ali2019time} for earlier work) that for any quantum state of the inverted harmonic oscillator the OTOC is analytically obtained as
\begin{align}
\langle \psi| [x(t),p(0)]^2 |\psi\rangle
= - \cosh^2 \Omega t.    
\label{eq:otociho}
\end{align}
Therefore it grows exponentially for our Gaussian state too, with the exponent $2\Omega$. If we consider the square root of this expression, as we know $-i [x(t),p(0)] \sim \delta x(t)/\delta x(0)$ which leads to the classical definition of the Lyapunov exponent, the quantum Lyapunov exponent defined by the OTOC is given just by $\Omega$.

The quantum chaos and its Lyapunov exponent is closely related to Krylov complexity.
The spread Krylov complexity of the inverted harmonic oscillator was analytically studied in \cite{balasubramanian2022quantum} for an analytically continued thermo field double state, and it grows exponentially in time with the Lyapunov exponent $\Omega$. The operator Krylov complexity for a Gaussian operator of $x$ was computed in \cite{hashimoto2023krylov} for the inverted harmonic oscillator and again it grows exponentially in time with the Lyapunov exponent $\Omega$.\footnote{See \cite{huh2024spread} for the spread Krylov complexity in the double-well model.} Since in \cite{hashimoto2026holography} a connection between the Wasserstein space and the Krylov complexity was pointed out, the universal behavior of the exponential growth in the inverted harmonic oscillator could have the same origin for quantum scrambling.

\item {\bf Half-probability contour length}.

In \cite{toda1986quantal,toda1987quantal},
the half-probability contour length, which is the length $L$ of the contour enclosing probability $1/2$ in the entire phase space, may serve as the distance measure that provides the Lyapunov exponent. Let us employ this definition and evaluate the length for our quantum state.

For a normalized Gaussian distribution in diagonalized coordinates $\vec\beta$ in two dimensions,
\begin{equation}
\rho(\beta_1,\beta_2)=\frac{1}{2\pi} e^{-(\beta_1^2+\beta_2^2)/2},
\end{equation}
the half-probability contour is a circle whose radius is given by $r_0=\sqrt{2 \log 2}$. 
Returning to $\alpha$-space with eigenvalues $\lambda_1\ge\lambda_2>0$ of $\Sigma$,
the semi-axes of the ellipse that defines the half-probability contour are
$a_{\rm ell}\equiv \sqrt{2\log2\,\lambda_1}$ and
$b_{\rm ell}\equiv \sqrt{2\log2\,\lambda_2}$. Using the ellipse-circumference formula, which is given by the Elliptic integral of the second kind,
\begin{equation}
L=4a_{\rm ell}\,E\!\left(\sqrt{1-\frac{b_{\rm ell}^2}{a_{\rm ell}^2}}\right)
=4\sqrt{2\log2}\,\sqrt{\lambda_1}\,
E\!\left(\sqrt{1-\frac{\lambda_2}{\lambda_1}}\right),
\end{equation}
and late-time limits
\begin{equation}
\lambda_1\sim\frac{1+a^2}{4a}\frac{\Omega}{\omega}e^{2\Omega t},
\qquad
\lambda_2\sim\frac14,
\qquad
E\!\left(\sqrt{1-\lambda_2/\lambda_1}\right)\sim E(1)=\frac{\pi}{2},
\end{equation}
we obtain the late time limit of the contour length $L$ as
\begin{equation}
L
\sim
2\pi\sqrt{2\log2}\,
\sqrt{\frac{1+a^2}{4a}\frac{\Omega}{\omega}}\,e^{\Omega t}.
\label{eq:half}
\end{equation}
Thus the contour length $L$ grows exponentially in time, and the quantum Lyapunov exponent defined by the half-probability contour length is $\Omega$.

\item {\bf Wasserstein distance}.

Finally let us evaluate the Wasserstein distance
between initial and time-evolved Husimi Q-representations. They are Gaussians. 
For two centered Gaussians with covariances $\Sigma_0=\Sigma(t=0)$ and $\Sigma(t)$, the 2-Wasserstein distance is given by 
\begin{equation}
W_2
=
\sqrt{\Tr\!\left(\Sigma_0+\Sigma(t)-2(\Sigma_0^{1/2}\Sigma(t)\Sigma_0^{1/2})^{1/2}\right)}.
\label{eq:2wsigma}
\end{equation}
From \eqref{eq:note-sigma-explicit} we know
\begin{equation}
\Tr\Sigma=\frac{(1+G_1)^2+G_2^2}{4G_1},
\end{equation}
and the 2-Wasserstein distance can be evaluated with the $2\times2$ matrix square root formula
\begin{equation}
A^{1/2}
=
\frac{A+\sqrt{\det A}\,\mathbf 1}
{\sqrt{\Tr A+2\sqrt{\det A}}}.
\end{equation}
From this, asymptotically at the late time,
\begin{equation}
W_2
\sim
\sqrt{\frac{1+G_2^2}{4G_1}}
\sim
\frac14 \sqrt{
\frac{1+(\Omega/\omega)^2}{\Omega/\omega}
\frac{1+a^2}{a}
}\,e^{\Omega t}.
\label{eq:w2iho}
\end{equation}
This shows an exponential growth, with the exponent $\Omega$.

For the 1-Wasserstein distance $W_1$, there is no known explicit formula for the distance between two-dimensional Gaussian distributions. However, asymptotically our Gaussian has a special form, which helps us to calculate $W_1$. The late time expression of the covariance matrix \eqref{eq:note-sigma-explicit} is
\begin{align}
    \Sigma \sim 
    e^{2\Omega t}\frac{1}{16}\frac{\omega}{\Omega}\frac{1+a^2}{a}
\begin{pmatrix}
1 & \Omega/\omega\\
\Omega/\omega & (\Omega/\omega)^2
\end{pmatrix}.
\end{align}
First, only the overall factor scales, and second, the determinant of this matrix vanishes, thus the distribution in the phase space is on a line ${\rm Im}\, \alpha = (\Omega/\omega)\,  {\rm Re} \, \alpha$. The Gaussian is so elongated by the scrambling at late times, and is approximated by a one-dimensional distribution. The 1-Wasserstein distance between one-dimensional Gaussians is analytically obtained, and in the current case, the distance between time $t_1$ and time $t_2$ is
\begin{align}
    W_1 \sim \frac{1}{4}\sqrt{\frac{2}{\pi}}\sqrt{\frac{(1+(\Omega/\omega)^2)}{\Omega/\omega}\frac{1+a^2}{a}}\bigg| e^{\Omega t_1}-e^{\Omega t_2}\bigg|.
    \label{eq:w1iho}
\end{align}
Therefore at late times, this $W_1$ has the same exponential growth as that of $W_2$ given in \eqref{eq:w2iho}, except only for the numerical overall factor.

\end{description}

We have seen three ways to look at the quantum Lyapunov exponent, and all agree to have the same $e^{\Omega t}$ growth. Furthermore, the half-probability contour length and the 2-Wasserstein distance share the same initial-condition dependence, $\sqrt{(1+a^2)/a}$. These agreements show that the Wasserstein distance between the initial and the time-evolved Husimi Q-representations may serve as a good measure for quantum chaos, even to quantify the quantum Lyapunov exponent.

In fact, as is obvious from Fig.~\ref{fig:note-p8-deformation}, the Wasserstein distance defined by the optimal transport can be well-approximated by the half-probability contour length, for the case when the distribution is single-modal. However, in general the energy eigenstates in chaotic quantum mechanics the Husimi Q-representation is multi-modal, thus it is expected that neither the Wasserstein distance nor the 
half-probability contour length can properly measure the exponential growth. The study in this section is solely for a single demonstration of the capability of the Wasserstein distance as a measure of quantum chaos, at the semiclassical level of the distribution of the probability.\footnote{
Another related discussion is on the resemblance between the OTOC formula obtained in \cite{wang2021quantum} and the 2-Wasserstein distance formula \eqref{eq:2wsigma}.
In \cite{wang2021quantum}, for coherent-state OTOC
$C(t)=\langle\alpha|[\hat x(t),\hat p(0)]^2|\alpha\rangle$, a propagator decomposition
$\langle\beta|U(t)|\alpha\rangle=e^{i\phi}\delta_{\beta,g_t(\alpha)}+f(\beta,\alpha,t)$
(where the first term is for the motion of the center $g_t$ of distribution while the second term $f$ describes the spreading of the wave function) leads to the expression
\begin{equation}
C(t)=
\sum_\gamma
\left(\Im\gamma-\Im\alpha\right)^2
\left(\Re g_t(\gamma)-\Re g_t(\alpha)\right)^2
\left|f(g_t\gamma,\alpha,t)\right|^2.
\end{equation}
We note that this includes a transport-cost-like structure, and looks like the half-probability contour length, as the spreading $|f|^2$ is included as a factor in this OTOC formula. Thus the optimal transport cost given by the Gaussian Wasserstein decomposition \eqref{eq:2wsigma} is of a similar kind. To look for more similarity would be interesting.
}

\section{Summary and discussions}

The main discovery in this paper is the dimensional reduction of the Wasserstein space (Sec.~\ref{sec:coupled-ho}), summarized in Fig.~\ref{fig:gram-linear}. The Wasserstein space is constructed by the embedding of the 1-Wasserstein distance into a Euclidean space, and the distance is measured by the optimal transport between Husimi Q-representation of the energy eigenstates of the coupled harmonic oscillator \eqref{eq:hamcho}. The clear correlation between the effective embedding dimensions and the coupling constant $g$ of the model shows that chaos reduces the effective dimensions of the Wasserstein space. The main idea of \cite{hashimoto2026holography} was to use the Wasserstein space as a holographic spacetime, and the conjecture was that the manifold hypothesis should select quantum systems allowing a holographic dual. This paper reveals the mechanism of the selection, which is quantum chaos. Quantum chaos provides dimensional reduction of the Wasserstein space to satisfy the manifold hypothesis for the holographic principle. 

Another discovery made in Sec.~\ref{sec:fold} of this paper is the folding structure of the Wasserstein space when the system scrambles, given in Fig.~\ref{fig:oned-embedding-final}. The one-dimensional quantum scrambling models in Sec.~\ref{sec:fold}, the double-well model \eqref{eq:dwp} and the partly flattened oscillator model \eqref{eq:kinkyp}, the OTOC grows exponentially at the energy equal to the flat part of the potential, and the energy eigenstates at that energy produce a folding structure in the Wasserstein space. As we have shown in Sec.~\ref{subsec:separatrix}, the Lyapunov exponent present in the separatrix in the phase space is the origin of this folded structure, due to the recombination of the classical periodic orbits.

The dimensional reduction shown in Sec.~\ref{sec:coupled-ho} needs to be explained by a dynamical mechanism, and we speculate that it may be multiple occurrences of this folding in Sec.~\ref{sec:fold}. In fact, separatrices are often the origin of chaos, which is also the case for quantum situations. Therefore, our findings on the chaotic dimensional reduction and the scrambling giving the folded structure in the Wasserstein space, obtained from the optimal transport between Husimi Q-representations of the energy eigenstates, are expected to be closely related to each other, and may serve as the keys to the conjecture that the Wasserstein spaces are the holographic spacetimes.

As is expected from the results described above, in the course of this work, important notions in quantum chaos have naturally appeared: quantum scars (Sec.~\ref{sec:scar}) and quantum Lyapunov exponents (Sec.~\ref{sec:lyap}). In Sec.~\ref{sec:scar}, a branch structure in the Wasserstein space in the right panel of Fig.~\ref{fig:embed-g1100-g1} was discovered. It suggests some special property for the states joining the branch, and indeed in Fig.~\ref{fig:scar} the relevant states are regarded as quantum scar states, that enjoy the flat direction of the potential $g x^2y^2$ of the coupled harmonic oscillator system at higher energy. Thus the optimal transport provides a novel method to spot quantum scars.
And in Sec.~\ref{sec:lyap}, quantum Lyapunov exponents characterizing the quantum chaos are studied to check the consistency and intimacy between the standard and modern methodology using OTOCs, an old method using the half-probability contour length, and the optimal transport. For the inverted harmonic oscillator which magnifies the separatrix in the double-well model, all three methods (\eqref{eq:otociho}, \eqref{eq:half}, \eqref{eq:w2iho} and \eqref{eq:w1iho}) provide the exponential growth and agree to share the same value for the quantum Lyapunov exponent. The quantum scars and the quantum Lyapunov exponents are key concepts in quantum chaos, and in this paper our optimal transport method is shown to accommodate these concepts within the scheme, which is encouraging for the bigger scope \cite{hashimoto2026holography} of the relationship between the holographic principle and the optimal transport with the manifold hypothesis.

A technical remark is in order concerning the coupled-oscillator analysis in Sec.~\ref{sec:coupled-ho}.
%
%
The distance used there is not the exact Wasserstein distance \eqref{eq:W1-exact}, but its entropically regularized approximation \eqref{entropic-regularization}, evaluated through the Sinkhorn form \eqref{eq:sinkhorn-form} and the iteration \eqref{eq:sinkhorn-iter}.
As shown in App.~\ref{app:numerical-precision}, the dominant numerical uncertainty comes from the phase-space resolution, while the dependence on the regularization parameter, basis truncation, and iteration count is comparatively mild; see Tables~\ref{tab:validation-resolution}--\ref{tab:error-budget}. Furthermore, the higher-precision rerun with the parameter set \eqref{eq:hp-params} preserves the qualitative tendency of the Gram spectra, although the overall eigenvalue scale changes; see Fig.~\ref{fig:hp-gram-comparison} and Table~\ref{tab:hp-fit}. Therefore, the present computations support the robustness of the qualitative structure of the Wasserstein geometry, while a more systematic extrapolation in the phase-space grid and basis cutoff would be desirable before attributing quantitative significance to fitted parameters such as the decay constant $B$.

A second point concerns the physical interpretation of the dimensional reduction observed in Sec.~\ref{sec:coupled-ho}. Our transport analysis is performed in the symmetry sector $(P_x,P_y,S_{xy})=(+,+,+)$ and uses the lowest 12 states in that sector. As discussed around Fig.~\ref{fig:husimi-four-g-b}, this window is not restricted to the immediate vicinity of the ground state of the unreduced spectrum, but its relation to the OTOC regime studied in \cite{akutagawa2020out} is still indirect. For this reason, the most conservative reading of our result is that increasing non-integrability is accompanied by a stronger compression of the Wasserstein space. This tendency is already nontrivial, and it is encouraging that it remains stable under the numerical checks presented in App.~\ref{app:numerical-precision}. It would be important in future work to enlarge the state window, compare different symmetry sectors, and evaluate additional diagnostics of chaos in the same energy range. The same caution applies to the Wasserstein branch studied in Sec.~\ref{sec:scar}: the combined evidence of Fig.~\ref{fig:embed-g1100-g1}, Fig.~\ref{fig:scar}, and Figs.~\ref{fig:poincare}--\ref{fig:poincare2} makes the scar-like interpretation plausible, but a more direct quantitative characterization of the branch states would further sharpen its physical meaning.

Finally, let us describe the importance of this paper from broader viewpoints. 
From the perspective advocated in \cite{hashimoto2026holography}, the main significance of the present paper is that it provides a first nontrivial consistency check of the idea that holographically relevant quantum systems may be characterized by a dimensional reduction of an emergent Wasserstein geometry. In this sense, the present work does not merely add another application of optimal transport to quantum states. Rather, it shows that several structures usually discussed separately in quantum chaos --- the separatrices (leading to the folded structure of the Wasserstein space), the similarity and typicality among the states in the coupled oscillator (leading to the dimensional reduction of the Wasserstein space), the quantum scar (leading to the branch structure), and the Lyapunov growth (which is captured by $W_1$ and $W_2$) --- can all be viewed in a common geometric language based on the optimal transport of Husimi distributions. This unified viewpoint is valuable precisely because it allows one to check the grand picture behind the proposal of \cite{hashimoto2026holography}: namely, that the manifold hypothesis for holography should not be regarded as an abstract kinematical assumption, but may instead arise dynamically from the chaotic organization of quantum states in phase space. Remember that the original AdS/CFT conjecture \cite{Maldacena:1997re} assumes the large $N$ limit and the strong coupling limit for the ${\cal N}=4$ Supersymmetric $SU(N)$ Yang-Mills theory; from our perspective, the strong coupling limit would be necessary for holography as it ensures the dimensional reduction of the Wasserstein space for the manifold hypothesis, and the large $N$ limit would be relevant as it ensures that the branching structure due to quantum scars can be treated as ignorable atypical states. Although the present study is still limited to a simple quantum-mechanical model and therefore cannot establish the conjecture in full generality, it already suggests that optimal transport is sensitive to structures that are central both to quantum chaos and to holography. For this reason, extending the same analysis to a wider class of chaotic systems and, ultimately, to genuinely holographic quantum field theories would constitute a direct test of the central physical picture proposed in
\cite{hashimoto2026holography}.

\section*{Acknowledgments}

We would like to thank S.~Heusler, Y.~Hirono, K.~Ito, T.~Kawamoto, K.~Saito and A.~Tanaka for insightful discussions.
Writing and running the numerical codes used in this paper and generation of the pdf figures were supported by OpenAI Codex and Claude Code.
The work of K.~H.~was supported in part by JSPS KAKENHI Grant No.~JP22H01217, JP22H05111 and JP22H05115.
The work of N.~T.~was supported in part by JSPS KAKENHI Grant No.~JP21H05189, JP22H05111 and 25K07282.
The work of K.~Y.~was supported in part by JSPS KAKENHI Grant No.~JP22H05115, 25K07313 and the Asahipen Hikari Foundation.


\clearpage

\appendix
\section{2-Wasserstein results of Secs.~\ref{sec:oned-models} and \ref{sec:coupled-ho}}
\label{sec:2-w}

In this appendix, we present our numerical results using the 2-Wasserstein distance, instead of the 1-Wasserstein distance which we have used in Sec.~\ref{sec:oned-models} and Sec.~\ref{sec:coupled-ho}. We find that our main conclusions are unchanged.

\begin{figure}[t]
\centering
\subfigure[Harmonic oscillator]{
\includegraphics[width=0.45\textwidth]{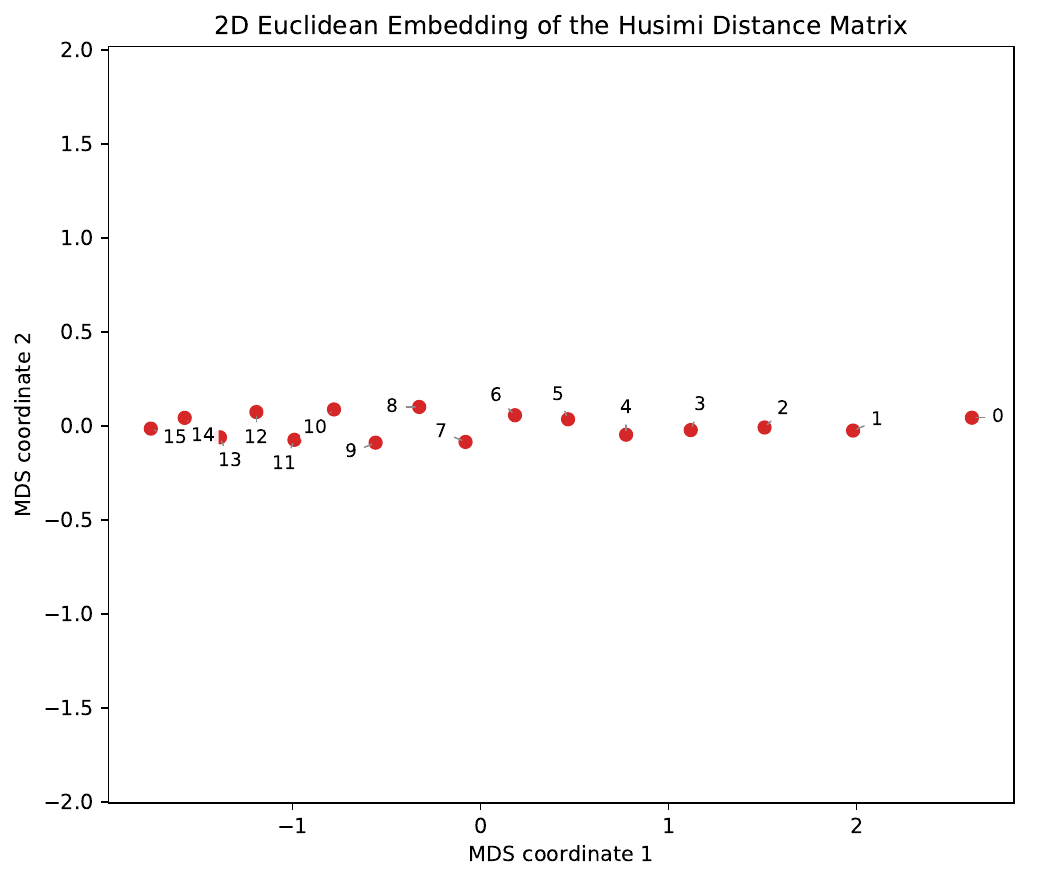}
}
\subfigure[Double-well]{
\includegraphics[width=0.45\textwidth]{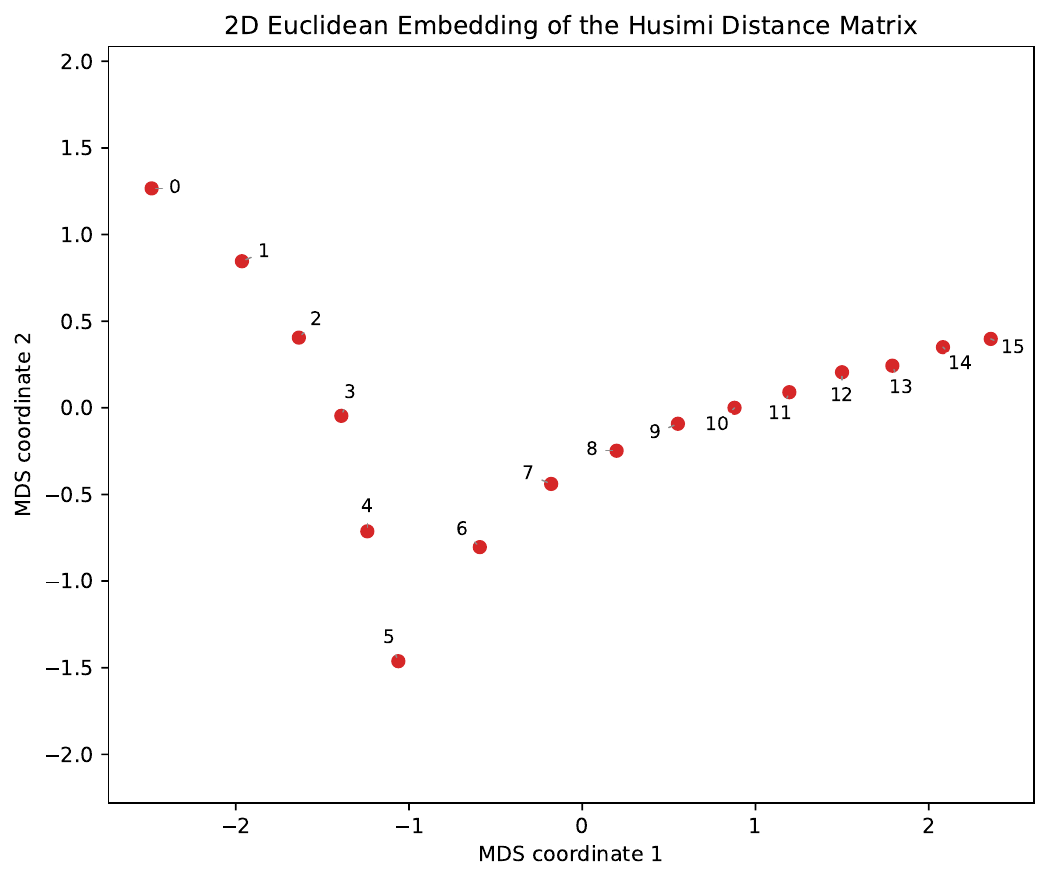}
}
\subfigure[Partly flattened oscillator]{
\includegraphics[width=0.45\textwidth]{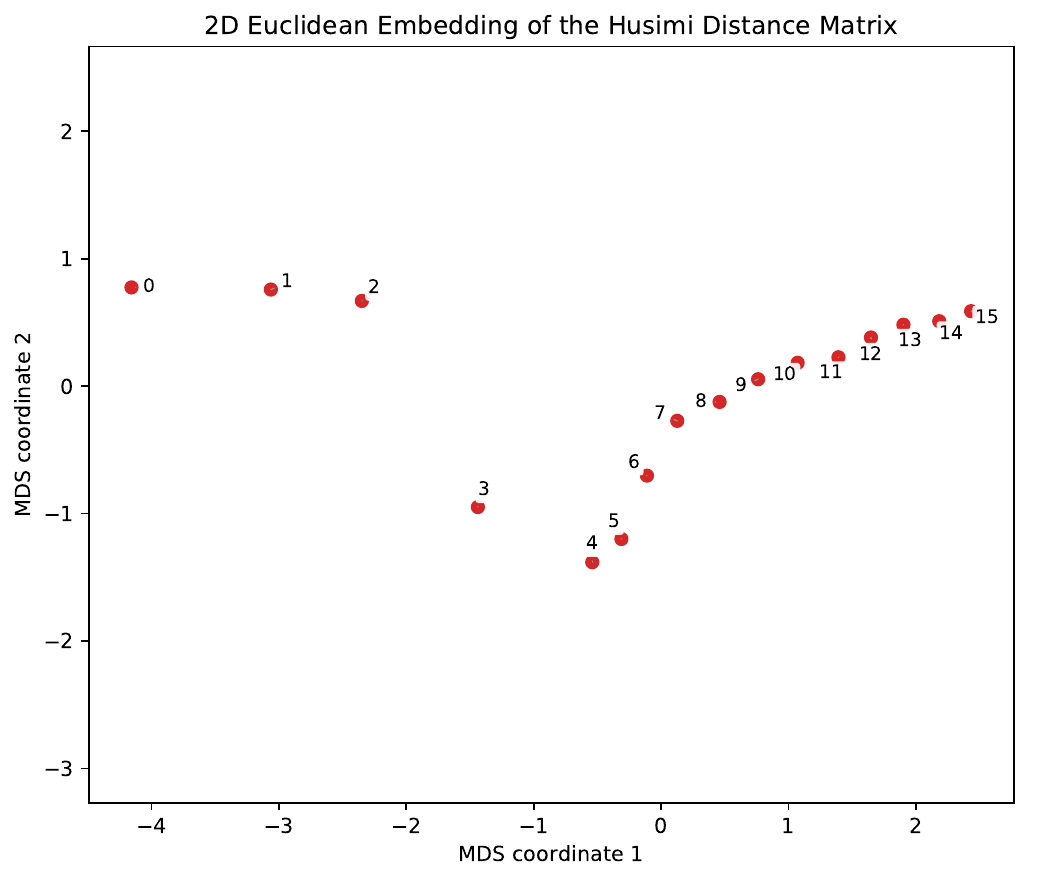}
}
\caption{Final 2-dimensional Euclidean embeddings of the 2-Wasserstein distance matrices.}
\label{fig:oned-embedding-final2}
\end{figure}

For the one-dimensional models, the numerical study we have performed in Sec.~\ref{sec:oned-models}, now with the 2-Wasserstein distance, results in Table \ref{tab:one-dimensional-w2-gram-eigenvalues-top52} for the Gram matrix eigenvalues,\footnote{The Gram matrix eigenvalues for any 2-Wasserstein distance between distributions in one-dimension should be non-negative. However, note that in our case the Husimi distributions are on two-dimensions, so the eigenvalues can be negative.} which can be compared with the 1-Wasserstein result, Table \ref{tab:one-dimensional-w1-gram-eigenvalues-top5}. The folding of the Wasserstein space is again observed for the 2-Wasserstein distance, see Fig.~\ref{fig:oned-embedding-final2}, which can be compared with the 1-Wasserstein result of Fig.~\ref{fig:oned-embedding-final}.
We find that the 1-Wasserstein and the 2-Wasserstein results are quite similar to each other.

\begin{table}[th]
\centering
\begin{tabular}{lrrrrr}
\hline
Potential & 1 & 2 & 3 & 4 & 5 \\
\hline
Harmonic oscillator & 26.05 & 0.06 & 0.04 & 0.04 & 0.03 \\
Double-well & 35.58 & 6.43 & 0.26 & 0.24 & -0.14 \\
Partly flattened oscillator & 47.13 & 12.40 & 2.57 & 0.98 & 0.17 \\
\hline
\end{tabular}
\caption{Top five Gram-matrix eigenvalues for the one-dimensional 2-Wasserstein calculations, ordered by decreasing absolute value.}
\label{tab:one-dimensional-w2-gram-eigenvalues-top52}
\end{table}

\begin{figure}[t]
\centering
\includegraphics[width=0.9\textwidth]{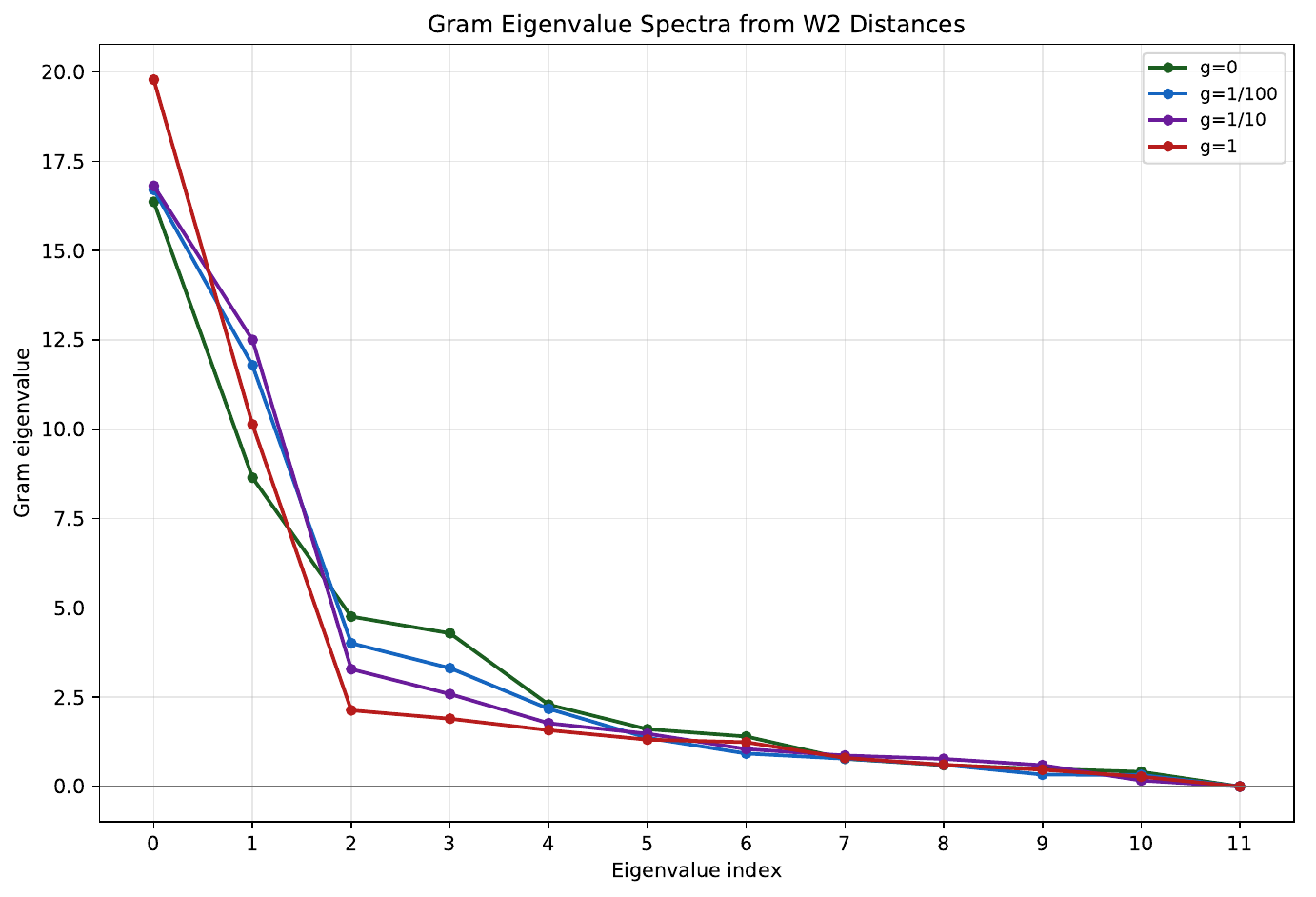}
\caption{Linear-scale Gram eigenvalue spectra for $g=0,1/100,1/10,1$ for the 2-Wasserstein distance.}
\label{fig:gram-linear2}
\end{figure}

\begin{figure}[t]
\centering
\includegraphics[width=0.85\textwidth]{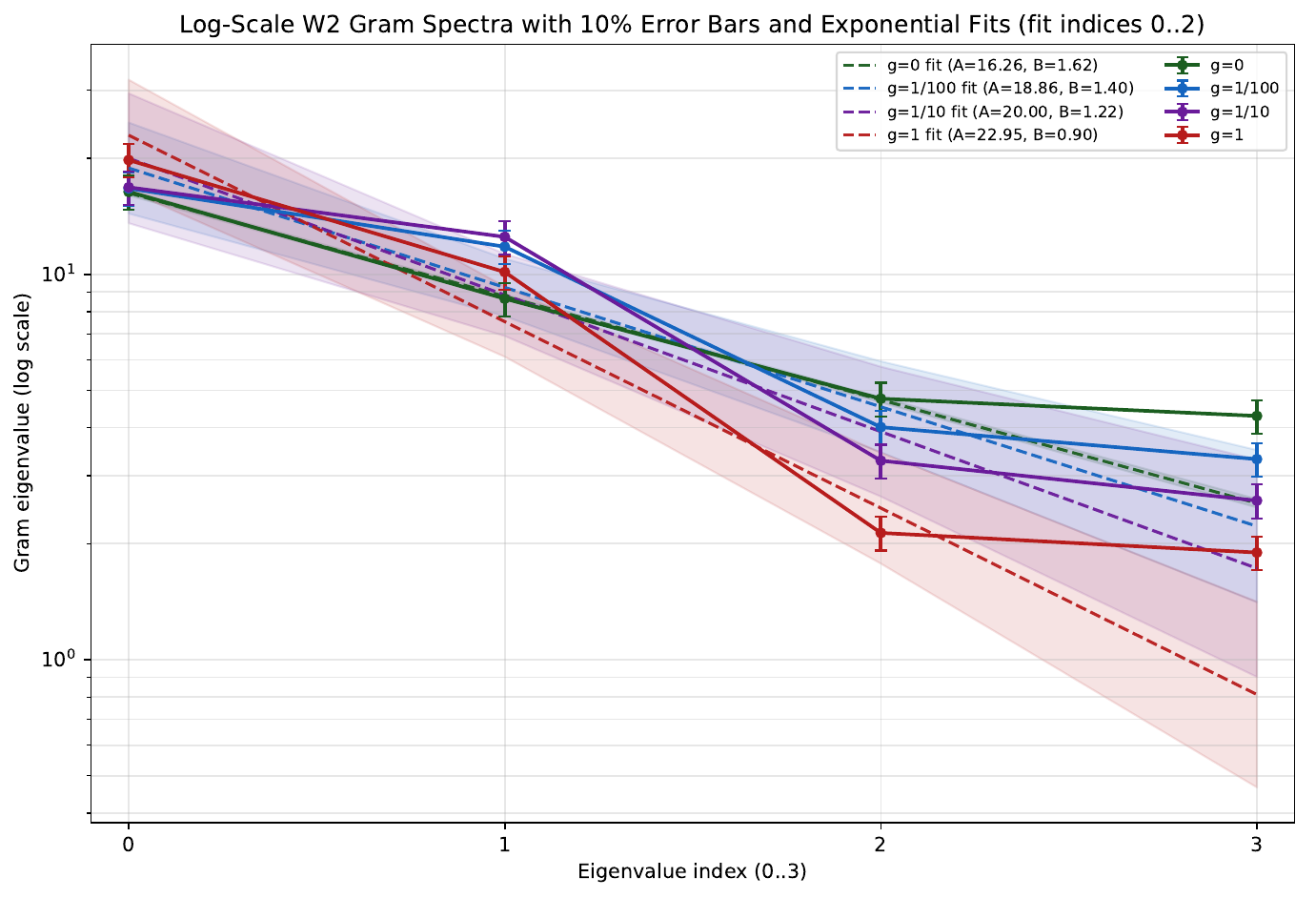}
\caption{Log-scale spectra (indices $0$--$3$ shown) with exponential-fit curves and $10\%$ error bars, for the 2-Wasserstein distance. Fits are performed using $k=0,1,2$.}
\label{fig:gram-logfit2}
\end{figure}

For the coupled harmonic oscillator which we have studied in Sec.~\ref{sec:coupled-ho}, the 2-Wasserstein result of the Gram eigenvalue decay is presented in Fig.~\ref{fig:gram-linear2}, and its logarithmic fit is given in Fig.~\ref{fig:gram-logfit2}. The fitted values of the decay constant $B$ is summarized in Table \ref{tab:gram-fit2}. As one can see, all the results show that our main conclusions on the dimensional reduction with the chaoticity are intact even with the 2-Wasserstein distance.

\begin{table}[th]
\centering
\begin{tabular}{c|ccc}
\hline
$g$ & $A$ & $B$ & $R^2$ \\
\hline
$0$ & $16.2553$ & $1.6183$ &  $0.9996$ \\
$1/100$ & $18.8629$ & $1.4021$ & $0.9198$ \\
$1/10$ & $19.9954$ & $1.2247$ & $0.8808$ \\
$1$ & $22.9475$ & $0.8979$ & $0.9495$ \\
\hline
\end{tabular}
\caption{Exponential fit parameters for Gram-spectrum decay, $\lambda_k\sim A e^{-k/B}$, fitted at $k=0,1,2$, for the 2-Wasserstein distance.}
\label{tab:gram-fit2}
\end{table}

\section{Numerical precision of the coupled harmonic oscillator computation}
\label{app:numerical-precision}

The coupled harmonic oscillator analysis in Sec.~\ref{sec:coupled-ho} relies on a discrete
numerical approximation to the $1$-Wasserstein distance between Husimi Q-representations.
This appendix systematically validates each parameter against the baseline choice and
presents a higher-precision cross-check for all four coupling values
$g \in \{0,\,1/100,\,1/10,\,1\}$. Sec.~\ref{subsec:error-summary} then summarises the
resulting picture and its implications for the physical conclusions of the paper.

The validation results in this appendix would also apply to the numerical calculations in Sec.~\ref{sec:oned-models} for the one-dimensional models, which are numerically less demanding compared to the two-dimensional model in Sec.~\ref{sec:coupled-ho}.

\subsection{Numerical accuracy of Sinkhorn algorithm}
\label{subsec:sinkhorn}

The Sinkhorn algorithm and its entropically-regularized version are introduced in Sec.~\ref {subsec:otwdm}.
The accuracy of the computed $W_1$ distances is governed by four parameters.
\begin{enumerate}
    \item \emph{Phase-space grid resolution} $n_q=n_p$: the four-dimensional phase space is discretized into $n_q^2 \times n_p^2$ grids with $n_q = n_p$ (see Eq.~\eqref{resolution}).
the 4D grid has ${\rm nq}^4$ points;
a coarser grid forces the optimal transport plan to take steps no smaller than the grid
spacing, systematically overestimating the true continuous $W_1$ distance.
\item \emph{One-dimensional number basis size} $N_{\rm basis}$: the Husimi Q-distribution is computed
from a truncated even-parity harmonic-oscillator basis with $N_{\rm basis}$ states per axis;
too few states misrepresent the wavefunction and distort the resulting distributions.
\item \emph{Regularization strength} $\varepsilon$: the entropy term in \eqref{entropic-regularization}
biases the regularized Wasserstein distance $W_1^{(\varepsilon)}$ away from the exact $W_1$; this bias vanishes as
$\varepsilon\to 0$, but the Gibbs kernel $K_{ij}=e^{-C_{ij}/\varepsilon}$ underflows in
double precision below $\varepsilon\approx 0.03$ when the 4D Euclidean cost $C_{ij}$ is $\mathcal{O}(1)$, setting a
practical lower bound.
\item \emph{Sinkhorn iteration count} $N_{\mathrm{iter}}$: more iterations bring the marginal
residuals $\|K Q_1-Q_2\|$ closer to zero; in practice, the transport cost itself stabilizes
far earlier than the dual variables do, so even a modest iteration count suffices.
\end{enumerate}


\subsection{Parameter sensitivity studies}
\label{subsec:validation}

All validation studies use the same physical system: $g = 1$ coupling, lowest $n_{\rm states}=12$
even-parity states.
The numerical errors due to grid resolution ($n_q$) and Sinkhorn convergence ($\varepsilon$ and $N_\text{iter}$) are intrinsic to the computational method, so the conclusions are expected to apply to the other coupling $g$ values studied in the paper.
For the basis number $N_\text{basis}$, it is expected that a larger $N_\text{basis}$ is necessary to maintain a good numerical accuracy in the case of a larger coupling $g$. The validation for $g=1$ will then clarify the convergence with respect to $N_\text{basis}$ for the most stringent case considered in this work.
The baseline parameter set is (see Eqs.~\eqref{resolution} and \eqref{hyperparameters})
\begin{equation}
  n_q = n_p = 9,\quad
  \varepsilon = 0.08,\quad
  N_{\rm iter} = 500,\quad
  N_{\rm basis} = 10.
  \label{baseline}
\end{equation}
Each parameter is varied in turn while the others are held fixed.
We report the following three diagnostics for each comparison.
\begin{itemize}
  \item The top three Gram eigenvalues $\lambda_0,\lambda_1,\lambda_2$.
  \item The pairwise relative deviation defined by
\[
\Delta_{ij}(\%) =
100 \times
\frac{|D^{\rm test}_{ij}-D^{\rm base}_{ij}|}
{D^{\rm base}_{ij}},
\qquad i<j ,
\]
where \(D^{\rm test}\) is the distance matrix for the test setting parameters and 
\(D^{\rm base}\) is that for the baseline setting~\eqref{baseline}.
The columns ``Mean \(\Delta\%\)'' and ``Max \(\Delta\%\)''
denote, respectively, the mean and maximum of \(\Delta_{ij}(\%)\)
over the 66 off-diagonal pairs.\footnote{Since this is a relative error, the maximum value is often dominated by pairs with small baseline distance; we therefore use it together
with the mean deviation and the Pearson correlation coefficient.}
  \item The Pearson correlation coefficient \(r\) between the baseline and test
  distance matrices.  Explicitly, denoting the averages over the 66
  off-diagonal pairs by \(\overline{D}^{\rm base}\) and
  \(\overline{D}^{\rm test}\), we define
\[
r =
\frac{
\sum_{i<j}
\left(D^{\rm base}_{ij}-\overline{D}^{\rm base}\right)
\left(D^{\rm test}_{ij}-\overline{D}^{\rm test}\right)
}{
\sqrt{
\sum_{i<j}
\left(D^{\rm base}_{ij}-\overline{D}^{\rm base}\right)^2
}
\sqrt{
\sum_{i<j}
\left(D^{\rm test}_{ij}-\overline{D}^{\rm test}\right)^2
}
}\, .
\]
This coefficient quantifies how well the relative structure of the
pairwise distances is preserved: \(r\simeq 1\) means that pairs which
are far apart in the baseline calculation remain far apart in the test
calculation, and similarly for nearby pairs.  Since \(r\) is insensitive
to an overall rescaling of all distances, we use it together with the
relative deviations and the Gram eigenvalues.
\end{itemize}

\subsubsection{Phase-space resolution}
\label{subsec:validation-resolution}

The 4D phase-space grid contains $(n_q \times n_p)^2 = n_q^4$ points.
Table~\ref{tab:validation-resolution} shows how results change as the grid is refined from
$\text{nq}=9$ to $13$.

\begin{table}[h]
\centering
\begin{tabular}{lrrrrrr}
\hline
Setting
& $\lambda_0$ & $\lambda_1$ & $\lambda_2$
        & Mean $\Delta\%$ & Max $\Delta\%$ & Pearson $r$ \\
\hline
$n_q=9$ (baseline) 
& 19.340 & 8.189 & 1.034 & --- & --- & --- \\
$n_q=11$           
& 17.057 & 6.121 & 1.106 & 9.5 & 23.4 & 0.9847 \\
$n_q=13$           
& 16.168 & 4.926 & 0.794 & 14.3 & 34.9 & 0.9808 \\
\hline
\end{tabular}
\caption{Phase-space resolution sensitivity ($g=1$, $\varepsilon=0.08$, $N_\text{iter}=500$, $N_\text{basis}=10$).
  }
\label{tab:validation-resolution}
\end{table}


The Gram eigenvalues \(\lambda_0\) and \(\lambda_1\) decrease
systematically as the grid is refined.  This is consistent with the
expected discretization effect: on a coarse grid, the transport plan
cannot resolve displacements smaller than the grid spacing and therefore
tends to overestimate the continuous \(W_1\) cost.  Quantitatively, the
mean relative difference from the \(n_q=9\) baseline is \(9.5\%\) for
\(n_q=11\) and \(14.3\%\) for \(n_q=13\).  Thus the dominant effect of
grid refinement is an overall downward shift of the distance scale,
while the Pearson correlation \(r\simeq 0.98\) shows that the relative
structure of the pairwise distances is well preserved.

\subsubsection{Sinkhorn regularization parameter}
\label{subsec:validation-epsilon}

Table~\ref{tab:validation-epsilon} shows the dependence on $\varepsilon$.

\begin{table}[h]
\centering
\begin{tabular}{lrrrrrr}
\hline
Setting 
& $\lambda_0$ & $\lambda_1$ & $\lambda_2$
        & Mean $\Delta\%$ & Max $\Delta\%$ & Pearson $r$ \\
\hline
$\varepsilon=0.16$
& 19.659 & 8.426 & 1.075 & 1.2 & 2.6  & 1.0000 \\
$\varepsilon=0.08$ (baseline) 
& 19.340 & 8.189 & 1.034 & --- & --- & --- \\
$\varepsilon=0.04$, $N_\text{iter}=1000$      
& 19.255 & 8.140 & 1.024 & 0.2 & 0.8  & 1.0000 \\
\hline
\end{tabular}
\caption{Sinkhorn regularization parameter sensitivity ($g=1$, $n_q=9$, $N_\text{basis}=10$). The baseline iteration number $N_\text{iter} = 500$ is used for $\varepsilon = 0.16$ and $0.08$, while it is increased to $N_\text{iter} = 1000$ for $\varepsilon = 0.04$ to achieve a better convergence of the Sinkhorn iteration.}
\label{tab:validation-epsilon}
\end{table}

%
The comparison of the baseline setting $\varepsilon = 0.08$ with $\varepsilon = 0.04$ shows a mean distance error of only $0.2\%$ and
Pearson $r = 1.0000$, confirming that $\varepsilon = 0.08$ (actually a coarser setting $\varepsilon = 0.16$) is well into the converged regime.\footnote{For the smallest value \(\varepsilon=0.04\), we increase the number of
Sinkhorn iterations from the baseline value \(N_{\rm iter}=500\) to
\(N_{\rm iter}=1000\).  This is because the Sinkhorn scaling problem
becomes more ill-conditioned as \(\varepsilon\) decreases, and a smaller
regularization parameter typically requires more iterations for comparable
convergence.  Thus the \(\varepsilon=0.04\) run should be regarded as a
better-converged small-\(\varepsilon\) reference calculation, rather than
as a strict one-parameter variation with all other numerical settings
fixed.}
We note that $\varepsilon \lesssim 0.02$ is inaccessible in double precision because
$e^{-C_{\rm max}/\varepsilon}$ underflows for the 4D Euclidean cost, which reaches
$C_{\rm max} \approx 38$.

\subsubsection{One-dimensional number basis truncation}
\label{subsec:validation-basis}

The Husimi distributions are computed in a truncated even-parity harmonic-oscillator basis
with $N_{\rm basis}$ states per axis (quantum numbers $n = 0, 2, 4,  \ldots, n_\text{max}$, where $n_{\rm max} = 2(N_{\rm basis}-1)$).
Table~\ref{tab:validation-basis} shows the sensitivity to this truncation.

\begin{table}[h]
\centering
\begin{tabular}{lrrrrrr}
\hline
Setting 
& $\lambda_0$ & $\lambda_1$ & $\lambda_2$
        & Mean $\Delta\%$ & Max $\Delta\%$ & Pearson $r$ \\
\hline
$N_\text{basis}=8$
& 22.039 &  8.544 & 1.254 &  6.9 &  92.7 & 0.9772 \\
$N_\text{basis}=10$ (baseline)
& 19.340 & 8.189 & 1.034 & --- & --- & --- \\
$N_\text{basis}=15$
& 19.182 &  8.113 & 1.023 &  0.6 &   3.5 & 0.9997 \\
\hline
\end{tabular}
\caption{Basis truncation sensitivity ($g=1$, $n_q=9$, $\varepsilon=0.08$, $N_\text{iter}=500$).}
\label{tab:validation-basis}
\end{table}

The comparison between $N_\text{basis}=10$ and $N_\text{basis}=15$ shows a mean
error of only $0.6\%$ and Pearson $r = 0.9997$, confirming that $N_\text{basis}=10$ is well converged.
$N_\text{basis}=6$ is necessary at least to give the correct result in the uncoupled case $g=0$ (and also for $g>0$), as noted in Sec.~\ref{sec:coupled-HO_Husimi-Q}.
$N_\text{basis}=8$ is slightly larger than this minimum and gives a decent match with the baseline result (mean error $6.9\%$ of the pair distances), while it still shows a maximum relative error of $92.7\%$ in the individual pair distances.

\subsubsection{Sinkhorn iteration count}
\label{subsec:validation-iter}

To assess the convergence of the vectors $u,v$ appearing in the Sinkhorn iterations, we monitor the maximum relative update of the vector $u$ between successive iterations,
\[
  r^{(t)} \equiv \max_k \frac{|u_k^{(t)} - u_k^{(t-1)}|}{u_k^{(t-1)}},
\]
and record the value at the final step, $r^{(N_\text{iter})}$, as the
per-pair convergence residual.
All $N_\text{iter}$ iterations are always executed without early stopping.
A pair is declared converged when $r^{(N_\text{iter})} < 10^{-3}$.
The convergence rate (fraction of the 66 pairs satisfying this criterion)
grows from $4.5\%$ at $N_\text{iter}=500$ to $90.9\%$ at
$N_\text{iter}=5000$.  Despite this low formal convergence rate, the transport costs themselves
stabilize much earlier: Table~\ref{tab:validation-iter} shows that doubling or quadrupling the
iteration count changes the distances by at most $0.7\%$.
This is a standard feature of Sinkhorn-type iterations, in which the
transport cost stabilizes faster than the scaling vectors $u$, $v$
themselves.\footnote{%
More precisely, near the Sinkhorn fixed point, the transport cost
converges to its limit at twice the asymptotic rate of the dual
scalings, so an $\mathcal{O}(\delta)$ residual deviation of $(u,v)$
propagates into $D_{nm}$ only at $\mathcal{O}(\delta^2)$.  Consequently
the distance matrix --- and with it the leading Gram eigenvalues and the
exponential-fit constants $A$, $B$ derived from it --- is far more
precise than the formal convergence rate based on
$r^{(N_\text{iter})}<10^{-3}$ would suggest.}

\begin{table}[h]
\centering
\begin{tabular}{lrrrrrrr}
\hline
$N_\text{iter}$
& $\lambda_0$ & $\lambda_1$ & $\lambda_2$
        & Conv.\,rate & Mean $\Delta\%$ & Max $\Delta\%$ & Pearson $r$ \\
\hline
$500$  (baseline) 
& 19.340 & 8.189 & 1.034 &  4.5\% & ---  & ---  & --- \\
$1000$            
& 19.339 & 8.189 & 1.034 & 34.8\% & 0.0  & 0.7  & 1.0000 \\
$2000$            
& 19.339 & 8.189 & 1.034 & 63.6\% & 0.0  & 0.7  & 1.0000 \\
$5000$
& 19.339 & 8.189 & 1.034 & 90.9\% & 0.0  & 0.7  & 1.0000 \\
\hline
\end{tabular}
\caption{Sinkhorn iteration count sensitivity ($g=1$, $n_q=9$, $\varepsilon=0.08$, $N_\text{basis}=10$).
  ``Conv.\,rate'' is the fraction of the $66$ pairs with final scaling-vector residual $r^{(N_\text{iter})} < 10^{-3}$.}
\label{tab:validation-iter}
\end{table}

To localize where the residual $0.7\%$ change of Table~\ref{tab:validation-iter}
sits, we ran a targeted element-wise comparison.  We first compute all $66$
pairs at $N_\text{iter}=500$ to obtain $D_{500}$ and the per-pair residual
$r^{(500)}$, then select $16$ representative pairs covering the regions
expected to be most sensitive --- the five with the largest $r^{(500)}$, the
five with the smallest $D_{500}$, the three largest, and three near the
median distance --- and recompute these at $N_\text{iter}=5000$ as a
near-converged reference.
Of these $16$ pairs, only five show any change above $0.001\%$, and all
five have $D < 1$; the maximum relative deviation is about $0.75\%$ at the
closest pair ($D \approx 0.51$), while every pair with $D \ge 1.5$ agrees
with the $N_\text{iter}=5000$ reference to four decimal places.
The sign is also consistent across the affected pairs: $N_\text{iter}=500$
slightly \emph{underestimates} these short-distance entries, which is
opposite to the phase-space-resolution effect that overestimates the
distance scale on a coarse grid.

\subsubsection{Summary of error estimates}
\label{subsec:error-budget}

Table~\ref{tab:error-budget} summarizes the four numerical error sources
and their estimated effects on the pairwise distance matrix.  We also
show the Pearson correlation coefficient \(r\) between the baseline and
test distance matrices, which diagnoses how well the relative structure
of the pairwise distances is preserved.  When several test settings are
available for one source, we quote the largest observed deviation and
the Pearson coefficient for the corresponding comparison.

\begin{table}[h]
\centering
\begin{tabular}{lcc}
\hline
Source & Estimated distance deviation & Pearson \(r\) \\
\hline
Phase-space resolution
& \(\sim 14\%\) systematic
& 0.9808
\\
Sinkhorn \(\varepsilon\)
& \(0.2\%\)
& 1.0000
\\
Basis truncation
& \(0.6\%\)
& 0.9997
\\
Sinkhorn \(N_{\rm iter}\)
& \({\le}0.75\%\) (near pairs only)
& 1.0000
\\
\hline
\end{tabular}
\caption{
Estimated deviations of the baseline results from the corresponding
higher-accuracy or more-converged parameter choices.  The quoted
Pearson coefficient is computed from the independent off-diagonal
entries of the baseline and test distance matrices.
}
\label{tab:error-budget}
\end{table}

The dominant source of numerical uncertainty is the phase-space grid
resolution, motivating the higher-precision computation in the next
subsection.  The Pearson correlations between the baseline and test
distance matrices are close to unity, indicating that the relative
structure of the pairwise distances is well preserved under these
parameter variations.

\subsection{Higher-precision results for all four coupling values}
\label{subsec:high-precision}

To provide a higher-accuracy cross-check of the main-text results, we rerun the computation
for all four coupling values $g \in \{0, 1/100, 1/10, 1\}$ with the improved parameter set
\begin{equation}
  n_q = n_p = 13, \quad
  N_{\rm basis} = 10, \quad
  \varepsilon = 0.08, \quad N_{\rm iter} = 500,
  \label{eq:hp-params}
\end{equation}
upgrading only the phase-space grid from $9^4 = 6{,}561$ to $13^4 = 28{,}561$ points,
while keeping the number basis ($N_{\rm basis}=10$) and the Sinkhorn parameters
unchanged.

Table~\ref{tab:hp-gram} compares the top Gram-matrix eigenvalues for $g=1/100$ and $g=1$.
The overall scale decreases by $\sim\!26\%$ for $g=1/100$ and $\sim\!16\%$ for $g=1$,
while the qualitative structure is mostly preserved.

\begin{table}[h]
\centering
\begin{tabular}{lccccccc}
\hline
 & \multicolumn{3}{c}{$g=1/100$} & & \multicolumn{3}{c}{$g=1$} \\
\cline{2-4}\cline{6-8}
 & $\lambda_0$ & $\lambda_1$ & $\lambda_2$ & & $\lambda_0$ & $\lambda_1$ & $\lambda_2$ \\
\hline
Baseline~~~ ($n_q=9$) & 16.4443 & 10.0684 & 2.1760 & & 19.3400 & 8.1893 & 1.0341 \\
High-prec.\ ($n_q=13$) & 12.252 & 6.099 & 1.406 & & 16.168 & 4.926 & 0.794 \\
\hline
\end{tabular}
\caption{Top three Gram-matrix eigenvalues at baseline and higher-precision parameters,
  for $g=1/100$ and $g=1$. The results for $g=1$ are the same as those in Table~\ref{tab:validation-resolution}.}
\label{tab:hp-gram}
\end{table}

\begin{figure}[t]
\centering
\includegraphics[width=0.85\textwidth]{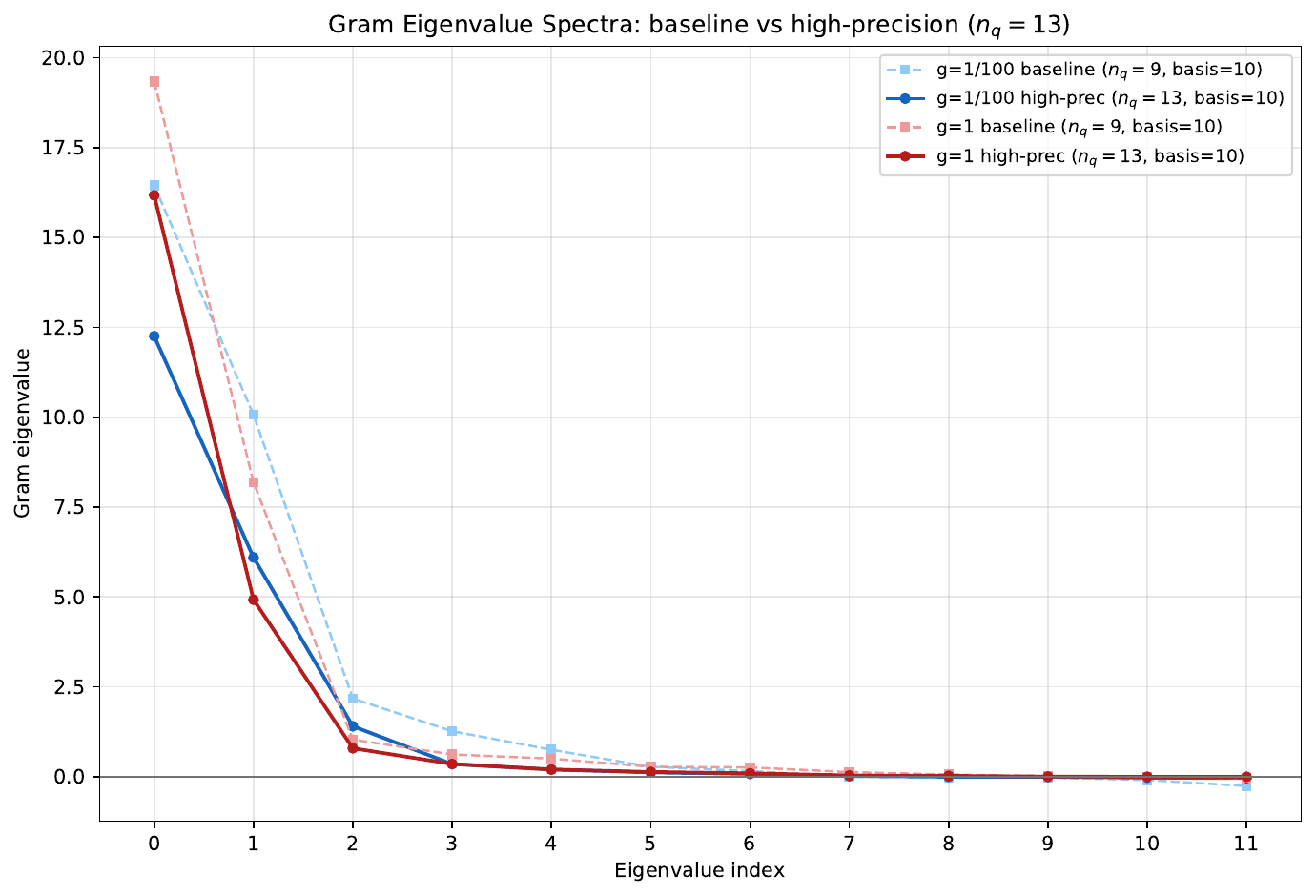}
\caption{Gram eigenvalue spectra for $g=1/100$ (blue) and $g=1$ (red) at baseline parameters
  ($n_q=9$, dashed, squares) and higher-precision parameters ($n_q=13$, solid, circles).}
\label{fig:hp-gram-comparison}
\end{figure}

\begin{figure}[th]
\centering
\includegraphics[width=0.85\textwidth]{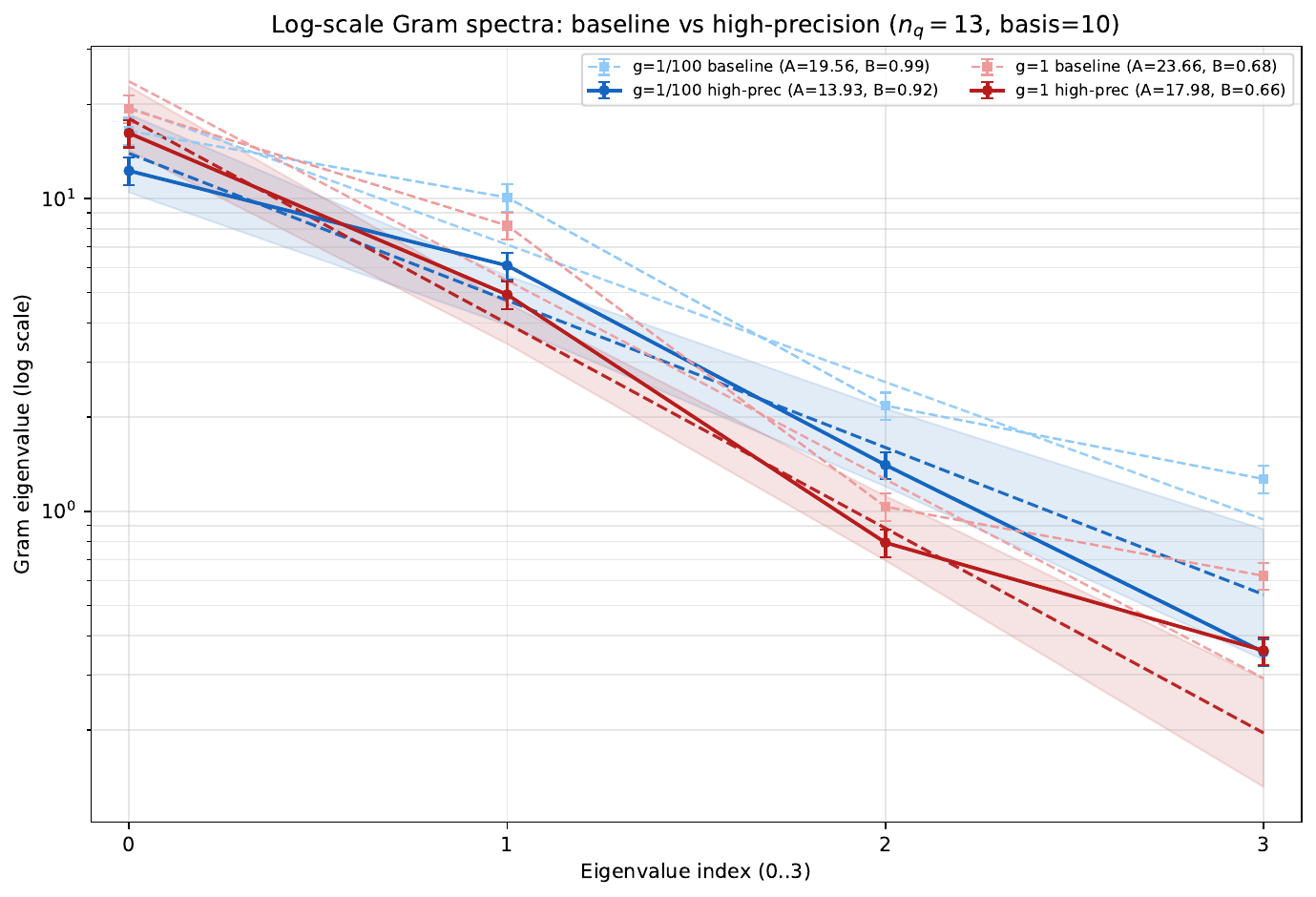}
\caption{Log-scale Gram spectra with 10\% error bars and exponential fits,
  baseline ($n_q=9$, dashed) vs higher-precision ($n_q=13$, solid), for $g=1/100$ and $g=1$.}
\label{fig:hp-gram-logfit}
\end{figure}

Figure~\ref{fig:hp-gram-comparison} compares the full Gram eigenvalue spectra; the overall
spectral shape is preserved, with a nearly uniform downward shift at higher resolution.
Figure~\ref{fig:hp-gram-logfit} shows the log-scale spectra with exponential fits
$\lambda_k \sim A e^{-k/B}$ (fit range $k=0,1,2$, 10\% relative error bars);
the fitted parameters are collected in Table~\ref{tab:hp-fit}.

\begin{table}[h]
\centering
\begin{tabular}{lccc|ccc}
\hline
 & \multicolumn{3}{c|}{Baseline ($n_q=9$)}
 & \multicolumn{3}{c}{High-prec.\ ($n_q=13$)} \\
$g$ & $A$ & $B$ & $R^2$ & $A$ & $B$ & $R^2$ \\
\hline
$0$     & 15.453 & 1.373 & 0.969 & 10.712 & 1.347 & 0.876 \\
$1/100$ & 19.561 & 0.989 & 0.818 & 13.930 & 0.924 & 0.960 \\
$1/10$  & 21.035 & 0.867 & 0.656 & 14.489 & 0.860 & 0.993 \\
$1$     & 23.661 & 0.683 & 0.847 & 17.979 & 0.664 & 0.985 \\
\hline
\end{tabular}
\caption{Exponential fit parameters $\lambda_k \sim A e^{-k/B}$ (fit at $k=0,1,2$, with
  10\% relative error bars) at baseline and higher-precision resolution, for all four
  coupling values.}
\label{tab:hp-fit}
\end{table}

The central result in Table~\ref{tab:hp-fit} is the robustness of the ordering
$B(g=0) > B(g=1/100) > B(g=1/10) > B(g=1)$: the monotonic decrease of the spectral
decay constant with increasing coupling, which characterises the growing compactness of
the Wasserstein geometry in the chaotic regime, is fully preserved at higher resolution.
Quantitatively, the baseline values $B = 1.373, 0.989, 0.867, 0.683$ shift to
$B = 1.347, 0.924, 0.860, 0.664$ at $n_q = 13$, a change of less than $4\%$ in each case.
Notably, the fit quality $R^2$ (see~\eqref{eq:gram-R2} for the definition) is uniformly high at higher precision and
improves substantially for the three non-integrable couplings: from $0.818$
to $0.960$ for $g=1/100$, from $0.656$ to $0.993$ for $g=1/10$, and from
$0.847$ to $0.985$ for $g=1$.  The integrable case $g=0$ is the exception,
with $R^2$ slightly decreasing from $0.969$ to $0.876$: the energy
degeneracies of the uncoupled 2D oscillator produce a near-plateau
$\lambda_1 \simeq \lambda_2$ in the Gram spectrum (cf.\ $\lambda_1 = 3.69$,
$\lambda_2 = 2.85$ at $n_q=13$), which the finer grid resolves more
sharply, reducing the quality of a single-exponential fit in the
integrable limit.  Across all couplings, $\lambda_k \sim Ae^{-k/B}$
remains a useful summary of the leading Gram spectrum, with the residual
deviation from exponential largest in the integrable limit, where chaos
has not lifted the degeneracies.

\begin{figure}[t]
\centering
\subfigure[$g=1/100$]{%
  \includegraphics[width=0.47\textwidth]{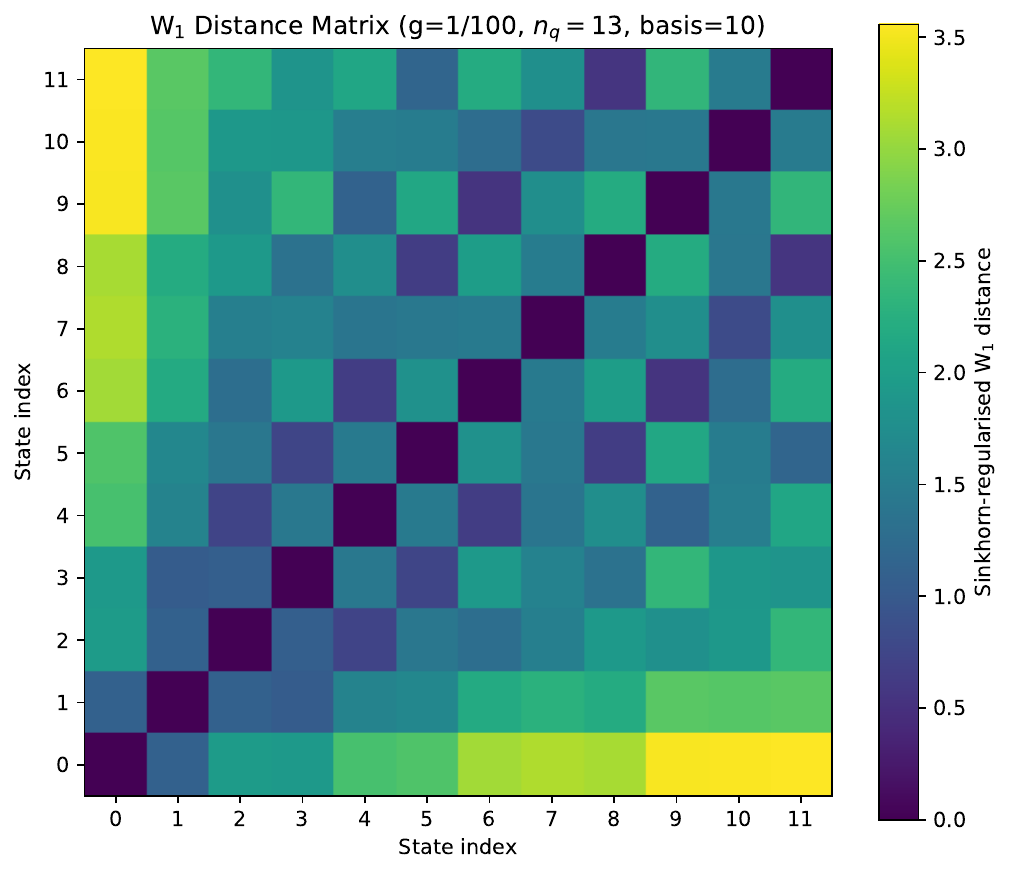}}
\subfigure[$g=1$]{%
  \includegraphics[width=0.47\textwidth]{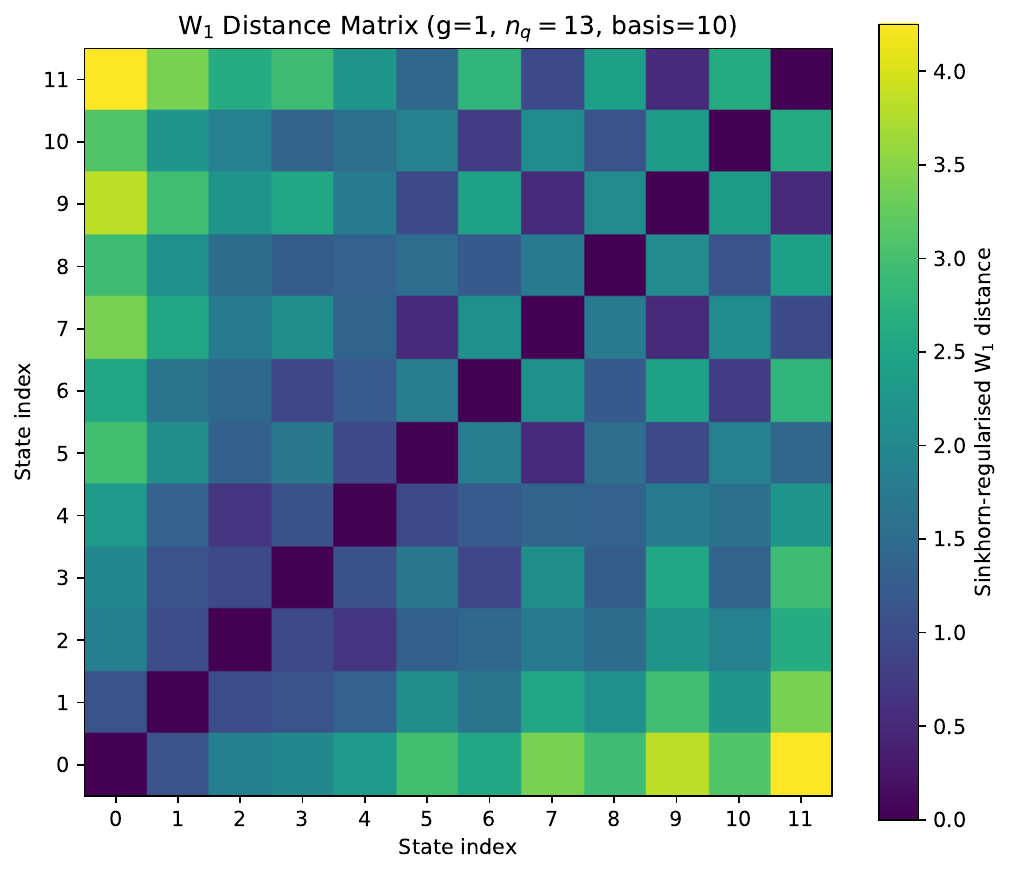}}
\caption{$W_1$ distance matrices at higher precision ($n_q=13$, $N_\text{basis}=10$),
  for $g=1/100$ (left) and $g=1$ (right).}
\label{fig:hp-distmats}
\end{figure}

\begin{figure}[th]
\centering
\subfigure[$g=1/100$]{%
  \includegraphics[width=0.47\textwidth]{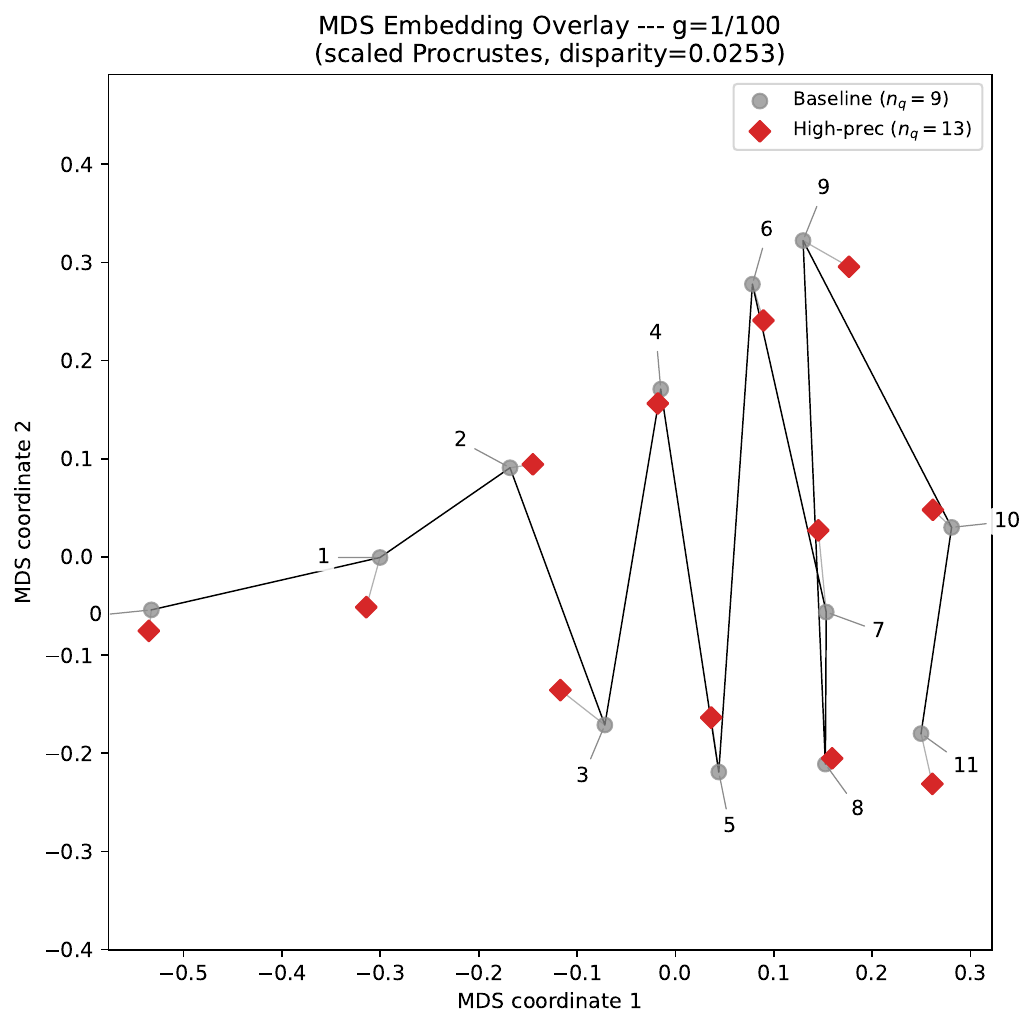}}
\subfigure[$g=1$]{%
  \includegraphics[width=0.47\textwidth]{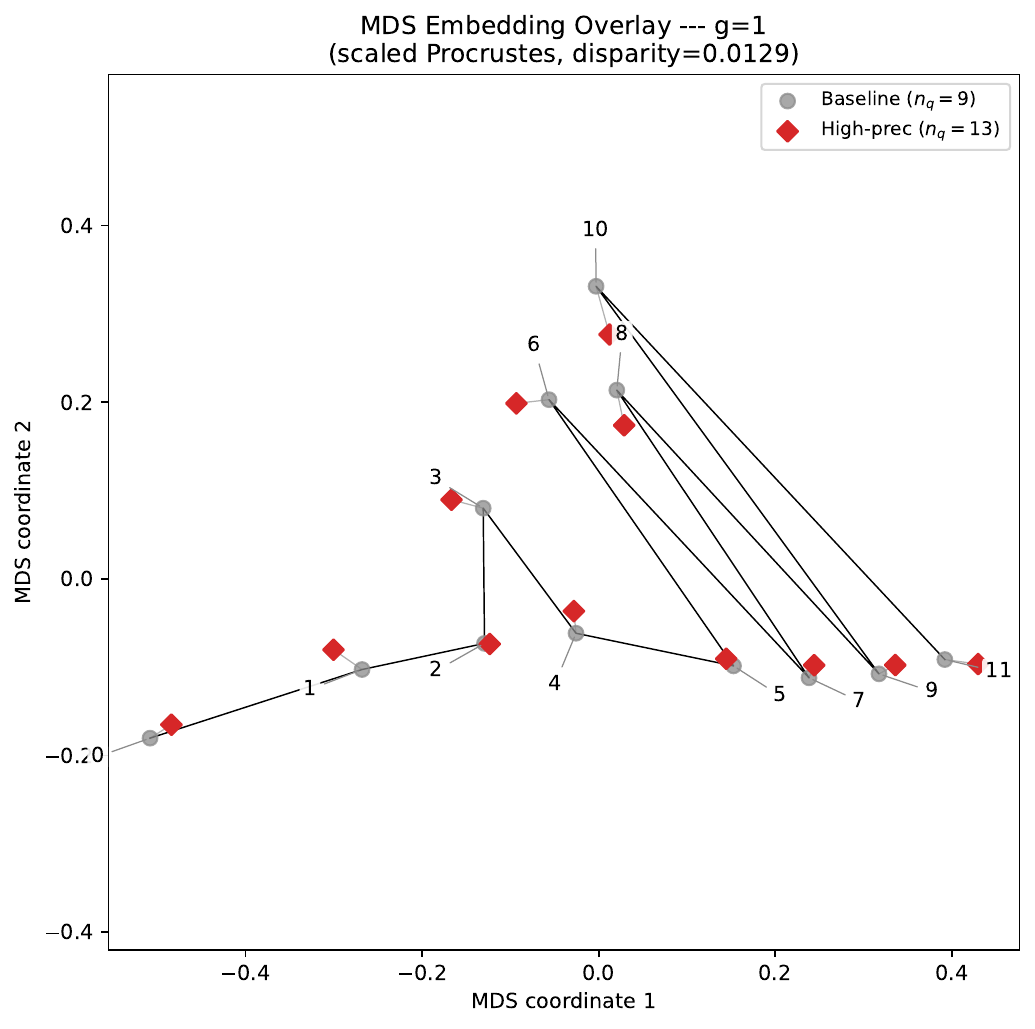}}
\caption{2D MDS embeddings of the baseline ($n_q=9$, grey circles) and higher-precision
  ($n_q=13$, red diamonds) distance matrices, aligned by full Procrustes (translation,
  rotation and uniform scaling), for $g=1/100$ (left) and $g=1$ (right).
  Thin lines connect corresponding states.
  Compare with Fig.~\ref{fig:embed-g1100-g1} in the main text.}
\label{fig:hp-embeddings}
\end{figure}

The distance matrices at higher precision are shown in Fig.~\ref{fig:hp-distmats};
Fig.~\ref{fig:hp-embeddings} overlays the baseline and higher-precision 2D MDS embeddings
after Procrustes alignment,\footnote{%
Procrustes alignment finds the orthogonal transformation (rotation, with reflection
allowed) that minimizes the Frobenius distance between two point sets sharing a common
point-to-point indexing, after first centering each set at its centroid.  In the
\emph{full} variant used here, each set is additionally rescaled to unit Frobenius
norm, so that only the relative geometry --- not the overall size --- is compared.
For the present pair (baseline $n_q=9$ vs higher-precision $n_q=13$) this rescaling
effectively magnifies the higher-precision embedding, whose absolute MDS scale is
smaller, by a factor of $\approx 1.20$ for $g=1/100$ and $\approx 1.14$ for $g=1$
relative to the baseline.}
to be compared with Fig.~\ref{fig:embed-g1100-g1} in the main text.

In both cases the Procrustes disparity is small ($0.025$ for $g=1/100$ and $0.013$
for $g=1$ under full Procrustes), confirming that the qualitative geometry of the
Wasserstein space is robust to the resolution upgrade.

\subsection{Summary of error analysis}
\label{subsec:error-summary}

\paragraph{Numerical error budget.}
Among the four precision parameters, the Sinkhorn and basis settings
($\varepsilon$, $N_{\rm iter}$, $N_{\rm basis}$) each contributes less than $1\%$ to the
pairwise distances in their validated ranges.
The sole non-negligible error is the phase-space grid resolution: at $n_q=9$, the discrete
OT computation overestimates the continuous $W_1$ distances by roughly $14\%$ on average
(up to $35\%$ for individual pairs), with a corresponding inflation of the Gram eigenvalues.
The higher-precision rerun at $n_q=13$ (Sec.~\ref{subsec:high-precision}) shows that the
inflation amounts to a $\sim\!16$--$26\%$ reduction of the top Gram eigenvalues, while the
qualitative spectral structure is fully preserved.

\paragraph{Robustness of the physical observable.}
The physical observable that carries the message of the paper is the spectral decay constant $B$.
$B$ shifts by less than $8\%$ between the baseline and higher-precision
runs, a change well within the $\sim\!14\%$ distance uncertainty.
Most importantly, the ordering $B(g=0)>B(g=1/100)>B(g=1/10)>B(g=1)$ is now verified
directly at both resolutions (Table~\ref{tab:hp-fit}),
confirming that the progressive dimensional reduction of the Wasserstein geometry with
increasing coupling is a genuine physical effect.

\paragraph{Physical interpretation.}
The exponential fit quality $R^2$ at higher resolution is high across all four
coupling values ($0.88$, $0.96$, $0.99$, $0.99$ for $g=0, 1/100, 1/10, 1$), and improves
substantially over the baseline for the three non-zero couplings (from $0.82$, $0.66$, $0.85$).
This is physically significant: it indicates that the form $\lambda_k \sim Ae^{-k/B}$
properly reflects an intrinsic property of the distribution of eigenstate separations in Wasserstein space that the coarser grid partially obscures.

Since $B$ decreases monotonically with $g$ and is numerically stable, it can serve as a
well-defined, resolution-independent probe of the onset of chaos: a large $B$ (slowly
decaying spectrum, higher effective dimensionality) corresponds to a near-integrable system
in which eigenstates are geometrically diverse in phase space, while a small $B$ (faster
decay) signals the collapse of this diversity that accompanies quantum chaos.
This picture is consistent with the manifold hypothesis discussed in the main text: quantum
chaos drives the eigenstate manifold toward a low-dimensional subspace of Wasserstein space,
and the numerical validation shows that this conclusion does not depend on the specific
discrete approximation used to compute $W_1$.

\paragraph{Extension to higher excitation levels.}
Let us also make cautionary remarks on extending the analysis above to higher excitation levels, for which the structure of the Wasserstein space has not been studied in this work.
Extending the analysis to higher excited states requires revisiting each of the four precision
parameters identified in Sec.~\ref{subsec:sinkhorn}.
The basis truncation study (Sec.~\ref{subsec:validation-basis}) shows that $N_{\rm basis}=10$
is sufficient for the lowest 12 states, but the required basis size grows with the maximum
quantum numbers of the states included; for strongly coupled, high-lying states the Husimi
distributions become more spread in phase space, demanding both a larger basis and, potentially,
a larger phase-space box.
The phase-space resolution systematic ($\sim\!14\%$ at $n_q=9$) identified as the dominant
error source is also likely to worsen for higher excited states.
Their broader Husimi support requires a correspondingly larger phase-space box, and
retaining $n_q=9$ in step would enlarge the grid spacing $\Delta q$ and coarsen the
resolution per Husimi feature.
Furthermore, the Gram matrix of a larger set of states may generally contain more substantial
negative eigenvalues, which would reflect a richer and more genuinely non-Euclidean structure of the
Wasserstein geometry — a physically meaningful signal, but one that requires care in the
MDS embedding step.
A sensible strategy for extending the analysis is therefore to first survey the dependence of
$B$ on $n_{\rm states}$ at the baseline resolution, verifying basis
convergence state-by-state for each new batch of levels, and then to apply the
higher-precision parameters to the physically most interesting coupling values.
Such an analysis of higher excitation levels may yield new insights into the structure of the Wasserstein space, but we defer it to future studies.

\bibliographystyle{JHEP}
\bibliography{For_paper1}

@article{heusler2007periodic,
  title={Periodic-orbit theory of level correlations},
  author={Heusler, Stefan and M{\"u}ller, Sebastian and Altland, Alexander and Braun, Petr and Haake, Fritz},
  journal={Physical review letters},
  volume={98},
  number={4},
  pages={044103},
  year={2007},
  publisher={APS}
}

@article{muller2004semiclassical,
  title={Semiclassical foundation of universality in quantum chaos},
  author={M{\"u}ller, Sebastian and Heusler, Stefan and Braun, Petr and Haake, Fritz and Altland, Alexander},
  journal={Physical review letters},
  volume={93},
  number={1},
  pages={014103},
  year={2004},
  publisher={APS}
}

@article{muller2005periodic,
  title={Periodic-orbit theory of universality in quantum chaos},
  author={M{\"u}ller, Sebastian and Heusler, Stefan and Braun, Petr and Haake, Fritz and Altland, Alexander},
  journal={Physical Review E—Statistical, Nonlinear, and Soft Matter Physics},
  volume={72},
  number={4},
  pages={046207},
  year={2005},
  publisher={APS}
}

@article{sieber2001correlations,
  title={Correlations between periodic orbits and their r{\^o}le in spectral statistics},
  author={Sieber, Martin and Richter, Klaus},
  journal={Physica Scripta},
  volume={2001},
  number={T90},
  pages={128--133},
  year={2001}
}

@article{matinyan1981classical,
  title={Classical Yang-Mills mechanics. nonlinear color oscillations},
  author={Matinyan, SG and Savvidy, GK and Ter-Arutyunyan-Savvidy, NG},
  journal={Sov. Phys. JETP},
  volume={80},
  pages={830--838},
  year={1981}
}

@article{matinyan1981stochasticity,
  title={Stochasticity of classical Yang-Mills mechanics and its elimination by using the Higgs mechanism},
  author={Matinyan, SG and Savvidi, GK and Ter-Arutyunyan-Savvidi, NG},
  journal={JETP Lett.(Engl. Transl.);(United States)},
  volume={34},
  number={11},
  year={1981},
  publisher={Erevan Physics Institute}
}

@article{savvidy1984classical,
  title={Classical and quantum mechanics of non-Abelian gauge fields},
  author={Savvidy, GK},
  journal={Nuclear Physics B},
  volume={246},
  number={2},
  pages={302--334},
  year={1984},
  publisher={Elsevier}
}

@article{deutsch1991quantum,
  title={Quantum statistical mechanics in a closed system},
  author={Deutsch, Josh M},
  journal={Physical review a},
  volume={43},
  number={4},
  pages={2046},
  year={1991},
  publisher={APS}
}

@article{srednicki1994chaos,
  title={Chaos and quantum thermalization},
  author={Srednicki, Mark},
  journal={Physical review e},
  volume={50},
  number={2},
  pages={888},
  year={1994},
  publisher={APS}
}

@article{heller1984bound,
  title={Bound-state eigenfunctions of classically chaotic Hamiltonian systems: scars of periodic orbits},
  author={Heller, Eric J},
  journal={Physical Review Letters},
  volume={53},
  number={16},
  pages={1515},
  year={1984},
  publisher={APS}
}

@article{turner2018weak,
  title={Weak ergodicity breaking from quantum many-body scars},
  author={Turner, Christopher J and Michailidis, Alexios A and Abanin, Dmitry A and Serbyn, Maksym and Papi{\'c}, Zlatko},
  journal={Nature Physics},
  volume={14},
  number={7},
  pages={745--749},
  year={2018},
  publisher={Nature Publishing Group UK London}
}

@article{shiraishi2017systematic,
  title={Systematic construction of counterexamples to the eigenstate thermalization hypothesis},
  author={Shiraishi, Naoto and Mori, Takashi},
  journal={Physical review letters},
  volume={119},
  number={3},
  pages={030601},
  year={2017},
  publisher={APS}
}

@article{larkin1969quasiclassical,
  title={Quasiclassical method in the theory of superconductivity},
  author={Larkin, Anatoly I and Ovchinnikov, Yu N},
  journal={Sov Phys JETP},
  volume={28},
  number={6},
  pages={1200--1205},
  year={1969}
}

@misc{Kitaev2015,
  author       = {Alexei Kitaev},
  title        = {A simple model of quantum holography},
  howpublished = {Talks at the Kavli Institute for Theoretical Physics},
  month        = feb,
  year         = {2015},
  note         = {February 12, April 7, and May 27, 2015}
}

@article{sekino2008fast,
  title={Fast scramblers},
  author={Sekino, Yasuhiro and Susskind, Leonard},
  journal={Journal of High Energy Physics},
  volume={2008},
  number={10},
  pages={065--065},
  year={2008}
}

@article{maldacena2016bound,
  title={A bound on chaos},
  author={Maldacena, Juan and Shenker, Stephen H and Stanford, Douglas},
  journal={Journal of High Energy Physics},
  volume={2016},
  number={8},
  pages={106},
  year={2016},
  publisher={Springer}
}

@article{huh2024spread,
  title={Spread complexity in saddle-dominated scrambling},
  author={Huh, Kyoung-Bum and Jeong, Hyun-Sik and Pedraza, Juan F},
  journal={Journal of High Energy Physics},
  volume={2024},
  number={5},
  pages={1--27},
  year={2024},
  publisher={Springer}
}

@article{balasubramanian2022quantum,
  title={Quantum chaos and the complexity of spread of states},
  author={Balasubramanian, Vijay and Caputa, Pawel and Magan, Javier M and Wu, Qingyue},
  journal={Physical Review D},
  volume={106},
  number={4},
  pages={046007},
  year={2022},
  publisher={APS}
}

@article{hashimoto2023krylov,
  title={Krylov complexity and chaos in quantum mechanics},
  author={Hashimoto, Koji and Murata, Keiju and Tanahashi, Norihiro and Watanabe, Ryota},
  journal={Journal of High Energy Physics},
  volume={2023},
  number={11},
  pages={1--41},
  year={2023},
  publisher={Springer}
}

@article{morita2022extracting,
  title={Extracting classical Lyapunov exponent from one-dimensional quantum mechanics},
  author={Morita, Takeshi},
  journal={Physical Review D},
  volume={106},
  number={10},
  pages={106001},
  year={2022},
  publisher={APS}
}

@article{wang2021quantum,
  title={Quantum chaos and physical distance between quantum states},
  author={Wang, Zhenduo and Wang, Yijie and Wu, Biao},
  journal={Physical Review E},
  volume={103},
  number={4},
  pages={042209},
  year={2021},
  publisher={APS}
}

@article{zyczkowski1993generalize,
  title={How to generalize the Lapunov exponent for quantum mechanics},
  author={{\.Z}yczkowski, Karol and Wiedemann, Harald and S{\l}omczy{\'n}ski, Wojciech},
  journal={Vistas in astronomy},
  volume={37},
  pages={153--156},
  year={1993},
  publisher={North-Holland}
}

@article{takahashi1986wigner,
  title={Wigner and Husimi functions in quantum mechanics},
  author={Takahashi, Kin'ya},
  journal={Journal of the Physical Society of Japan},
  volume={55},
  number={3},
  pages={762--779},
  year={1986},
  publisher={The Physical Society of Japan}
}

@article{toda1987quantal,
  title={Quantal lyapunov exponent},
  author={Toda, M and Ikeda, K},
  journal={Physics Letters A},
  volume={124},
  number={3},
  pages={165--169},
  year={1987},
  publisher={Elsevier}
}

@article{toda1986quantal,
  title={Quantal lyapunov exponent},
  author={Toda, M and Ikeda, K},
  journal={Bussei Kenkyu},
  volume={46},
  number={2},
  pages={202--204},
  year={1986},
  publisher={}
}

@article{shibata2020onsager,
  title={Onsager’s scars in disordered spin chains},
  author={Shibata, Naoyuki and Yoshioka, Nobuyuki and Katsura, Hosho},
  journal={Physical Review Letters},
  volume={124},
  number={18},
  pages={180604},
  year={2020},
  publisher={APS}
}

@article{dahlqvist1990existence,
  title={Existence of stable orbits in the x 2 y 2 potential},
  author={Dahlqvist, Per and Russberg, Gunnar},
  journal={Physical review letters},
  volume={65},
  number={23},
  pages={2837},
  year={1990},
  publisher={APS}
}

@article{marcinek1994yang,
  title={Yang-Mills classical mechanics revisited},
  author={Marcinek, Roman and Pollak, Eli and Zakrzewski, Jakub},
  journal={Physics Letters B},
  volume={327},
  number={1-2},
  pages={67--69},
  year={1994},
  publisher={Elsevier}
}

@article{santhanam1998chaos,
  title={Chaos and exponentially localized eigenstates in smooth Hamiltonian systems},
  author={Santhanam, MS and Sheorey, VB and Lakshminarayan, A},
  journal={Physical Review E},
  volume={57},
  number={1},
  pages={345},
  year={1998},
  publisher={APS}
}

@article{pullen1981comparison,
  title={Comparison of classical and quantum spectra for a totally bound potential},
  author={Pullen, RA and Edmonds, AR},
  journal={Journal of Physics A: Mathematical and General},
  volume={14},
  number={12},
  pages={L477--L484},
  year={1981}
}

@book{bir1994chaos,
  title={Chaos and gauge field theory},
  author={Bir, TS and Matinyan, Sergei G and others},
  volume={56},
  year={1994},
  publisher={World Scientific}
}

@article{Hashimoto:2017oit,
    author = "Hashimoto, Koji and Murata, Keiju and Yoshii, Ryosuke",
    title = "{Out-of-time-order correlators in quantum mechanics}",
    eprint = "1703.09435",
    archivePrefix = "arXiv",
    primaryClass = "hep-th",
    doi = "10.1007/JHEP10(2017)138",
    journal = "JHEP",
    volume = "10",
    pages = "138",
    year = "2017"
}

@article{Hashimoto:2020xfr,
    author = "Hashimoto, Koji and Huh, Kyoung-Bum and Kim, Keun-Young and Watanabe, Ryota",
    title = "{Exponential growth of out-of-time-order correlator without chaos: inverted harmonic oscillator}",
    eprint = "2007.04746",
    archivePrefix = "arXiv",
    primaryClass = "hep-th",
    reportNumber = "OU-HET-1064",
    doi = "10.1007/JHEP11(2020)068",
    journal = "JHEP",
    volume = "11",
    pages = "068",
    year = "2020"
}

@article{hashimoto2026holography,
  title={Holography and Optimal Transport: Emergent Wasserstein Spacetime in Harmonic Oscillator, SYK and Krylov Complexity},
  author={Hashimoto, Koji and Tanahashi, Norihiro},
  journal={arXiv preprint arXiv:2604.17649},
  year={2026}
}

@article{akutagawa2020out,
  title={Out-of-time-order correlator in coupled harmonic oscillators},
  author={Akutagawa, Tetsuya and Hashimoto, Koji and Sasaki, Toshiaki and Watanabe, Ryota},
  journal={Journal of High Energy Physics},
  volume={2020},
  number={8},
  pages={13},
  year={2020},
  publisher={Springer}
}

@article{ali2019time,
  title={Time evolution of complexity: a critique of three methods},
  author={Ali, Tibra and Bhattacharyya, Arpan and Haque, S Shajidul and Kim, Eugene H and Moynihan, Nathan},
  journal={Journal of High Energy Physics},
  volume={2019},
  number={4},
  pages={1--43},
  year={2019},
  publisher={Springer}
}

@article{fefferman2016testing,
  title={Testing the manifold hypothesis},
  author={Fefferman, Charles and Mitter, Sanjoy and Narayanan, Hariharan},
  journal={Journal of the American Mathematical Society},
  volume={29},
  number={4},
  pages={983--1049},
  year={2016}
}

@article{bengio2013representation,
  title={Representation learning: A review and new perspectives},
  author={Bengio, Yoshua and Courville, Aaron and Vincent, Pascal},
  journal={IEEE transactions on pattern analysis and machine intelligence},
  volume={35},
  number={8},
  pages={1798--1828},
  year={2013},
  publisher={IEEE}
}

@article{husimi1940some,
  title={Some formal properties of the density matrix},
  author={Husimi, K{\^o}di},
  journal={Proceedings of the Physico-Mathematical Society of Japan. 3rd Series},
  volume={22},
  number={4},
  pages={264--314},
  year={1940},
  publisher={The Physical Society of Japan, The Mathematical Society of Japan}
}

@article{hooft1993dimensional,
  title={Dimensional reduction in quantum gravity},
  author={Hooft, Gerardt},
  journal={arXiv preprint gr-qc/9310026},
  year={1993}
}

@article{Susskind:1994vu,
    author = "Susskind, Leonard",
    title = "{The World as a hologram}",
    eprint = "hep-th/9409089",
    archivePrefix = "arXiv",
    reportNumber = "SU-ITP-94-33",
    doi = "10.1063/1.531249",
    journal = "J. Math. Phys.",
    volume = "36",
    pages = "6377--6396",
    year = "1995"
}

@book{villani2008optimal,
  title={Optimal transport: old and new},
  author={Villani, C{\'e}dric and others},
  volume={338},
  year={2008},
  publisher={Springer}
}

@article{Maldacena:1997re,
    author = "Maldacena, Juan Martin",
    title = "{The Large N limit of superconformal field theories and supergravity}",
    eprint = "hep-th/9711200",
    archivePrefix = "arXiv",
    reportNumber = "HUTP-97-A097, HUTP-98-A097",
    doi = "10.4310/ATMP.1998.v2.n2.a1",
    journal = "Adv. Theor. Math. Phys.",
    volume = "2",
    pages = "231--252",
    year = "1998"
}

@inproceedings{cuturi2013sinkhorn,
  title     = {Sinkhorn distances: Lightspeed computation of optimal transport},
  author    = {Cuturi, Marco},
  booktitle = {Advances in Neural Information Processing Systems},
  volume    = {26},
  year      = {2013}
}

\end{document}